\def\version{February 24, 2003}

\documentclass[12pt]{article}
\usepackage{latexsym}
\usepackage{amsmath}
\usepackage{amsfonts}
\usepackage{amssymb}
\usepackage{epsfig}
\usepackage{verbatim}

\newtheorem{lemma}{Lemma}[section]
\newtheorem{theorem}{Theorem}[section]
\newtheorem{maintheorem}{Theorem}

\newtheorem{corollary}{Corollary}[section]

\newtheorem{remark}{Remark}[section]

\numberwithin{equation}{section}

\def\pr{\Bbb P}
\def\ex{\Bbb E}
\def\I{\Bbb I}
\def\var{\text{Var}}
\def\a{\alpha}
\def\be{\beta}

\def\sig{\sigma}

\def\eps{\varepsilon}
\def\k{\kappa}
\def\la{\lambda}
\def\bss{\boldsymbol{\sigma}}

\def\qed{$\blacksquare$}

\newcommand{\bigo}[1]{{O}(#1)}

\def\bst{\boldsymbol{\tau}}

\begin{document}
\begin{titlepage}
\title{Phase Diagram for the Constrained Integer Partitioning Problem.}

\author{C.~Borgs%
\thanks{Microsoft Research, 1 Microsoft Way, Redmond, WA 98052}
\and J.~T.~Chayes$^*$
\and S.~Mertens%
\thanks{Institut f\"ur Theoretische Physik,
Otto-von-Guericke Universit\"at, D-39016 Magdeburg, Germany} \and
B.~Pittel%
\thanks{Department of Mathematics, Ohio State
University, Columbus, Ohio {\rm 43210}}
\thanks{Research of B. Pittel supported by
Microsoft during his visit in March-June, 2002, and by the NSF in
July- December, 2002} }
\date{\version}
\maketitle \thispagestyle{empty}

\begin{abstract}
{\footnotesize We consider the problem of partitioning $n$
integers into two subsets of given cardinalities such that the
discrepancy, the absolute value of the difference of their sums,
is minimized. The integers are i.i.d.~random variables chosen
uniformly from the set $\{1,\dots,M\}$. We study how the typical
behavior of the optimal partition depends on $n,M$ and the bias
$s$, the difference between the cardinalities of the two subsets
in the partition. In particular, we rigorously establish this
typical behavior as a function of the two parameters
$\kappa:=n^{-1}\log_2M$ and $b:=|s|/n$ by proving the existence of
three distinct ``phases'' in the $\kappa b$-plane, characterized
by the value of the discrepancy and the number of optimal
solutions:  a ``perfect phase'' with exponentially many optimal
solutions with discrepancy $0$ or $1$; a ``hard phase'' with
minimal discrepancy of order $Me^{-\Theta(n)}$; and a ``sorted
phase'' with an unique optimal partition of order $Mn$, obtained
by putting the $(s+n)/2$ smallest integers in one subset. Our
phase diagram covers all but a relatively small region in the
$\kappa b$-plane. We also show that the three phases can be
alternatively characterized by the number of basis solutions of
the associated linear programming problem, and by the fraction of
these basis solutions whose $\pm 1$-valued components form optimal
integer partitions of the subproblem with the corresponding
weights. We show in particular that this fraction is
one
in the sorted phase, and
exponentially small in both the perfect and hard phases,
and strictly exponentially smaller in the
hard phase than in the perfect phase.
Open problems are
discussed, and numerical experiments are presented.}
\end{abstract}

\end{titlepage}

\bibliographystyle{plain}
\section{Introduction}
\label{sec:Intro} Phase transitions in random combinatorial
problems have been the subject of much recent attention.
The random optimum partitioning problem is the only NP-hard
problem for which the existence of a sharp phase transition has
been rigorously established, as have many detailed properties of
the transition (\cite{BCP2}, see \cite{BCP1} for a short
overview). Here we study a constrained version of the random
optimum partitioning problem, and extend some of the results of
\cite{BCP2} to that case.

The integer optimum partitioning problem is a classic problem of
combinatorial optimization in which a given set of $n$ integers is
partitioned into two subsets in order to minimize the absolute
value of the difference between the sum of the integers in the two
subsets, the so-called {\it discrepancy}. Notice that for any
given set of integers, the discrepancies of all partitions have
the same parity, namely that of the sum of the $n$ integers.  We
call a partition {\em perfect} if its discrepancy is $0$, when
this sum is even, or $1$, when this sum is odd. The decision
question is whether there exists a perfect partition.  In the
uniformly random version, an instance is a given a set of $n$
i.i.d.~integers drawn uniformly at random from $\{1, 2, \dots,
M\}$.  We will sometimes use the notation $m=\log_2M$; notice that
each of the random integers has $m$ binary bits.  Previous work
had established a sharp transition as a function of the parameter
$\kappa:=m/n$, characterized by a dramatic change in the
probability of a perfect partition.   For $M$ and $n$ tending to
infinity in the limiting ratio $\kappa = m/n$, the probability of
a perfect partition tends to  $0$ for $ \kappa < 1$, while the
probability tends to $1$ for $\kappa > 1$. This result was
suggested by the work of one of the authors \cite{Mer1} and proved
in a paper by the three other authors \cite{BCP2}. See \cite{Hay}
for a beautiful introduction to the optimum partitioning phase
transition and some of its properties.

Here we consider a constrained variant of the problem in which we
require that the two subsets have given cardinalities; we say that
the difference of the two cardinalities is the {\it bias}, $s$, of
the partition. We establish the phase diagram of the random
constrained integer partitioning problem as a function of the two
parameters $\kappa:=m/n$ and $b:=|s|/n$.  In the language of
statistical physics, $b$ would be called the magnetization, and
the problem considered here, where $b$ is constrained to assume a
particular value, would be called the ``microcanonical'' integer
partitioning problem.  Microcanonical problems are known to be
much more difficult than their unconstrained analogues,
particularly in the case of random systems.

Let us first review previous rigorous and nonrigorous work on the
random optimum partitioning problem. A good deal of rigorous work
has been done for the unconstrained random partitioning problem
with random numbers drawn from a compact interval in $\Bbb R$,
which can be interpreted informally as the limiting case of $m\gg
n$. Karmarkar and Karp \cite{KK} gave a linear time algorithm for
a suboptimal solution with a typical discrepancy of order at most
$O(n^{-c\log n})$ for some constant $c >0$. Confirming a
conjecture by Karmakar and Karp, Yakir \cite{Y} proved that the
expected discrepancy delivered by this algorithm is indeed
$n^{-\theta(\log n)}$. The optimum solution was studied by
Karmarkar, Karp, Lueker and Odlyzko \cite{KKLO} who proved that
the typical minimum discrepancy is much smaller, with the median
 of order $\theta(2^{-n}{n^{1/2}} )$.  More recently, Lueker
\cite{Lue} proved exponential bounds for the expected minimum
discrepancy. Loosely speaking, these results correspond to $m$ far
exceeding $n$, and hence $\kappa\to\infty$, thus well above the
phase transition of the unconstrained problem which occurs for
$\kappa = 1$.

There have also been (nonrigorous) studies of optimum partitioning
in the theoretical physics and artificial intelligence
communities, where the possibility of a phase transition was
examined.  Fu \cite{Fu} noted that the minimum discrepancy is
analogous to the ground state energy of an infinite-range, random
antiferromagnetic spin model, but concluded incorrectly that the
model did not have a phase transition. Gent and Walsh \cite{GW}
studied the problem numerically and introduced the parameter
$\kappa = m/n$.  They noticed that the number of perfect
partitions falls off dramatically at a transition point estimated
to be close to $\kappa = 0.96$. Ferreira and Fontanari studied the
random spin model of Fu, and used statistical mechanical methods
to get estimates of the optimum partition \cite{FF1} and to
evaluate the average performance of simple heuristics \cite{FF2}.
Ferreira and Fontanari \cite{FF1} also considered the constrained
optimum partitioning problem, noted that the constrained problem
is analogous to putting the random antiferromagnet in an external
field, and observed that the problem becomes much easier when the
bias parameter $b$ satisfies $b>\sqrt{2}-1$.

Returning to the unconstrained problem, one of the authors of this
paper used statistical mechanical methods and the parameterization
of Gent and Walsh to derive a compelling, but nonrigorous argument
for a phase transition at $\kappa = 1$, and also derived many of
the properties of the transition \cite{Mer1}. In a later work,
this author \cite{Mer2} analyzed Fu's model by using a heuristic
approximation known in statistical mechanics as Derrida's random
energy model \cite{Der}, and obtained the limiting distribution of
the $k$-th smallest discrepancy.

Motivated by the statistical mechanics analysis in \cite{Mer1} and
\cite{Mer2}, the other three authors of this paper undertook an
extensive rigorous study of the random integer partitioning
problem \cite{BCP2}. They established the existence of a
transition at $\kappa_c=1$ below which the probability of a
perfect partition tends to one with $n$ and $m$, and above which
it tends to zero, and also gave the finite-size scaling window of
the transition: namely, in terms of the more detailed
parametrization $m=\kappa_n$ with $ \kappa_n = 1 - \log_2 n/(2n) +
\lambda_n/n, $ the probability of a perfect partition tends to
$1$, $0$, or a computable $\lambda$-dependent constant strictly
between $0$ and $1$, depending on whether $\lambda_n$ tends to
$-\infty$, $\infty$, or $\lambda \in (-\infty,\infty)$,
respectively. The work also calculated the distribution of the
number of perfect partitions, the distribution of the minimum
discrepancy, and the joint distribution of the $k$ smallest
discrepancies, which give the entropy, the ground state energy and
the bottom of the energy spectrum, respectively.  In particular,
the paper  \cite{BCP2} provided a rigorous justification of the
Derrida-type approximation both inside and above the scaling
window, insofar as the joint distribution of $k$ (finite) smallest
discrepancies is concerned.

The location of the phase transition for the unconstrained problem
immediately yields a one-dimensional phase diagram as a function
of $\kappa$: For $\kappa \in (0,\kappa_c)$ with $\kappa_c = 1$,
the system is in a ``perfect phase'' in which the probability of a
perfect partition tends to $1$ as $M$ and $n$ tend to infinity in
the fixed function $\kappa$.  For $\kappa \in (\kappa_c,\infty)$,
the probability of a perfect partition tends to $0$, and moreover,
there is an unique optimal partition.  We call this the ``hard
phase,'' since for $\kappa > \kappa_c$, it is presumably
computationally difficult to find the optimal partition.

\begin{figure}[htb]
  \begin{center}
 \includegraphics[width=0.8\columnwidth]{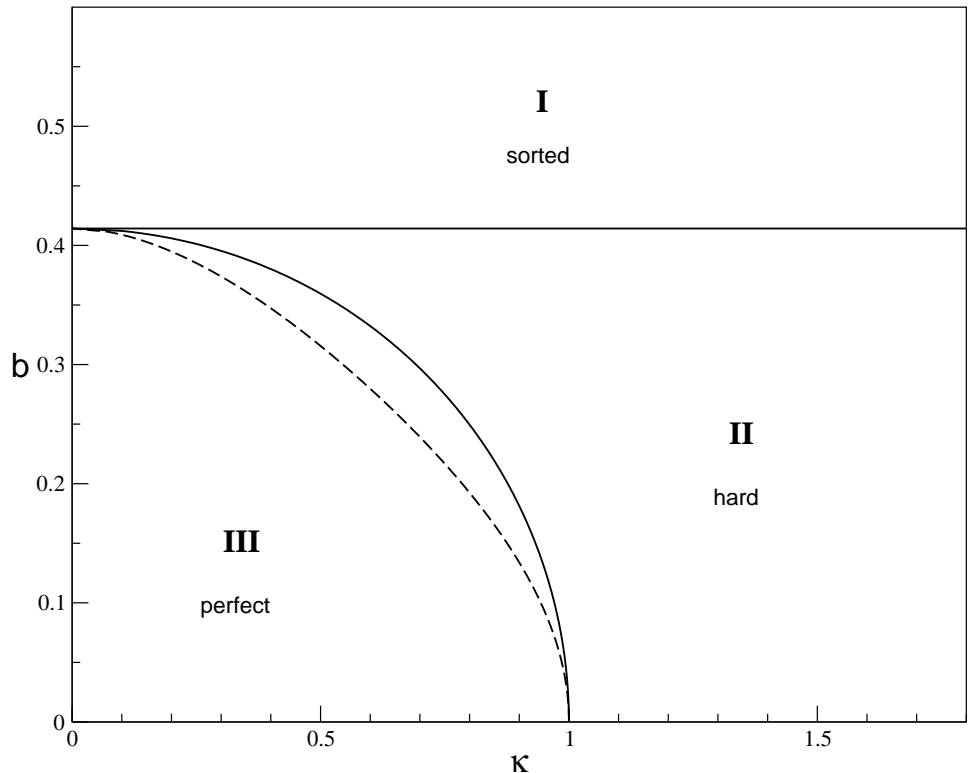} 
  \caption{Phase diagram of the constrained integer partitioning
            problem.
  \label{fig:phasediagram}}
  \end{center}
\end{figure}

In this work, we consider the constrained optimum partitioning
problem with bias $s$ and extend the phase diagram to the
two-dimensional $\kappa b$-plane.  See Figure
\ref{fig:phasediagram}. In addition to the extensions of the
perfect and hard phases, we establish the existence of a new phase
which we call the ``sorted phase.''

The sorted phase is easy to understand.  One way to meet the bias
constraint is to take the $(s+n)/2$ smallest integers and put them
in one subset of the partition.\footnote{Note that the task of
finding this partition is even easier than the task of sorting the
$n$ integers, which would take, on average, $\theta(n\log n)$
comparisons. Instead, the $(s+n)/2$ smallest integers can be found
in strictly linear time in $n$.} It is not difficult to see that
the resulting ``sorted partition'' is optimal if the total weight
of this subset is at least half of the sum of all $n$ integers.
We define the sorted phase as the subset of the $\kappa b$-plane
where the sorted partition is optimal. We prove that the sorted
phase is given by the condition
\begin{equation}
b > b_c :=\sqrt{2}-1,
\end{equation}
see region III in Figure \ref{fig:phasediagram}. Moreover, we show
that the minimal discrepancy in this phase is of the order $Mn$.
The region $b >\sqrt{2}-1$ is precisely where Ferreira and
Fontanari \cite{FF1} observed that the corresponding statistical
mechanical problem becomes ``self-averaging.''

Our analysis of the perfect and hard phases for $b< b_c$ is much
more difficult.  In this region, we use integral representations
for the number of partitions with a given discrepancy and bias;
these representations generalize those used in \cite{BCP2}.  The
asymptotic analysis of the resulting two-dimensional random
integrals leads to saddle point equations for a saddle point
described in terms two real parameters $\eta$ and $\zeta$. For
discrepancies of order $o(M)$  (including, in particular, the case
of perfect partitions), the saddle point equations determining
$\zeta$ and $\eta$ are:
\begin{equation}
\begin{aligned}
&\int\limits_0^1x\tanh(\zeta x+\eta)\,dx=0,\\
&\int\limits_0^1\tanh(\zeta x+\eta)\,dx=-b.
\end{aligned}
\label{saddlepoint}
\end{equation}
The solution $(\zeta,\eta)$ of these equations can be used to
define the
two
convex curves in Figure \ref{fig:phasediagram}.
To this end, let%
\footnote{It turns out the solutions of the saddle point equations
\ref{saddlepoint} are just the stationary points of the function
$L(\zeta,\eta)$}
\begin{equation}
L(\zeta,\eta):=b\eta+\int\limits_0^1\log(2\cosh(\zeta x+\eta))\,dx
\label{L}
\end{equation}
\begin{equation}
\rho(\zeta,\eta) :=1-\frac{\tanh(\zeta+\eta)-\tanh(\eta)}
{2\zeta}. \label{rho-new}
\end{equation}
For $(\zeta,\eta)$ a solution of \eqref{saddlepoint}, we then
define
\begin{equation}
\begin{aligned}
\kappa_-(b):&=-\log_2\rho(\zeta,\eta),\\
\kappa_c(b):&={\textstyle\frac 1{ \log 2}}L(\zeta,\eta).\\
\end{aligned}
\label{kappac}
\end{equation}
From bottom to top, the
two convex curves joining $(0,b_c)$ and
$(1,0)$ in Figure \ref{fig:phasediagram} are then given by $\kappa
= \kappa_-(b)$ and $\kappa = \kappa_c(b)$.

In this paper, we prove that, in the region $\kappa<\kappa_-(b)$,
with probability
tending to one as $n$ tends to infinity (or, more succinctly, with high
probability, w.h.p.) there exist
perfect partitions; see region I in Figure \ref{fig:phasediagram}.
Moreover the number of perfect partitions is about $2^{(\kappa_c -
\kappa)n}$ in this ``perfect phase.'' We also prove that
w.h.p.~there are no perfect partitions in the region $b < b_c$ and
$\kappa > \kappa_c(b)$.
As in the unconstrained problem, we call
this the ``hard phase.'' Our results leave open the question of
what happens in the narrow region
$\kappa_-<\kappa<\kappa_c$,
and also whether the optimal partition is unique in the hard phase;
see the final section for a discussion of this and other open
questions.

We are also able to prove that these phase transitions correspond
to qualitative changes in the solution space of the associated
linear programming problem (LPP).  In the actual optimum
partitioning problem, each integer is put in one subset or the
other.  The relaxed version is defined by allowing any fraction of
each integer to be put in either of the two partitions.  Using our
theorems on the typical behavior of integer partitioning problem
and some general properties of the LPP, we show the following.  In
the sorted phase, i.e.~for $b > b_c = \sqrt 2 - 1$, w.h.p.~the LPP
has a unique solution given by the sorted partition itself. For $b
< b_c$,~i.e. in the perfect and hard phases, w.h.p.~the relaxed
minimum discrepancy is zero, and the total number of optimal basis
solutions is exponentially large, of order
$2^{k_c(b)n+O_p({n^{1/2}})}$.
Finally, in the perfect and hard phases, we consider the fraction
of these basis solutions whose integer-valued components form an
optimal integer partition of the subproblem with the
corresponding subset of the weights.  We show that this fraction
is exponentially small.
Moreover, except for the crescent-shaped region between
$\kappa=\kappa_-(b)$ and $\kappa=\kappa_c(b)$, we show that the
fraction is strictly exponentially smaller in the hard phase than
in the perfect phase. This fraction thus represents some measure
of the algorithmic difficulty of the problem, see Remark
\ref{rem8.1}.


The outline of the paper is as follows.  In the next section, we
define the problem in detail, and precisely state our main
results. In Section~\ref{sec:proof-strategy}, we introduce our
integral representation and show how it leads to the relevant
saddle point equations.  We also give a brief heuristic derivation
of some of the phase boundaries. Section~\ref{sec:saddle-point}
contains a proof of existence and properties of the solution of
the saddle point equations for $b<b_c$.  In
Section~\ref{sec:perfect}, we establish an asymptotic formula for
the number of partitions with given discrepancy and bias in the
perfect phase. As a corollary, we obtain both the existence of
exponentially many perfect partitions for $\kappa<\kappa_-(b)$,
and a theorem on the distribution of the bias in the unconstrained
problem for $\kappa<1$. The analysis of the hard phase is done in
Section~\ref{sec:hard}.  In Section~\ref{sec:sorted}, we prove
that the sorted partitions are optimal for $b>b_c$.  We also show
why the sorted phase boundary coincides with the boundary for
existence of solutions of the saddle point equations.  In
Section~\ref{sec:LP}, we formulate the relaxed version of the
optimum partitioning problem, and establish our results on the
space of basis solutions of the LPP.  Finally, in
Section~\ref{sec:openprob}, we discuss open problems and a few
numerical experiments addressing some of these problems.

\section{Statement of Main Results}
\label{sec:stat-res}

Let $X_1,\dots,X_n$ be $n$ independent copies of a generic random
variable which is distributed uniformly on $\{1,\dots,M\}$. We are
interested in the case when $M$ grows exponentially with $n$, and
define $\kappa$ as the exponential rate, i.e.
\begin{equation}
\kappa=\frac 1n \log_2M.
\end{equation}
To avoid trivial counterexamples, we will always assume that
$\kappa$ stay bounded away from both $0$ and $\infty$ as
$n\to\infty$.
We will use
$\pr$ and $\ex$, with or without subindex $n$, to denote the
probability measure and the expectation induced by $\bold X=(X_1,
\dots, X_n)$.

A partition of integers into two disjoint subsets is coded by an
$n$-long binary sequence
$\bss=(\sig_1,\dots,\sig_n),\,\sig_j\in\{-1,1\}$; so the subsets
are $\{j:\sig_j=1\}$ and $\{j:\sig_j=-1\}$. Obviously $\bss$ and
$-\bss$ are the codes of the same partition.  Given a partition
$\bss$, we define its {\it discrepancy\/} (or energy), $d(\bold
X,\bss)$, and {\it bias\/} (or magnetization), $s(\bss)$, as
\begin{align}
d(\bold X,\bss)=&|\bss\cdot\bold X|, \text{ with } \bss\cdot\bold
X=\sum_{j=1}^n\sig_jX_j, \label{discr}
\\
s(\bss)=&\,\bss\cdot\bold e =|\{j:\sig_j=1\}|-|\{j:\sig_j=-1\}|.
\label{imbal}
\end{align}
Here $\bold e$ is the vector $(1,\dots,1)$. Clearly
$s(\bss)$ is an integer in $\{-n,\dots,n\}$,
so let $s\in\{-n,\dots,n\}$ and
define the bias density
\begin{equation}
b=\frac {|s|}{n}
\end{equation}
so that $b\in [0,1]$. Note that by symmetry it suffices to
consider
$s(\bss)\in\{0,\dots,n\}$,
so we will often take a
non-negative integer $s\in\{0,\dots,n\}$,
in which case $s = bn$. We define an {\it optimum partition}
as a partition $\bss$ that minimizes the discrepancy $d(\bold
X,\bss)$ among all the partitions with bias equal to $s$, and a
{\it perfect partition} as a partition $\bss$ with $|d(\bold
X,\bss)|\leq 1$.

Theorems~\ref{thm1}, \ref{thm2} and \ref{thm3} below  describe
our main results on the phases labelled I, II, and III in Figure 1
in the introduction.  In the statement of these theorems we will
use the parameters $\zeta,\eta,\kappa_c(b)$ and
$\kappa_-(b)$ defined in \eqref{saddlepoint} --
\eqref{kappac}. Before getting to principal results, we must begin
with an existence statement for the parameters $\zeta,\eta$.

\begin{maintheorem}\notag
\label{thm0} Let $b<b_c$, where $b_c=\sqrt{2}-1$. Then the saddle
point equations \eqref{saddlepoint} have a unique solution
$(\zeta,\eta)=(\zeta(b),\eta(b))$.
\end{maintheorem}
This theorem is proved in Section~\ref{sec:saddle-point}.

Let
\begin{equation}
Z_n(\ell,s)=Z_n(\ell,s;\bold X)
\end{equation}
denote the random number of partitions $\bss$ with $\bss\cdot\bold
X=\ell$ and $\bss\cdot\bold e=s$.  Since $s(\bss)$ has the same
parity as $n$, and $d(\bold X,\bss)$ has the same parity as
$\sum_{j=1}^nX_j$, we will only consider values of $s$ which have
the same parity as $n$, and values of $\ell$ which have the same
parity as $\sum_{j=1}^nX_j$.  In the theorems in this section and
in much of the rest of the paper, we will not state these
restrictions explicitly.

Our central goal is to use the saddle point solution in order to
bound the $Z_n(\ell,s)$ for various given values of $\ell$ and
$s$. To formulate our results in a compact, yet unambiguous form,
we use a shorthand $a_n<a$ ($a_n>a$, resp.) instead of $\limsup
a_n<a$ ($\liminf a_n>a$, resp.), even when the $n$-dependence of
$a_n$ is only implicit, as in $\kappa=n^{-1}\log_2 M$ and
$b=|s/n|$.  We will also use
the notation $f_n=O_p(g_n)$ and $f_n=o_p(h_n)$ if $f_n/g_n$ is
bounded in probability and $f_n/h_n$ goes to zero in probability,
respectively. Also, as is customary, we will say that an event
happens with high probability (w.h.p.) if the probability of this
event approaches $1$ as $n\to\infty$. In all our statements $n$,
$M$, $s$ and $\ell$ will be integers with $n\geq 1$, $M\geq 1$ and
$s\geq 0$. Our main results in the perfect phase are summarized in
the next theorem and remark.
\begin{maintheorem}
\label{thm1} Let $\ell=o(Mn^{1/2}), b<b_c$ and
$\kappa<\kappa_-(b)$. Then w.h.p.~$Z_n(\ell,s)\ge 1$ and
\begin{equation}
Z_n(\ell,s)=2^{[\kappa_c(b)-\kappa]n}
e^{S_n {n^{1/2}}+o({n^{1/2}})},
\label{Z-asymp}
\end{equation}
where $S_n$ converges in probability to a Gaussian with mean zero
and variance $\sigma^2= \var(\log(2\cosh(\zeta U+\eta)))$, with
$U$ uniformly distributed on $[0,1]$. Consequently, w.h.p., there
exist exponentially many perfect partitions, with $\ell=0$ if
$\sum_jX_j$ is even, and $|\ell|=1$ if $\sum_jX_j$ is odd.
\end{maintheorem}
\begin{remark}
\label{REM1}
Under the conditions of Theorem~\ref{thm1}, we
actually prove a much more accurate estimate.
Namely, we show that there
are $2\times 2$ positive definite matrices
$R$ and $K$
with deterministic entries,
and a constant
$q<1$ such that, with probability $1-O(q^{\log^2 n})$,
\begin{equation}
Z_n(\ell,s)=\exp\left(\zeta\frac{\ell}{M}+\eta
s+\sum_{j=1}^n\log(2 \cosh(\zeta X_j/M+\eta))\right)
\frac{\exp(-\frac{1}{4}\boldsymbol\tau_n
R^{-1}\boldsymbol\tau^\prime_n)} {\pi Mn\sqrt{\mbox{det
}R}}(1+o(1)) .\label{Znls}
\end{equation}
Here $\boldsymbol\tau_n$ is a two-dimensional random vector which
converges in probability to a Gaussian vector $\boldsymbol \tau$
with zero mean and covariance matrix $K$.  See
Theorem~\ref{thm5.1} in Section~\ref{sec:asymptotic-Z_n}.

We also prove a corollary relating the distribution of the
bias in the unconstrained problem to the distribution of the
bias between heads and tails in fair coin flips;
see Subsection~\ref{sec:BiasDist}.
\end{remark}
The proof of Theorem~\ref{thm1} and Remark~\ref{REM1}
is given in Section~\ref{sec:perfect}.

Note that the above expression for $Z_n(\ell,s)$ is much more
complicated than its analogue in the unconstrained case, see
equation (2.6) in \cite{BCP2}.  Both the sum in the first
exponent and the entire second exponent represent fluctuations
which were not present in the unconstrained case, and which make
the analysis of the perfect phase much more difficult here; see
also Remark~\ref{REM4} below.

Our next theorem, which describes our main results on the hard
phase, has two parts:  The first shows that there are no perfect
partitions above $\kappa = \kappa_c(b)$, and the second gives a
bound on the number of optimum partition for $\kappa >\kappa_-$.
%
%
To state
the theorem, let $d_{opt}=d_{opt}(n;s)$ denote the
discrepancy of the optimal partition,
and let $Z_{opt}=Z_{opt}(n;s)$ denote the number
of optimal partitions.

\begin{maintheorem}
\label{thm2} Let $b<b_c$.
\begin{enumerate}
\item If $\kappa>\kappa_c(b)$, then
there exists a $\delta>0$ such that
with probability $1-O(e^{-\delta\log^2 n})$
there are no perfect partitions, and moreover
\begin{equation}
\label{dmin>}
d_{opt} \ge
2^{n[\kappa-\kappa_c(b)]-O_p(n^{1/2})}.
\end{equation}

\item If $\kappa>\kappa_-(b)$
and $\eps>0$, then there exists a constant
$\delta>0$ such that
\begin{equation}
\label{dmin<}
d_{opt} \le
2^{n[\kappa-\kappa_-(b)+\eps]},
\end{equation}
and
\begin{equation}
\label{Zopt}
Z_{opt}\leq 2^{n[\kappa_c(b)-\kappa_-(b)+\eps]},
\end{equation}
both with probability
$1-O(e^{-\delta\log^2 n})$.
\end{enumerate}
\end{maintheorem}
This theorem is proved in Section~\ref{sec:hard}. However,
perhaps somewhat surprisingly, the proof of the upper bound in
\eqref{dmin<} runs parallel to the proof of Theorem~\ref{thm1}
that established existence of perfect partitions for
$\kappa<\kappa_-(b)$.

\begin{remark}
\label{REM2} We believe that the bound in
\eqref{dmin>} is actually sharp.
If we {\em assume} that this is the case,
in fact, even if we assume that the
weaker bound
\begin{equation}
\label{dmin-conj}
d_{opt} =
2^{n(\kappa-\kappa_c+o_p(1))}
\end{equation}
holds w.h.p.~whenever $\kappa>\kappa_c$, then we can significantly
improve the upper bound \eqref{Zopt}. Indeed, under the assumption
\eqref{dmin-conj}, $Z_{opt}$ grows subexponentially with
$n$ whenever $\kappa>\kappa_c(b)$, see Remark~\ref{rem6.3}~(iii).
\end{remark}

The optimum partition problem is much simpler for $b>b_c$.  Our
main result on the sorted phase is the following theorem,
which is proved in Section~\ref{sec:sorted}.
\begin{maintheorem}
\label{thm3} Let $b>b_c$. Then w.h.p.~the optimal partition is
uniquely obtained by putting $(s+n)/2$ smallest integers $X_j$ in
one part, and the remaining $(n-s)/2$ integers into another part.
W.h.p., $d_{opt}$ is asymptotic to
$\frac{Mn}{4}\bigl[(1+b)^2-2\bigr]$, i.e., of order $Mn$.
\end{maintheorem}
By this theorem, for $b$ sufficiently large, the partition is
determined by the decreasing order of weights $X_j$, but not by
the actual values of $X_j$.

Until now, all our statements have been about the likely
properties of the optimum partition $\bss$, whose components are
allowed to assume only two values, $-1$ and $+1$. In an
interesting twist, these results shed light on the likely
properties of the related linear programming problem (LPP): find
the minimum value
of $d$ subject to linear constraints
\begin{eqnarray*}
&-d\le\sum_{j=1}^n\sig_jX_j,\quad\sum_{j=1}^n\sig_jX_j\le d,\\
&\sum_{j=1}^n\sig_j=s,\\
&-1\le\sig_j\le 1,\quad (1\le j\le n).
\end{eqnarray*}
We denote this minimum value of $d$ by $d^{LPP}_{opt}$.
Our last theorem indicates that the LPP inherits the phase diagram
of the optimum partition problem, and moreover provides, however
incomplete, some way to rate the three regions according the
algorithmic difficulty of the optimal partition problem. To make
this precise, we define $F_n(\kappa,b)$  to be the fraction of
basis solutions $\bss$ with the property that the $\pm 1$-valued
components $\sig_i$ form an optimal partition of the corresponding
subproblem with weights $X_i$.  Henceforth, we will call this
the ``optimal subpartition property.''
\begin{maintheorem}
\label{thm4}
\begin{enumerate}
\item \label{thm4a} If $b>b_c$, then w.h.p.~the sorted partition
is a unique solution of the LPP, and thus
$d^{LPP}_{opt}=\Theta(Mn)$ and $F_n(\kappa,b)=1$.

Let $b<b_c$.
\item \label{thm4b} Then w.h.p.~$d^{LPP}_{opt}=0$.  In
addition, w.h.p.~there are
$2^{[\kappa_c(b)+o(1)]n}$
basis
solutions, each having either none or exactly two components
$\sigma_i \neq \pm 1$.
\item\label{thm4c}
W.h.p.~$F_n(\kappa,b) = 2^{-[\kappa +o(1)]n }$
for
$\kappa<\kappa_-(b)$, and
$2^{-[\kappa_c(b)+o(1)]n}\leq
F_n(\kappa,b) \leq  2^{-[\kappa_-(b)+o(1)]n}$ for
$\kappa>\kappa_-(b)$.
\end{enumerate}
\end{maintheorem}
\begin{remark}
\label{REM3} (i) If one assume that the number of optimal partitions
$Z_{opt}$ in the hard phase grows subexponentially
with probability at least $1-o(n^{-2})$ (see Remark~\ref{REM2} for
a motivation of this assumption), our upper bound on the fraction
$F_n(\kappa,b)$ in the hard phase can be improved to match the
lower bound, yielding $F_n(\kappa,b)=2^{-n[\kappa_c(b)+o(1)\,]}$
in the hard phase, see Remark~\ref{rem8.1}~(i).

(ii) If, on the other hand, the asymptotics of
Theorem~\ref{thm5.1} hold up to $\kappa_c$, more precisely, if one
assumes that for $b<b_c$ and $\kappa<\kappa_c(b)$
\begin{equation}
Z_n(\ell,s)=2^{n [\kappa_c(b)-\kappa+o(1))]}
\end{equation}
holds with probability least $1-o(n^{-2})$, then a bound of the
form $F_n(\kappa,b) = 2^{-n[\kappa +o(1)]}$ can be extended to all
$\kappa<\kappa_c$, see Remark~\ref{rem8.1}~(ii).
\end{remark}

We close this section with a few additional remarks and an
additional theorem on the expected number of perfect partitions.
We start with a discussion of Theorem~\ref{thm1}.

\begin{remark}\label{REM4}
While \eqref{Z-asymp} has only been proved for
$\kappa<\kappa_-(b)$, an {\em upper bound} of the same form can be
shown to hold for all $\kappa$, see Theorem~\ref{thm6.1}. Due to
the random fluctuations of the Gaussian term, this upper bound is
of order $e^{-\Theta( {n^{1/2}})}$ with positive probability as soon
as $\kappa \ge \kappa_c(b) - O(n^{-1/2})$, so that in this regime,
there are no perfect partitions with probability bounded away from
zero.  Note, however, that as $b \to 0$, the variance
$\sigma^2= \var(\log(2\cosh(\zeta U+\eta)))$ of the
Gaussian term tends to zero.

We expect that for all $b\in (0,b_c)$
fluctuations of this kind
persist around the true threshold, whether it is actually equal to
$\kappa_c(b)$, or whether it is some other value
$\tilde\kappa_c(b)<\kappa_c(b)$.  For the constrained partition
problem with $b\in (0,b_c)$,
we therefore expect a scaling window of width at least
$n^{-1/2}$ in which the probability of perfect partitions lies
strictly between $0$ and $1$. This is to be contrasted with the
unconstrained case, where we had a very narrow scaling window of
width $\Theta (n^{-1})$ about the transition point $\kappa_c$, see
\cite{BCP2}.

\end{remark}

Next let us  consider the statements of Theorem~\ref{thm2}. Here
again the situation is much more complicated than in the
unconstrained case.  By Theorem~\ref{thm1} and the lower bound in
Theorem~\ref{thm2}a, the minimum discrepancy changes from being at
most one to being exponentially large as $\kappa$ crosses the
interval $[\kappa_-,\kappa_c]$. However, we also prove (see
Section~\ref{sec:expected}) that the {\it expected\/}
number of perfect partitions remains exponentially large until
$\kappa$ reaches a value strictly exceeding $\kappa_c$. This is
the content of the following theorem and remark.

\begin{maintheorem}
\label{thm5} Let $\ell\in\{-1,0,1\}$ and $b \in (0,1)$. Then
\begin{equation}
\lim_{n\to\infty}\left[n^{-1}\log\ex(Z_n(\ell,s))
-R(\kappa,b)\right]=0
\label{1/nlog=R}
\end{equation}
where
\begin{equation}
R(\kappa,b)=H((1+b)/2)+\la b+\log(\la^{-1}\sinh\la)-\kappa\log
2,\label{Rkb=}
\end{equation}
with $H(u)=u\log (1/u)+(1-u)\log(1/(1-u))$, and $\la$ satisfying
$
\coth\la={\la}^{-1}-b.
$
\end{maintheorem}

\begin{figure}[htb]
  \begin{center}
 \includegraphics[width=10cm]{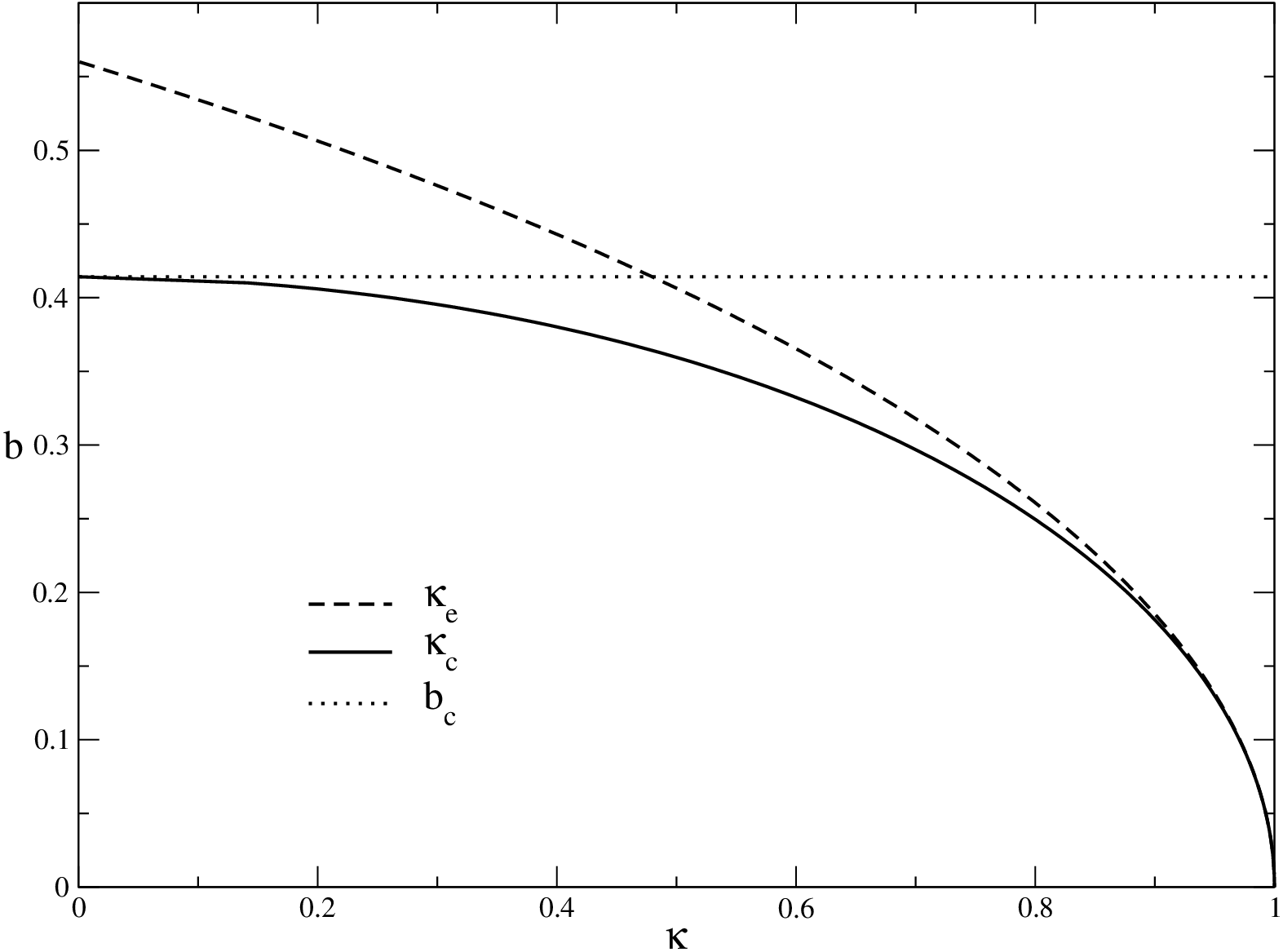}
  \caption{The curves  $\kappa=\kappa_c(b)$,
  $\kappa=\kappa_e(b)$ and $b=b_c$.
  \label{fig:kappa-e}}
  \end{center}
\end{figure}

\begin{remark}\label{REM5}
Graphing the curve $R(\kappa,b)=0$, i.e.
\begin{equation}
\kappa=\kappa_e(b):=\frac{H((1+b)/2)+\la
b+\log(\la^{-1}\sinh\la)}{\log 2},
\end{equation}
we see that it lies strictly above $\kappa=\kappa_c(b)$, except at
the only common point $\kappa=1,b=0$, see Fig.~\ref{fig:kappa-e}.
In particular, the curve intersects the $b$-axis at
$b=0.56\dots>b_c=0.41\dots$. Thus for the points $(\kappa,b)$
between the curves $\kappa=\kappa_c(b)$ and
$\kappa=\kappa_{e}(b)$,
the expected number of perfect partitions grows
exponentially, while w.h.p.~there are no perfect partitions at
all. This complex behavior did not manifest itself in the
unconstrained optimum partitioning problem  \cite{BCP2}.
\end{remark}


We close this section with a discussion of the results of
Theorem~\ref{thm4}.


\begin{remark}
\label{REM6} Clearly, the fraction $[F_n(\kappa,b)]^{-1}$ is the
expected number of times one has to generate a uniformly random
basis solution of the LPP to get a basis solution with the optimal
subpartition property. In absence of a better candidate,
$[F_n(\kappa,b)]^{-1}$ seems to be a possible measure of
algorithmic difficulty of the integer partition problem. Our
theorem says that w.h.p.~this measure is lowest for the
sorted phase ($b>b_c$) where $[F_n(\kappa,b)]^{-1}=1$), next
lowest for the perfect phase ($b<b_c$ and $\kappa<\kappa_-(b)$)
where $[F_n(\kappa,b)]^{-1}=2^{\kappa n+o(n)}$, and indeed hardest
in the hard phase ($b<b_c$ and $\kappa>\kappa_c(b)$) where
$[F_n(\kappa,b)]^{-1}\geq 2^{(\kappa_-(b)+o(1))n}$.

If we make the additional assumptions of Remark~\ref{REM3}, we
have that $[F_n(\kappa,b)]^{-1}=2^{\kappa n+o(n)}$ for $b<b_c$ and
$\kappa<\kappa_c(b)$, while $[F_n(\kappa,b)]^{-1}=2^{\kappa_c
n+o(n)}$ for $\kappa>\kappa_c$.  This gives an easy-hard-easy
picture along any curve $b=b(\kappa)$, $\kappa\ge 0$, that crosses
the lines $\kappa=\kappa_c(b)$ and $b=b_c$, provided that
$b(\kappa)$ is strictly increasing in $\kappa$. Indeed, at a point
$(\kappa,b)$ on such a curve $[F_n]^{-1}$ grows exponentially in
$n\kappa$, as long as $\kappa<\kappa_c(b(\kappa))$, i.e.~until the
curve $b=b(\kappa)$ intersects $\kappa=\kappa_c(b)$. For larger
values of $\kappa$, $[F_n]^{-1}$ grows at the exponential rate
$n\kappa_c(b(\kappa))<n\kappa$. Once the curve $b=b(\kappa)$
crosses the horizontal line $b=b_c$, the problem becomes even
easier with $[F_n]^{-1}=1$.  In fact, above this line, the problem
becomes easy in the usual sense, since the rounded and the
fractional problem are identical, and sorted partitions can be
found in linear time.
\end{remark}

\section{Preliminaries and Outline of Proof Strategy}
\label{sec:proof-strategy}

In this section, we define our notation, review the heuristics of
the proof, and point out why naive extensions of the unconstrained
analysis of \cite{BCP2} fail in the constrained case.

\subsection{Sorted Partitions}

We first discuss our strategy to prove that in region III, the
optimal partition is sorted and has discrepancy of order $Mn$. To
this end, we consider $n$ weights $X_1,\dots,X_n$, chosen
uniformly at random from $\{1,\dots,M\}$, and reorder them in such
a way that their sizes are increasing, $X_{\pi(1)}\leq
X_{\pi(2)}\leq \dots\leq X_{\pi(n)}$, where $\pi(1), \dots,
\pi(n)$ is a suitable permutation of $1,\dots,n$. Since $M$ is
assumed to grow exponentially with $n$, we have, in particular,
$n^2=o(M)$, which implies that w.h.p.~no two weights are equal. So
w.h.p.~the permutation $\pi$ is unique and
$X_{\pi(1)}<X_{\pi(2)}<\cdots< X_{\pi(n)}$.

Given a bias $s>0$, (with $s\equiv n(\mbox{mod }2$)), we need to
find an optimum partition that puts $k=(s+n)/2$ integers in
one part, and the remaining $n-k$ integers into another part. One
such feasible partition is obtained if we select the $k$
smallest integers for the first part; we call it the sorted
partition. It is coded by the $\bss$, with $\sigma_{\pi(i)}=1$ for
$i\le k$ and $\sigma_{\pi(i)}=-1$ for $i>k$. If the total weight
of $(n-k)$ largest weights is, at most, the total weight of $k$
smallest weights, then it is intuitively clear that the sorted
partition is optimal. More precisely: if

\begin{equation}
\delta_s(\boldsymbol X)
=\sum_{j=1}^{k}X_{\pi(j)}-\sum_{j=k+1}^nX_{\pi(j)}\geq 0
\label{3.1}
\end{equation}
then the sorted partition is the unique, optimal partition, and
the minimal discrepancy is
\begin{equation}
d_{opt} = \delta_s(\boldsymbol X). \label{3.2}
\end{equation}
See Section \ref{sec:sorted-partitions} for a formal proof.%
\footnote{If $\delta_s(\boldsymbol X)=-1$, the sorted partition is
still optimal (it is, in fact, perfect). But in general, it is not
the unique optimum partition.}

To determine the phase boundary of the phase III, we thus have to
determine the region of the phase diagram in which w.h.p.~the
sorted partition meets the condition \eqref{3.1}. Leaving the
probabilistic technicalities out of our heuristic discussion, let
us replace the condition \eqref{3.1} by its mean version, namely
$\ex(\delta_s(\mathbf X))\ge 0$. Consider an arbitrary
$b\in(0,1]$. Let $x_0=(1+b)/2$ and $M_0=\lfloor x_0
M\rfloor$. For a typical set of weights $X_1,\dots, X_n$, let us
consider the sorted partition with $\sigma_j=1$ for $X_j\leq M_0$,
and $\sigma_j=-1$ for $X_j> M_0$. Since the probability that
$X_j\leq M_0$ is equal to $\tilde x_0=M_0/M =x_0+O(M^{-1})$, we
get that the expected number of weights $X_j$ with
$X_j\leq M_0$ is $n \tilde x_0$, implying that the expected bias
is $2n\tilde x_0 -n = n b +O(n/M)$. The expected discrepancy can
be calculated in a similar manner, giving the expression
\begin{equation}
\begin{aligned}
\ex\bigg[ \sum_j X_j\I\Big(X_j\leq \lfloor x_0 M\rfloor\Big) &-
\sum_j X_j\I\Big(X_j> \lfloor x_0 M\rfloor\Big) \bigg]
\\
&= \frac nM\bigg[M_0(1+M_0)-\frac{M(1+M)}2\bigg]
\\
&=\Big[x_0^2-\frac 12 + O(M^{-1})\Big]Mn
\\
&=\bigg[\bigg(\frac {b+1}2\bigg)^2-\frac 12+O(M^{-1})\bigg]Mn.
\end{aligned}
\label{3.3}
\end{equation}
So, $\ex(\delta_s(\mathbf X))$ is large positive, of order $Mn$,
iff $(b+1)^2/4-1/2>0$, or equivalently $b>b_c=\sqrt{2}-1$. In
Section~\ref{sec:sorted-partitions} we prove the condition
$b>b_c$ is both necessary and sufficient for $\delta_s(\mathbf X)$
to be, w.h.p., positive, of order $Mn$.  In language of statistical
mechanics, we show that, for $b>b_c$,
$\delta_s(\mathbf X)$ is ``self-averaging,''
i.e.~its distribution is sharply concentrated around
$\ex(\delta_s(\mathbf X))$.

\begin{remark}
\label{rem3.1} On the heuristic level presented here, the above
arguments can easily be generalized to an arbitrary distribution
for the weights $X_1,\dots,X_n$, as long as these weights are
independent copies of a generic (discrete) variable $X$ with a
reasonably well behaved probability distribution.  Assuming, e.g.,
that the variable $X/M$ has a limiting distribution with density
$\mu$, one obtains that the critical value of $b$ is given by
$b_c=b_c(\mu)=2\int_0^{x_0}\mu(x)\,dx -1$, where $x_0$ is
determined by the equation
$\int_0^{x_0}x\mu(x)\,dx=\int_{x_0}^\infty x \mu(x)\, dx$. However,
we have not tried to extend all our results to this more
general $\mu$-density case.
\end{remark}

\subsection{Integral Representations}

Let us now turn to the much more difficult region ${b}<b_c$.
Without loss of generality, we may take $s\geq 0$, so that
$b=s/n$.

Let $Z_n(\ell,s)=Z_n(\ell,s;\bold X)$ denote the total number of
partitions $\bss$ such that $\bss\cdot\bold X=\ell$ and
$\bss\cdot\bold e=s$.  Guided by the results of \cite{BCP2}, one
might hope to prove that, as the parameter $\kappa= n^{-1}\log_2
M$ is varied, the model undergoes a phase transition between a
region with exponentially many perfect partitions and a region
with no perfect partitions.  Since perfect partitions correspond
to $\ell=0$ or $\ell=\pm 1$, we will be mainly interested in
$Z_n(\ell,s)$ for $|\ell|\leq 1$, while $s$ will typically be
chosen proportional to $n$.

 A starting point in \cite{BCP2} was an integral
(Fourier-inversion) type formula for $Z_n(\ell)= Z_n(\ell;\bold
X)$, the total number of $\bss$'s such that $\bss\cdot\bold
X=\ell$, namely
\begin{equation}
Z_n(\ell) =\frac{2^n}{\pi}\int\limits_{x\in (-\pi/2,\pi/2]}
\cos(\ell x)\prod_{j=1}^n \cos(xX_j)\,dx. \label{3.4}
\end{equation}
We need to derive a two-dimensional counterpart of that formula
for $Z_n(\ell,s)$.  To this end, let us first recall that by
\eqref{imbal}, $s=2|\{j:\sig_j=1\}|-n$, so that a generic value
$s$ of $s(\bss)$ must meet the condition $n+s\equiv 0(\text{mod
}2)$. In a similar way, we get that $\bss\cdot\bold X$ has the
same parity as the sum $\sum_jX_j$.  Keeping this in mind, we have
that on the event $\{\sum_jX_j\equiv \ell(\text{mod }2)\}$, for
$n+s\equiv 0(\text{mod }2)$,
\begin{equation}
\Bbb I(\bss\cdot\bold X=\ell,\,\bss\cdot\bold e=s)
=\frac{1}{\pi^2}\iint\limits_ {x,y\in
(-\pi/2,\pi/2]}e^{i(\bss\cdot\bold X-\ell)x}\,
 e^{i(\bss\cdot\bold e-s)y}
\,dxdy, \label{3.5}
\end{equation}
thus extending (4.6) in \cite{BCP2}. Multiplying both sides of the
identity by $2^n$, and summing over all $\bss$, we obtain
\begin{equation}
\begin{aligned}
Z_n(\ell,s)&=\frac{2^n}{\pi^2}\iint\limits_{x,y\in
(-\pi/2,\pi/2]}e^{-i(\ell x+sy)}
\prod_{j=1}^n\cos\bigl(xX_j+y\bigr)\,dxdy.
\\
&=2^n\,\pr_{1/2} \Bigl(\bss\cdot\bold X=\ell,\,\bss\cdot\bold
e=s\big|\big. \boldsymbol X\Bigr),
\end{aligned}
\label{3.6}
\end{equation}
where $\bss=(\sigma_1,\dots,\sigma_n)$ is a sequence of
i.i.d.~Bernoulli random variables with probability of
$\sigma_i=\pm 1$ equal to $1/2$.

We would like to estimate the
asymptotics of the integral in \eqref{3.6}, which is equivalent to
proving a local limit theorem for the conditional probability in
\eqref{3.6}.  In general, to compute -- via local limit theorems
-- the probability that some
random variable $A$ takes the value $a$,
it must be the case that the corresponding expectation of $A$ is
near $a$. Thus the analogue of the representation \eqref{3.6} for
the unconstrained problem was well adapted to the analysis of
perfect partitions.  Indeed, in that case, we wanted to estimate
$\pr_{1/2} (|\bss\cdot\bold X| \leq 1|\bold X)$,
and we had $\ex_{1/2}(\bss\cdot\bold X|\bold X)=0$.
However, in the constrained case, this strategy cannot be
expected to work for $b>0$, since $s=bn$ is very far from
the expectation of $\bss\cdot\bold e$, namely
$\ex_{1/2}(\bss\cdot\bold e|\bold X)=0$.

To resolve this substantial difficulty, we
introduce a {\it two-parameter\/} family of distributions for
$\sig_j$
as follows:
Given $\xi,\eta\in \Bbb R$, let
$\bss=(\sig_1,\dots,\sig_n)$ be a sequence of random variables
such that, conditioned on $\bold X$, $\sig_1,\dots,\sig_n$ are
mutually independent, and
\begin{equation}
\pr(\sig_j=1|\bold X)=P(\xi X_j+\eta), \qquad \qquad
\pr(\sig_j=-1|\bold X)=1-P(\xi X_j +\eta), \label{3.10}
\end{equation}
where
\begin{equation}
P(u):=\frac{e^{-u}}{2\cosh u}. \label{3.8}
\end{equation}
In terms of these random variables,  $Z_n(\ell,s)$ can be
rewritten as
\begin{equation}
\begin{aligned}
Z_n(\ell,s) &= e^{ nL_n(\xi,\eta;\bold X)}\,
\pr\bigl(\bss\cdot\bold X = \ell,\,\bss\cdot \bold e=s|\bold
X\bigr) \phantom{\frac {A_A}{A_A}}
\\
&= e^{nL_n(\xi,\eta;\bold X)}\,\frac{1}{\pi^2}
\iint\limits_{x,y\in (-\pi/2,\pi/2]}e^{-i(\ell x+sy)} \ex\bigl(
   \exp(i(x\bss\cdot\bold X+y\bss\cdot\bold e))|\bold X
   \bigr)
\,dxdy,
\end{aligned}
\label{3.11}
\end{equation}
where
\begin{equation}
L_n(\xi,\eta;\bold X):= \frac{\ell\xi}n + \frac{s\eta}n + \frac
1n\sum_{j=1}^n \log(2\cosh(\xi X_j+\eta)). \label{3.9}
\end{equation}
Indeed, fix $\xi,\eta\in \Bbb R$.  Then $Z_n(\ell,s)$ can be
rewritten as
\begin{equation}
\label{3.7}
\begin{aligned}
Z_n(\ell,s) &=\sum_{\bst\in\{-1,+1\}^n} \Bbb I(\bst\cdot\bold
X=\ell,\,\bst\cdot\bold e=s)
\\
&=\sum\limits_{{\bst: \bst\cdot\bold X=\ell,\atop\bst\cdot\bold
e=s} }\, e^{\xi(\ell-\bst\cdot\bold X)+\eta(s-\bst\cdot\bold e)}
=e^{\xi\ell+\eta s} \sum\limits_{{\bst: \bst\cdot\bold
X=\ell,\atop\bst\cdot\bold e=s} }\, \prod_{j=1}^n e^{-(\xi
X_j+\eta)\tau_j}
\\
&= \biggl[ e^{\xi\ell+\eta s} \prod_{j=1}^n (2\cosh(\xi X_j+\eta))
\biggr] \sum\limits_{{\bst: \bst\cdot\bold
X=\ell,\atop\bst\cdot\bold e=s} }\, \prod_{j^\prime=1}^n
P\Big((\xi X_{j^\prime}+\eta)\tau_{j^\prime} \Big)
\\
&=e^{nL_n(\xi,\eta;\mathbf X)}\sum\limits_{\tau:\tau\cdot\mathbf
X=\ell, \atop \tau\cdot\mathbf e=s}\prod_{j=1}^n\Bbb
P(\sigma_j=\tau_j|\mathbf X)
\\
&=e^{nL_n(\xi,\eta;\mathbf X)}\Bbb P(\bss\cdot\mathbf
X=\ell,\bss\cdot \mathbf e=s|\mathbf X),
\end{aligned}
\end{equation}
since $P(-u)=1-P(u)$, see equation \eqref{3.8}.

\subsection{Saddle Point Equations and their Solution}

Given $\xi,\eta$, we now face the problem of determining an
asymptotic value of the {\it local\/} probability in \eqref{3.11}.
This will obviously be easier if the chosen parameters $\ell$ and
$s$ are among the more likely values of $\bss\cdot\bold X$ and
$\bss\cdot\bold e$, respectively. A natural choice is to take
$\ell$ and $s$ equal to their expected values, that is
\begin{equation}
\ex\bigl(\sig\cdot\bold X|\bold X\bigr) =\ell,\quad
\ex\bigl(\bss\cdot\bold e|\bold X \bigr)=s, \label{3.12}
\end{equation}
or explicitly (using \eqref{3.8}, \eqref{3.10})
\begin{equation}
\begin{aligned}
\sum_{j=1}^nX_j\tanh (\xi X_j+\eta)=-\ell,
\\
\sum_{j=1}^n\tanh(\xi X_j+\eta)=-s.
\end{aligned}
\label{3.13}
\end{equation}

Note that the equations \eqref{3.13}
also arise naturally in an apparently different approach
to estimate the integral in
\eqref{3.6}, the ``method of steepest descent.''  In our context,
this corresponds to a complex shift of the integration path,
i.e., to changing the path of integration for $x$ to
the complex path from $-\pi/2+i\xi$ to $-\pi/2+i\xi$, and the path
of integration for $y$ to the complex path from $-\pi/2+i\eta$ to
$-\pi/2+i\eta$, where $\xi$ and $\eta$ are determined by a
suitable saddle point condition.  For general $\xi$ and $\eta$,
this leads to \eqref{3.11}, while the saddle point conditions turn
our to be nothing but \eqref{3.13}. In fact, this is how we first
obtained \eqref{3.11} and \eqref{3.13}.

Both approaches raise the question of uniqueness and existence of
a solution to the saddle point equations\eqref{3.13}.  In this
context, it is useful to realize that the conditions
\eqref{3.13} can be rewritten as
\begin{equation}
\label{3.13equiv} \frac{\partial L_n(\xi,\eta;\mathbf
X)}{\partial\xi}=0,\quad \frac{\partial L_n(\xi,\eta;\mathbf
X)}{\partial\eta}=0.
\end{equation}
Therefore any solution $(\xi,\eta)$ is a stationary point of the
strictly convex function $L_n(\xi,\eta;\mathbf X)$.  If a solution
exists, it is therefore the unique {\it minimum\/} point of $L_n$.
Using the first equation in \eqref{3.11}, we see also that
$(\xi,\eta)$ {\it maximizes\/} the local probability $\Bbb
P(\bss\cdot\mathbf X=\ell, \bss\cdot e=s|\mathbf X)$,
and hence makes it easier to do an asymptotic analysis.
This
observation justifies our choice of $\xi,\eta$.

In the actual proof,
we modify this approach a little since the
solution $\xi=\xi(\mathbf X)$,
$\eta=\eta(\bold X)$ does not lend itself to
a rigorous analysis of
$\Bbb P(\bss\cdot \mathbf X=\ell, \bss\cdot e=s|\mathbf X)$.
Instead, we will resort to
"suboptimal" $\xi=\zeta/M,\,\eta$, where $\zeta,\eta$ are
nonrandom constants, and $(\zeta,\eta)$ is a solution of nonrandom
equations, obtained by replacing the (scaled) sums in \eqref{3.13}
with their weak-law limits, see equations \eqref{3.17} below. This
way we will be able to establish an
explicit
asymptotic formula for $Z_n(\ell,s)$, which will ultimately lead us
to determine the phase boundaries.


In Section~\ref{sec:saddle-point}, we will show that these
deterministic equations have a (unique) solution
$\zeta=\zeta(b),\,\eta=\eta(b)$ iff $b<b_c=\sqrt{2}-1$, the same
$b_c$ that determines the sorted phase.  In other words, the
threshold $b_c$ plays two seemingly unrelated roles: both as a
threshold value of $b$ for solvability of the deterministic saddle
point equations \eqref{3.17}, and as a threshold for the sorted
partition being optimal.  On an informal level, the reason for
the coincidence is as follows: For simplicity, suppose that the
weights $X_j$ are all distinct, so that $X_1<\cdots<X_n$ after
reordering.  As $b$ approaches the point where the solutions
$(\zeta,\eta)$ to the saddle point equations \eqref{3.17} stop
existing, these solutions actually diverge, one tending to $+
\infty$ and the other to $- \infty$. According to equations
\eqref{3.10} and \eqref{3.8}, this in turn means that
$\pr(\sig_j=1|\bold X)$ tends to zero or one, depending on whether
$j < j_o$ or $j > j_o$, where $j_o = |\{j : \sigma_j = -1\}|$ is
the cutoff of the sorted partition for $\bold X$ with bias $s =
nb_c$.  Hence, the product measure $\Bbb P(\bss\cdot\mathbf
X=\ell,\bss\cdot \mathbf e=s|\mathbf X)$ tends to a delta function
on the (unique) sorted partition which is the solution to the
number partitioning problem for $\bold X$ at $b=b_c$.
See Subsection~\ref{sec:Sort-Saddle} for details.

\subsection{Asymptotic behavior of $Z_n(\ell,s)$.}

Proceeding with our heuristic discussion, let us simply assume that
the equations \eqref{3.13} do have a solution $\xi=\xi(\mathbf
X),\eta=\eta(\mathbf X)$. Then we may hope that, with this choice
of the parameters $\xi=\xi(\bold X)$, $\eta=\eta(\bold X)$, we
have a reasonable chance to prove---at least for the likely values
of $\bold X$---a local limit theorem for the {\it conditional\/}
probability in \eqref{3.11}, namely that w.h.p.
\begin{equation}
\pr\bigl(\bss\cdot\bold X=\ell,\bss\cdot\bold e=s|\bold X)\sim
\frac{2}{\pi\sqrt{\text{det }Q}}, \label{3.14}
\end{equation}
where
\begin{equation}
Q=
\begin{pmatrix}
\var(\bss\cdot X) & \text{cov}(\bss\cdot\bold X,\bss\cdot\bold e)
\\
\text{cov}(\bss\cdot\bold X,\bss\cdot\bold e) &
\var(\bss\cdot\bold e)
\end{pmatrix}.
\label{3.15}
\end{equation}
Here the (co)variances are conditioned on $\mathbf X$, so, e.g.,
$Q_{11}=\mbox{Var}(\bss\cdot\mathbf X|\mathbf X)$.
If \eqref{3.14} holds then by \eqref{3.7}, w.h.p.,
\begin{equation}
Z_n(\ell,s) \sim e^{nL_n(\xi,\eta;\bold X)}\,
\frac{2}{\pi\sqrt{\text{det }Q}} = e^{nL_n(\xi,\eta;\bold X)}\,
\frac{2}{\pi nM\sqrt{\text{det }R^{(n)}}} , \label{3.16}
\end{equation}
where $R^{(n)}$ is the matrix with matrix elements
$R^{(n)}_{11}=\frac 1{nM^2}\var(\bss\cdot X)$,
$R^{(n)}_{12}=R^{(n)}_{21}= \frac 1{nM}\text{cov}(\bss\cdot\bold
X,\bss\cdot\bold e)$ and $R^{(n)}_{22}=\frac 1n
\var(\bss\cdot\bold e)$.

Note that, in the limit $M\to\infty$, $X_j/M$ are independent,
uniform random variables in $[0,1]$. We therefore expect that as
$M,n \to \infty$ with $\kappa = n^{-1}log_2 M$ fixed, both
$\zeta(\bold X):= M\xi(\bold X)$ and $\eta(\bold X)$ are close, in
probability, to the deterministic $\zeta,\eta$, defined as the
roots of the averaged version of the ``saddle point equations''
\eqref{3.13}, namely
\begin{equation}
\begin{aligned}
\int\limits_0^1x\tanh(\zeta x+\eta)\,dx=& -\frac{\ell}{Mn},\\
\int\limits_0^1\tanh(\zeta x+\eta)\,dx=&-b,\quad b=\frac{s}{n}.
\end{aligned}
\label{3.17}
\end{equation}
Recall that, without loss of generality, we have taken $s\geq 0$, so
$b\geq 0$.

Furthermore, approximating $\xi(\bold X)$ and $\eta(\bold X)$ by
$M\zeta$ and $\eta$, respectively and using the bound $|d\cosh
u/du|\le 1$, it is easy to see that, because of the weak law of
large numbers, w.h.p.~
\begin{align} L_n(\xi(\bold X),\eta(\bold
X);\bold X)&= \frac{1}{n}\sum_{j=1}^n\log\left(e^{\ell\xi(\bold
X)+s\eta(\bold X)} 2\cosh(\xi(\bold X) X_j+\eta(\bold X))\right)
\notag
\\
&\sim \frac{\ell}{Mn}\zeta+b\eta+ \int\limits_0^1\log(2\cosh(\zeta
x+\eta))\,dx, \label{3.18}
\end{align}
and similarly for the matrix elements of $R^{(n)}$,
\begin{equation}
\label{3.19} R^{(n)}_{ij}\sim
\int\limits_0^1x^{2-(i+j)}(1-\tanh^2(\zeta x+\eta))\,dx.
\end{equation}
Putting everything together, we thus may hope to prove that for
$|\ell|\leq 1$ and $M$ growing exponentially with $n$, (i.e.
$\log_2M\sim \kappa n$ for some $n$-independent $\kappa$), we have
w.h.p.~
\begin{equation}
\begin{aligned}
\frac 1n \log Z_n(\ell,s) &\sim \int\limits_0^1\log(2\cosh(\zeta
x+\eta))\,dx +b\eta-\kappa
\\
&=\kappa_c(b)-\kappa,
\end{aligned}
\label{3.20}
\end{equation}
suggesting that for $\kappa<\kappa_c(b)$ there are exponentially
many perfect partitions, while for $\kappa>\kappa_c(b)$ there are
none.

However, this informal argument is too naive. Equation
\eqref{3.20} could not possibly hold for $\kappa>\kappa_c(b)$.
Indeed, $Z_n(\ell,s)$ is an integer, and thus cannot be
asymptotically equivalent to an exponentially small, yet positive
number. This means that a rigorous proof of \eqref{3.20} must be
based on the condition $\kappa<\kappa_c(b)$.  But our heuristic
discussion provides no clue as to how this condition might enter
the picture. Furthermore, our attempts to find such a proof are
stymied by mutual dependence of the random variables $\Bbb
P(\sigma_j=1|\mathbf X),\,(1\le j\le n)$, a consequence of the
fact that $(\xi(\mathbf X),\eta(\mathbf X))$ depends, in an
unwieldy manner, on the whole $\mathbf X$. This complicated
dependence of $(\xi(\mathbf X),\eta(\mathbf X))$ on $\mathbf X$
would have made it very hard to gain an insight into the random
fluctuations of the sum in \eqref{3.18}, even if we had found a
proof.

Fortunately, once we have informally connected $(\xi(\bold X),
\eta(\mathbf X))$ to $(\zeta,\eta)$ via
$\xi(\mathbf X)=(1+o_p(1))\zeta/M$, $\eta(\mathbf X)=(1+o_p(1))\eta$,
we may try to use the
{\it suboptimal\/} parameters $(\zeta/M,\eta)$ instead. The
corresponding random variables $\Bbb P(\sigma_j=1|\mathbf X)$ each
depend on their own $X_j$, and are thus mutually independent. A
key technical issue is whether the suboptimal parameters are good
enough to get an asymptotic formula for the corresponding
probability
$\Bbb P(\bss\cdot\mathbf X,\bss\cdot\mathbf X=s|\mathbf X)$, given
that now the random equations \eqref{3.13} may hold only approximately.
Our proof below shows that they are indeed sufficient.  With those
parameters, we will be able to get a sharp explicit approximation
for $\log Z_n(\ell,s)$, at least in the range
$\kappa<\kappa_-(b)$.

We prove the existence and uniqueness of the solution to
\eqref{3.17} in the next section, and then use them in
Section~\ref{sec:asymptotic-Z_n} to derive the asymptotics of
$Z_n(\ell,s)$.

\section{Solution of the Deterministic Saddle Point Equations}
\label{sec:saddle-point}

Based on the heuristic discussion of the last section, we {\it
define\/} $\bss$ as the random sequence $(\sig_1,\dots,\sig_n)$
such that, conditioned on $\bold X$, the $\sig_j$ are independent
and
\begin{equation}
\pr(\sig_j=1|\bold X)=P\left(\zeta\frac{X_j}{M}+\eta\right),\quad
\pr(\sig_j=-1| \bold X)=1-P\left(\zeta\frac{X_j}{M}+\eta\right),
\label{4.1}
\end{equation}
with $P(u)$ defined in \eqref{3.8}, and $(\zeta,\eta)$ is a
solution of the equations
\begin{equation}
\begin{aligned}
&\int\limits_0^1x\tanh(\zeta x+\eta)\,dx=0,\\
&\int\limits_0^1\tanh(\zeta x+\eta)\,dx=-b.
\end{aligned}
\label{4.2}
\end{equation}
These are the equations \eqref{3.17}, except that the right hand
side of the first equation is set $0$, since our focus is on
$\ell\ll Mn$.

However, does such a solution exist? A key observation here is
that the equations \eqref{4.2} mean that $(\zeta,\eta)$ is a
stationary point of the function
\begin{equation}
L(\zeta,\eta):=b\eta+\int\limits_0^1\log(2\cosh(\zeta x+\eta))\,dx
\label{4.3}
\end{equation}
which is the r.h.s.~in \eqref{3.18},
without the term $\zeta\ell/(Mn)$,
i.e., the (weak) limit of the function $L_n(\xi(\bold X),\eta(\bold
X);\bold X)$ defined in \eqref{3.9}.
Since $L(\zeta,\eta)$ is
strictly convex, it may have at most one stationary
point, and this point is a global minimum.  So a solution to
\eqref{4.2} exists iff $L(\zeta,\eta)$ attains its global infimum.

\begin{theorem}
\label{thm4.1} Let $0\leq b<b_c:=\sqrt{2}-1$.  Then:

\noindent (1) $L(\zeta,\eta)$ attains its infimum, hence there
exists a unique solution $(\zeta,\eta)=(\zeta(b),\eta(b))$ of the
equations \eqref{4.2}.

\noindent (2) The minimizers $\zeta(b),\eta(b)$ are
continuous functions, with
$\zeta(b)>0$, $\eta(b)<0$ and $\zeta(b)+\eta(b)>0$  whenever
$0<b<b_c$.

\noindent (3) $\hat{L} (b):=L(\zeta(b),\eta(b))$, the minimum of
$L(\zeta,\eta)$, decreases with $b$.  For $b\in (0,b_c)$, its
derivative is ${d\hat{L} (b)}/{db}=\eta(b)$.

\noindent (4) $\lim_{b\to b_c}\zeta(b)=\infty$, $\lim_{b\to
b_c}\eta(b)=-\infty$,
\begin{equation}
\lim_{b\to b_c}\frac{-\eta(b)}{\zeta(b)}=\frac{1}{\sqrt{2}},
\label{4.4}
\end{equation}
and $\lim_{b\to b_c} \hat{L}(b)= 0$.

\noindent (5) $\lim_{b\to 0}\zeta(b)=0$, $\lim_{b\to 0}\eta(b)=0$,
and $\lim_{b\to 0} \hat{L}(b)= \log 2$.
\end{theorem}

\begin{figure}[htb]
  \centering
  \includegraphics[width=0.45\columnwidth]{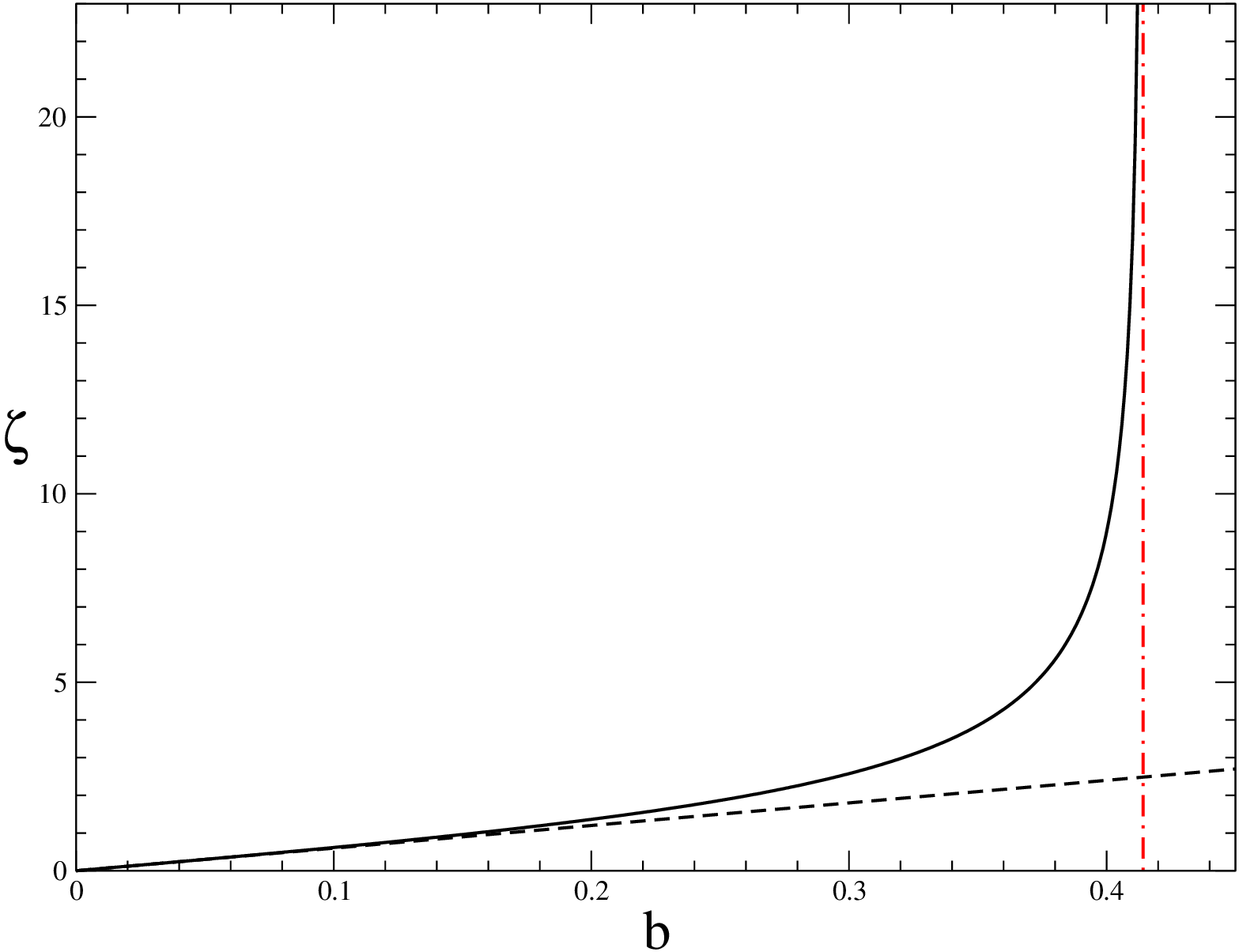}
  \hspace*{0.04\columnwidth}
  \includegraphics[width=0.45\columnwidth]{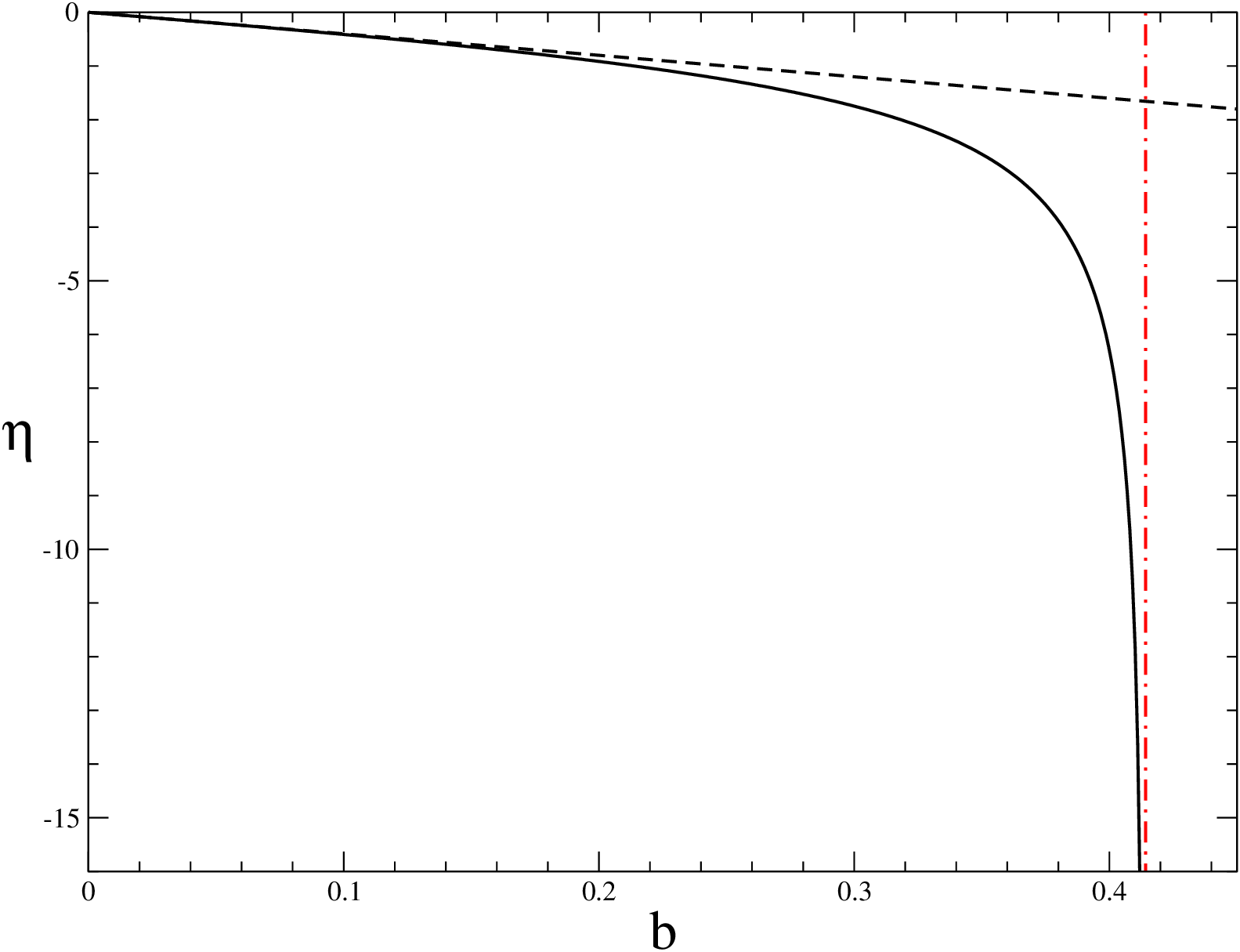}
  \caption{Numerical solution of eq.~\ref{4.2}. The dashed
    lines are the linearized solutions $\zeta(b) = 6b + \bigo{b^2}$
    and $\eta(b)=-4b+\bigo{b^2}$.
  \label{fig:zeta-eta}}
\end{figure}

\noindent {\bf Proof of Theorem~\ref{thm4.1}.\/}

\nobreak

\noindent (1)  Since
$\log(2\cosh u)\ge |u|$, we have
\begin{equation}
L(\zeta,\eta)\geq  \mathcal L(\zeta,\eta), \label{4.5}
\end{equation}
where
\begin{equation}
\mathcal L(\zeta,\eta)=b\eta+\int\limits_0^1|\zeta x+\eta|\,dx.
\end{equation}
It thus suffices to prove that
\begin{equation}
\liminf\limits_{\|(\zeta,\eta)\|_\infty\to\infty} \mathcal
L(\zeta,\eta)=\infty,
\end{equation}
where  $\|(\zeta,\eta)\|_\infty=\max\{|\zeta|,|\eta|\}$.

Since $\mathcal L$ is homogeneous, degree $1$, and continuous, it
suffices to prove that $\mathcal L(\zeta,\eta)>0$ for all
$(\zeta,\eta)$ such that $\|(\zeta,\eta)\|_\infty=1$.  To this
end, we will first show that
\begin{equation}
\min_{\|(\zeta,\eta)\|_\infty=1}\mathcal L(\zeta,\eta)=
\min_{-1\leq\eta\leq 0}\mathcal L(1,\eta). \label{4.min}
\end{equation}
Since ${\mathcal L}(\zeta,\eta)\ge {\mathcal L}(|\zeta|,-|\eta|)$,
it suffices to consider $\eta\leq 0$ and $\zeta\geq 0$. But
$\eta\leq 0$, $\zeta\geq 0$ and $\|(\zeta,\eta)\|_\infty=1$
implies that either $\zeta=1$ and $-1\leq\eta\leq 0$ or $\eta=-1$
and $0\leq \zeta\leq 1$. We thus have to show $\mathcal
L(\zeta,\eta)$ is bounded below by the right hand side of
\eqref{4.min} if $\eta=-1$ and $0\leq \zeta\leq 1$.  Indeed, under
these conditions, $|\zeta x +\eta|=1-\zeta x\geq 1-x=|x+\eta|$,
implying that $\mathcal L(\zeta,\eta) \geq \mathcal L(1,-1) $,
which is clearly bounded below by the right hand side of
\eqref{4.min}.

It remains to bound ${\mathcal L}(1,\eta)$ from below, for
$-1\le\eta\le 0$. Setting $\tilde x=-\eta$, we have
\begin{equation}
\mathcal L(1,\eta)=-b \tilde x +\int\limits_0^1|x-\tilde x|\,dx
=\frac 12 +\tilde x^2 -\tilde x(1+b).
\end{equation}
At $\tilde x=x_0:=(1+b)/2$ the right hand side attains its minimum
value $\frac12[1-(1+b)^2/2]$, which is bounded away form zero as
$b<b_c=\sqrt 2-1$. Thus we have proved that for $b\in
[0,\sqrt{2}-1)$, and all $\zeta,\,\eta$,
\begin{equation}
\mathcal L(\zeta,\eta)\ge c^*(b)\max\{|\zeta|,|\eta|\},
\end{equation}
where $c^*(b)=\frac12[1-(1+b)^2/2]>0$. Therefore $\mathcal
L(\zeta,\eta)$ attains its infimum, hence so does $L(\zeta,
\eta)$.  We conclude that \eqref{4.2} has a unique solution
solution $(\zeta,\eta)=(\zeta(b),\eta(b))$.

\medskip

\noindent (2) Suppose $\eta(b)\ge 0$. Using \eqref{4.2}, and
$y\tanh y>0$ if $y\neq 0$, one can show easily that $\eta(b)>0$
and $\zeta(b)<0$. But such a point $(\zeta(b),\eta(b))$ cannot be
a minimum point of $L(\zeta,\eta)$, since
$L(\zeta(b),\eta(b))>L(-\zeta(b),-\eta(b))$, due to the symmetry
of $\cosh y$, and the fact that $b\eta(b)>-b\eta(b)$. Therefore
$\eta=\eta(b)<0$, and
\begin{equation}
\zeta+\eta=\left.(\zeta x+\eta)\right|_{x=1}>0.
\end{equation}
Continuity will be proved along with (3), where we actually prove
continuous differentiability.
\medskip

\noindent  (3) The equations \eqref{4.2} are an explicit form of
$L_{\zeta}=0$, $L_{\eta}=0$. It is easy to show that the Jacobian
matrix $
\begin{pmatrix} L_{\zeta\zeta}&L_{\eta\zeta}
\\
L_{\zeta\eta}&L_{\eta\eta}\end{pmatrix} $ is nonsingular.
Therefore $\zeta(b)$, $\eta(b) $ are continuously differentiable,
and consequently
\begin{align}
\hat{L}_b(b) =&\Big[L_b(b,\eta,\zeta)
+L_{\zeta}(b,\zeta,\eta)\zeta_b(b)+L_{\eta}
(b,\zeta,\eta)\eta_b(b)\Big]_{\zeta=\zeta(b),\eta=\eta(b)}\\
=&L_b(b,\eta(b),\zeta(b))=\eta(b)<0.
\end{align}
Therefore $\hat{L}(b)$ decreases with $b$.

\medskip
\noindent (4) Pick $\zeta>0$, $x_0\in (0,1)$, and set
$\eta=-x_0\zeta$. Breaking the integral \eqref{4.3} into two
parts, $x\in [0,x_0]$ and $x\in [x_0,1]$, and using
\begin{equation}
\log(e^u+e^{-u})=u+\log(1+e^{-2u}),
\end{equation}
we get:
\begin{equation}
\begin{aligned}
L(\zeta,\eta)=&\frac{\zeta}{2}\bigl(2x_0^2-2x_0(1+b)+1\bigr)\\
&+\frac{1}{2\zeta}\int\limits_0^{2\zeta
x_0}\log(1+e^{-u})\,du+\frac{1}{2\zeta}
\int\limits_0^{2\zeta(1-x_0)}\log(1+e^{-u})\,du.
\end{aligned}
\label{4.22}
\end{equation}
At $x_0=(1+b)/2$ the quadratic polynomial attains its minimum
value $1-(1+b)^2/2$, which is positive since $b<b_c$. With this
$x_0$, choose $\zeta=\big(1-\frac{(1+b)^2}2\big)^{-1/2}$. Then
$\zeta\to\infty$ as $b\to b_c$, so that the integral terms add up
to $O(\zeta^{-1})$, which is also the order of the first term.
Hence $L(\zeta,\eta)\to 0$, and therefore $\lim_{b\to
b_c}\hat{L}(b)\le 0$. To see that $\lim_{b\to b_c}\hat{L}(b)= 0$,
we just note that $\hat{L}(b)=\min_{\eta,\zeta}L(\zeta,\eta) \geq
\min_{\eta,\zeta}\mathcal L(\zeta,\eta)=0$.

 To complete the proof, in \eqref{4.22} set $\zeta=\zeta(b)$ and
$x_0=-\eta(b)/\zeta(b)$, so that $\eta=\eta(b)$. (Note that
$x_0\in (0,1)$ as $\zeta(b)+\eta(b)>0$.) Since the quadratic
polynomial in \eqref{4.22} is non-negative if $b\leq b_c$, and
$\max(x_0,1-x_0)\ge 1/2$, we have that
\begin{equation}
\lim_{b\to b_c}\hat{L}(b)\ge\liminf_{b\to
b_c}\frac{1}{2\zeta(b)}\int\limits_0
^{\zeta(b)}\log(1+e^{-u})\,du>0,
\end{equation}
if $\liminf_{b\to b_c}\zeta(b)<\infty$. So, since $\lim_{b\to
b_c}\hat{L}(b)=0$, we conclude that $\zeta(b)\to\infty$ as
$b\to b_c$. But then, using \eqref{4.22} again,
\begin{align}
0= \lim_{b\to b_c}\frac{\zeta(b)}{2}\bigl[2x_0^2-2x_0(1+b)+1\bigr]
\Longrightarrow&
\lim_{b\to b_c}\bigl[2x_0^2-2x_0(1+b)+1\bigr]=0\\
\Longrightarrow& \lim_{b\to b_c}x_0=\frac{1}{\sqrt{2}},
\end{align}
the only root of $2x^2-2x\sqrt{2}+1=0$.

\medskip
\noindent(5) This statement is immediate from the fact that
$\zeta(b)$, $\eta(b) $ are continuously differentiable on
$(0,b_c)$ and the observation that for $b=0$, $L(\zeta,\eta)$
attains its minimum value of $\log 2$ at $(\zeta,\eta)=(0,0)$. \qed

\begin{remark}\label{rem:4.1}
Theorem~\ref{thm4.1} can easily be generalized to
the case of non-uniform i.i.d.~random variables $X_j$
provided
$X_j /M$ has a limiting density
$\mu(x),\,x\ge 0$. In that case, $b_c$ is replaced by
$b_c=b_c(\mu)$, as defined in Remark~\ref{3.1}, i.e.
\begin{equation}
\label{bcmux0} b_c=2\int_0^{x_0}\mu(x)\,dx-1,\quad
\int_0^{x_0}x\mu(x)\,dx=\int_{x_0} ^{\infty}x\mu(x)\,dx.
\end{equation}
The counterpart of $L(\zeta,\eta)$ is obviously
\begin{equation}
L(\zeta,\eta)=b\eta+\int_0^{\infty}\log (2\cosh(\zeta
x+\eta))\,\mu(x)\,dx.
\end{equation}
Remarkably, this extension requires only obvious changes
in the above proof.  For
instance, below is the proof of the item (1).  Note
that the surprising dual role of $b_c$, namely as both
the point at which the saddle point equation stops having
a solution and as the threshold for optimality of sorted
partitions, still holds for $b_c(\mu)$.  In fact, if
correctly interpreted, this dual role persists
{\em deterministically} for any fixed instance
$\{X_1, \dots, X_n\}$, see Subsection~\ref{sec:Sort-Saddle}.
\end{remark}

\noindent {\bf Proof of (1) for an arbitrary distribution $\mu$:}
First, we write
\begin{equation}
L(\zeta,\eta)\ge {{\mathcal L}}(\zeta,\eta)
=b\eta+\int_0^{\infty}|\zeta x+\eta|\, \mu(x)\,dx.
\end{equation}
Second, we bound
\begin{equation}
\min\limits_{\|(\zeta,\eta)\|_{\infty}=1}{\mathcal
L}(\zeta,\eta)\ge \min\left(\min\limits_{-1\le\eta\le 0}{\mathcal
L}(1,\eta),\, \min\limits_{0\le\zeta\le 1}{\mathcal
L}(\zeta,-1)\right).
\end{equation}
Furthermore,
\begin{equation}
\begin{aligned}
{\mathcal L}(\zeta,-1)&=-b+M(\zeta^{-1}),\nonumber,
\\
M(y)&:=\int_0^y(1-xy^{-1})\mu(x)\,dx+\int_y^{\infty}(xy^{-1}-1)
\mu(x)\,dx.
\end{aligned}
\end{equation}

Using \ref{bcmux0}, we see that
\begin{equation}
\min_{y\ge 0} M(y)=M(x_0)=b_c(\mu),
\end{equation}
hence
\begin{equation}
{\mathcal L}(\zeta,-1)\ge -b+b_c(\mu)>0,
\end{equation}
for $b<b_c(\mu)$. Likewise, if $\eta<0$, we set $\eta=-\tilde{x}$,
and easily obtain
\begin{equation}
{\mathcal L}(1,\eta)=\tilde{x}\bigl(-b+M(\tilde{x})\bigr) \ge
\tilde{x}(-b+b_c(\mu))>0,
\end{equation}
for $b<b_c(\mu)$. \qed

\section{The Perfect Phase}
\label{sec:perfect}

\subsection{Asymptotic Enumeration of Partitions
by Discrepancy and Bias} \label{sec:asymptotic-Z_n}

Now that we have proved existence of the solution $(\zeta,\eta)$
of \eqref{4.2} for ${b}<\sqrt{2}-1$, we are justified in using the
marginals \eqref{4.1}. It is critically important that each
$\pr(\sig_j=\pm 1|\bold X)$ depends only on its own $X_j$, so that
they are mutually independent. This would not have been the
case if we had used $(\xi(\bold X),\eta(\bold X))$, the solution of
the random equations \eqref{3.13}.

The corresponding version of \eqref{3.11} is
\begin{equation}
\begin{aligned}
Z_n(\ell,s) =\exp\bigg(\zeta\frac{\ell}{M}+s\eta +\sum_{j=1}^n
\log\Big(2\cosh\Big(\zeta\frac{X_j}{M}+\eta\Big)\Big)\bigg)
I_n(\bold X).
\end{aligned}
\label{5.1}
\end{equation}
Here $I_n(\bold X)$ is the random integral
\begin{equation}
I_n(\bold X)= \frac{1}{\pi^2}\iint\limits_{x,y\in (-\pi/2,\pi/2]}
e^{-i(\ell x+sy)}f(x,y;\bold X)\,dxdy ,
\label{5.2}
\end{equation}
where
\begin{equation}
\begin{aligned}
f(x,y;\bold X) :=&\ex\bigl(\exp(i(x\bss\cdot\bold
X+y\bss\cdot\bold e))|\bold X)
\\
=&\prod_{j=1}^n \bigl[p(X_j/M)e^{i\bigl(xX_j+y\bigr)}+
q(X_j/M)e^{-i\bigl(xX_j+y\bigr)}\bigr],
\end{aligned}
\label{5.3}
\end{equation}
with the shorthands $p(u)$ and $q(u)$ for $P(\zeta u+\eta)$ and
$1-P(\zeta u+\eta)$, where $(\zeta,\eta)$ is the solution of
\eqref{4.2}, and $P(\cdot)$ is given by \eqref{3.8}. Now that the
summands in \eqref{5.1} are simply i.i.d.~random variables, the
core of the problem is to determine the asymptotically
likely behavior of $I_n(\bold X)$.

\begin{theorem}
\label{thm5.1} Suppose that $\limsup |s|/n<b_c$,
$0<\liminf\kappa$,
$\limsup (\kappa-\kappa_-(b))<0$,
$\ell=o(Mn^{1/2})$ and $n=o(M^2)$.
Then there is a constant $\delta>0$ such that,
with probability
$1-O(e^{-\delta\log^2 n})$,
\begin{equation}
I_n(\bold X)=(1+o(1))\,
\frac{1}{\pi Mn\sqrt{\text{det}R}}\,
\exp\bigl(-\frac{1}{4}\boldsymbol{\tau}_n
R^{-1}\boldsymbol{\tau}_n^\prime\bigl).
\label{5.4}
\end{equation}
Here
\begin{equation}
R=\begin{pmatrix} 2\ex(U^2p(U)q(U))&2\ex(Up(U)q(U))
\\
2\ex(Up(U)q(U))&2\ex(p(U)q(U))\end{pmatrix},
 \label{5.5}
\end{equation}
and $\boldsymbol{\tau}_n$ is a two-dimensional random vector
which converges in probability to
a Gaussian vector $\boldsymbol\tau$, with
mean zero and covariance matrix
\begin{equation}
K=\begin{pmatrix} \var(U(p(U)-q(U))
&\text{cov}(U(p(U)-q(U)),\,p(U)-q(U))
\\
\text{cov}(U(p(U)-q(U)),\,p(U)-q(U)) &
\var(p(U)-q(U))\end{pmatrix}, \label{5.6}
\end{equation}
where $U$ is uniformly distributed on $[0,1]$. Furthermore,
$\|\boldsymbol\tau_{n}\|\leq \log n$ with probability at least
$1-O(e^{-\delta\\log^2n})$.

Consequently, with probability $1-O(e^{-\delta\log^2 n})$,
$Z_n(\ell,s)\ge 1$  and
\begin{equation}
\log Z_n(\ell,s)
=n [L(\zeta,\eta)-\kappa\log 2]+n^{1/2}S_n+o({n^{1/2}}),
\label{5.7}
\end{equation}
where
\begin{equation}
\label{5.8}
S_n=\frac 1{{n^{1/2}}}\sum_{j=1}^n
\bigg(\log\Big(2\cosh\Big(\zeta\frac{X_j}{M}+\eta\Big)\Big)
-\ex\Big[\log\Big(2\cosh\Big(\zeta\frac{X_j}{M}+\eta\Big)\Big)\Big]
\bigg)
\end{equation}
is asymptotically Gaussian with zero mean and
variance $\sigma^2= \var(\log(2\cosh(\zeta U+\eta)))$.
\end{theorem}

\begin{remark}
It is a consequence of Theorem~\ref{thm4.1} and
Theorem~\ref{thm5.1} that for all $b\in [0,b_c)$,
\begin{equation}
\label{5.9} \kappa_-(b)\leq \kappa_c(b),
\end{equation}
with
\begin{equation}
\label{5.10}
\kappa_-(0)=\kappa_c(0)=1
\quad\text{and}\quad
\kappa_-(b_c)=\kappa_c(b_c)=0.
\end{equation}
Indeed, let us assume that $\liminf (\kappa_-(b) -\kappa) >0$.  By
Theorem~\ref{thm5.1}, we then have that $ n^{-1}\log
Z_n(\ell,s)-\bigl[L(\zeta,\eta)-\kappa\log 2\bigr]\to 0$ in
probability.  In particular, since w.h.p.~$Z_n(\ell,s)\ge 1$, we
have $ \liminf(L(\zeta,\eta)-\kappa\log 2)=
\liminf(\kappa_c(b)-\kappa)\log 2\ge 0. $ So the condition
$\liminf (\kappa_-(b) -\kappa) >0$ implies that
$\liminf(\kappa_c(b)-\kappa)\ge 0$, which proves \eqref{5.9}.
Since $\zeta(0)=\eta(0)=0$, it follows from the definitions
\eqref{L} -- \eqref{kappac} that $\kappa_-(0)=\kappa_c(0)=1$.
Finally, by Theorem~\ref{thm4.1}, we have $\hat{L}(b_c-)=0$, or
$\kappa_c(b_c-)=1$. Numerical computations indicate that
$\kappa_-(b)<\kappa_c(b)$ for $0<b<b_c$, but the graphs of two
functions remain surprisingly close to each other. Our limited
attempts to prove this strict inequality have not succeeded.
\end{remark}

\noindent  {\bf Proof of Theorem~\ref{thm5.1}:\/} Pick a large,
but fixed, $B>0$. Split the integration into two parts, $|x|\le
B/M$ and $|x|>B/M$, and denote the corresponding integral
$I_{n1}(\bold X)$ and $I_{n2}(\bold X)$, respectively. Consider
$I_{n1}(\bold X)$ first. Begin with
\begin{equation}
\begin{aligned}
\Big|p(X_j/M)e^{i\bigl(xX_j+y\bigr)} +
&q(X_j/M)e^{-i\bigl(xX_j+y\bigr)}\Big|^2
\\
&=1-2p_jq_j(1-\cos(2\bigl(xX_j+y\bigr)))
\\
&\le\exp\Bigl(-2p_jq_j(1-\cos(2(xX_j+y)))\Bigr),
\end{aligned}
\end{equation}
where we have introduced the abbreviations $p_j=p(X_j/M)$ and
$q_j=q(X_j/M)$. Let
\begin{equation}
\mathcal S_n=\{j:\,X_j/M\le \pi/(3B)\}.
\end{equation}
Clearly, $j\in \mathcal S_n$ implies
\begin{equation}
2|xX_j+y|\le 2\left(\frac{B}{M}X_j+\frac{\pi}{2}\right)\le
\frac{5\pi}{3}<2\pi.
\end{equation}
Now, there exist a constant $c_1>0$ such that
\begin{equation}
1-\cos\a\ge c_1\a^2,\quad\a\in [-5\pi/3,5\pi/3].
\end{equation}
Since
\begin{equation}
2p_jq_j=\frac{1}{2\cosh^2(\zeta X_j/M+\eta)}\ge
\frac{1}{\max\{2\cosh^2(\eta),\,2\cosh^2(\zeta+\eta)\}},
\end{equation}
we thus have shown that there exist a $c_2>0$ such that
\begin{equation}
\begin{aligned}
|f(x,y;\bold X)| \le& \exp\left(-c_2\sum_{j\in\mathcal
S_n}\bigl(xX_j+y\bigr)^2\right)
\\
=& \exp\bigl(-c_2(x^2M_2+2xyM_1+y^2M_0)\bigr),
\end{aligned}
\label{5.17}
\end{equation}
where
\begin{equation}
M_k=M_k(B) =\sum_{j\in \mathcal S_n}X_j^{(k)}
=\sum_{j=1}^nX_j^{(k)}(B),\quad X_j^{(k)}(B):=X_j^k\Bbb
I_{\{X_j/M\le \pi/(3B)\}}.
\end{equation}
Since $X_j^{(k)}(B)$ are independent with
$X_j^{(k)}(B)/\ex[X_j^{(k)}(B)]=O(B)$, the event
\begin{equation}
\label{5.19}
A_n=\{M_k\in [0.99n\ex (X^{(k)}(B)),\,1.01n\ex
(X^{(k)}(B))],\,k=0,1,2\}
\end{equation}
has probability $1-O(e^{-\delta_1 n})$ for some $\delta_1>0$,
by, e.g., the Azuma-Hoeffding inequality.

We continue our computation on the event $A_n$.
It is easy to see that
\begin{equation}
x^2M_2+2xyM_1+y^2M_0\ge c_3n\bigl[(Mx)^2+y^2\bigr],\quad c_3>0.
\label{5.20}
\end{equation}
Set $r_n=n^{-1/2}\log n$. From \eqref{5.17} and \eqref{5.20}, it
follows that, for $c_4=c_2c_3$,
\begin{equation}
\begin{aligned}
\iint\limits_{\|(Mx,y)\|\ge r_n\atop |x|\leq B/M}|f(x,y;\bold X)|\,dx
dy\le& \frac{1}{Mn}\left(\,\,\iint\limits_{\|\bold z\|\ge \log n}
e^{-c_4\|\bold z\|^2}\,dz_1dz_2\right)
=O\left((Mn)^{-1}e^{-c_4\log^2 n}\right).
\end{aligned}
\label{5.21}
\end{equation}
For $\|(Mx,y)\|\le r_n$, we expand and exponentiate:
\begin{equation}
\begin{aligned}
p_je^{i\bigl(xX_j+y\bigr)}+&q_je^{-i\bigl(xX_j+y\bigr)}=
\\
&=1+i\bigl(xX_j+y\bigr)(p_j-q_j)
-\frac{1}{2}\bigl(xX_j+y\bigr)^2+O(|xX_j|^3+|y|^3)
\\
&=\exp\biggl(i\bigl(xX_j+y\bigr)(p_j-q_j)
-\frac{1}{2}\bigl(xX_j+y\bigr)^2
\\
&\qquad\quad +\frac{1}{2}\bigl(xX_j+y\bigr)^2
(p_j-q_j)^2+O(|xX_j|^3+|y|^3)\biggr)
\\
&=\exp\biggl(i\bigl(xX_j+y\bigr)(p_j-q_j)
\\
&\qquad\quad -2p_jq_j
(x^2X_j^2+2X_jxy+y^2)+O(|xX_j|^3+|y|^3)\biggr).
\end{aligned}
\end{equation}
Therefore,
\begin{equation}
\begin{aligned}
e^{-i(\ell x+sy)}f(x,y;\bold X)
=&e^{i(Mx)(T_{n1}-\ell/M)+iy(T_{n2}-s)}
\\
&\times \exp(-(Mx,y)Q(Mx,y)^\prime+O(nM^3|x|^3+n|y|^3)),
\end{aligned}
\label{5.23}
\end{equation}
where $(Mx,y)^\prime$ denotes the transpose of $(Mx,y)$,
and
\begin{equation}
\begin{aligned}
T_{n1}=&\sum_j(X_j/M)(p_j-q_j),\\
T_{n2}=&\sum_j(p_j-q_j),
\end{aligned}
\label{5.24}
\end{equation}
\begin{equation}
\begin{aligned}
Q_{11}=&\sum_j(X_j/M)^22p_jq_j,\\
Q_{12}=&\sum_j(X_j/M)2p_jq_j\\
Q_{22}=&\sum_j2p_jq_j.
\end{aligned}
\label{5.25}
\end{equation}

Here $T_{n1},T_{n2}$ are sums of i.i.d.~random variables, and it
is easy to show that the random vector
\begin{equation}
\label{5.26}
\boldsymbol{\tau}_n
=(\tau_{n1},\tau_{n2}):=n^{-1/2}(T_{n1}-\ex(T_{n1}),\,
T_{n2}-\ex(T_{n2}))
\end{equation}
is asymptotically Gaussian, with zero means, and the covariance
matrix $\{K_M(i,j)\}$, with
\begin{equation}
\begin{aligned}
K_M(1,1)=&\var(U_M(p(U_M)-q(U_M))),
\\
K_M(1,2)=&\text{cov}(U_M(p(U_M)-q(U_M)),p(U_M)-q(U_M)),
\\
K_M(2,2)=&\var(p(U_M)-q(U_M)),
\end{aligned}
\label{5.27}
\end{equation}
and $U_M$ distributed uniformly on $\{1/M,\dots,M/M\}$.
Consequently
\begin{equation}
K_M=K+O(M^{-1}),
\end{equation}
where $K$ is defined like $K_M$, with $U_M$ replaced by the
$[0,1]$-uniform $U$, which is the matrix $K$ defined in
\eqref{5.6}. We can use $K$ as the limiting covariance matrix
for $\boldsymbol{\tau}_n$.
Also, again by Azuma-Hoeffding, for
any $c>0$,
\begin{equation}
\label{tau-n-var}
\|\boldsymbol\tau_{n}\|\leq c\log n
\end{equation}
with probability $1-O(e^{-\delta_2\log^2 n})$
for some $\delta_2=\delta_2(c)>0$.
In addition,
\begin{equation}
\begin{aligned}
\ex(T_{n1})=&n\ex\bigl[(X/M)(p(X/M)-q(X/M))\bigr]
=n\ex(U(p(U)-q(U)))+O(n/M)
\\
=&O(n/M),\\
\ex(T_{n2}) =&n\ex\bigl[p(X/M)-q(X/M)\bigr]=n\ex(p(U)-q(U))+O(n/M)
\\
=&s+O(n/M).
\end{aligned}
\label{5.29}
\end{equation}
Using the equations \eqref{5.26} and \eqref{5.29}, we obtain
\begin{eqnarray*}
T_{n1}-\ell/M&=&n^{1/2}\bigl(\tau_{n1}+O(n^{1/2}/M+|\ell|/(Mn^{1/2}))
\bigr),\nonumber\\
T_{n2}-s&=&n^{1/2}\bigl(\tau_{n2}+O(n^{1/2}/M)\bigr).
\end{eqnarray*}
Both remainder terms are $o(1)$ since $n^{1/2}=o(M)$ and
$\ell=o(Mn^{1/2})$.
Since $Q_{ij}$ is a sum of $n$ bounded
i.i.d. random variables,
$Q_{ij}=\ex(Q_{ij})+O(n^{1/2}\log n)$ with probability
$1-O(e^{-\delta_3 log^2 n})$ for some $\delta_3>0$.
Approximating $\ex(Q_{ij})$ as
$\ex(Q_{ij})=n[R_{ij}+O(M^{-1})]=n[R_{ij}+o(n^{-1/2})]$, where
\begin{equation}
\begin{aligned}
R_{11}=&2\ex(U^2p(U)q(U)),\\
R_{12}=&2\ex(Up(U)q(U)),\\
R_{22}=&2\ex(p(U)q(U)),
\end{aligned}
\label{5.31}
\end{equation}
we get
\begin{equation}
Q=n(R+O(n^{-1/2}\log n))
\end{equation}
with probability
$1-O(e^{-\delta_3\log^2 n})$.

Consequently,
with probability $1-O(e^{-\delta_3\log^2 n})$,
\begin{multline}
\iint\limits_{\|(Mx,y)\|\le r_n}e^{-i(\ell x+sy)}f(x,y;\bold
X)\,dxdy
\\
=\iint\limits_{\|(Mx,y)\|\le r_n}
\exp\bigl(i(Mn^{1/2}x)(\tau_{n1}+o(1))
         +i(n^{1/2}y)(\tau_{n2}+o(1))\bigr)
\\
\exp\bigl(-(Mx,y)Q(Mx,y)^\prime +O(|Mx|^3n+|y|^3n)\bigr)\,dxdy
\\
=\frac{1}{Mn}\left[\frac{\pi}{\sqrt{\text{det }R}}\,
\exp\Bigl(-\frac{1}{4} \boldsymbol{\tau}
_nR^{-1}\boldsymbol{\tau}_n^{\prime}\Bigr)
+o(1) \right]. \label{5.32}
\end{multline}
Using \eqref{5.21}
and \eqref{tau-n-var},
we see that the r.h.s.~of \eqref{5.32}
also gives the asymptotics of
$I_{n1}(\bold X)$, the integral over all $(x,y)$ with $|x|\le
B/M$.

Let us turn now to $I_{n2}(\bold X)$. We want to show that for
some $\delta>0$ we have $I_{n2}=o(I_{n1})$ with probability
$1-O(e^{-\delta\log^2 n})$. First of all, by \eqref{5.2},
\eqref{5.3} and the definition of $I_{n2}(\bold X)$,
\begin{equation}
|I_{n2}(\bold X)|^2=\frac 1{\pi^4}\iiiint\limits_{|x_1|,|x_2|\in
[B/M,\pi/2]\atop y_1,y_2\in (-\pi/2,\pi/2]}
e^{-i\ell(x_1-x_2)-is(y_1-y_2)}F(\bold x,\bold y;\bold X)\,d\bold
xd\bold y,
\end{equation}
where
\begin{align}
F(\bold x,\bold y;\bold X) =& f(x_1,y_1;\bold
X)\,\overline{f(x_2,y_2;\bold X)}
\\
=&\prod_{j=1}^n\bigl[p_j^2e^{i(x_1^\prime X_j+y_1^\prime)}+
q^2_je^{-i(x_1^\prime X_j+y_1^\prime)}
\\
+&p_jq_j(e^{i(x_2^\prime X_j+y_2^\prime)}+e^{-i(x_2^\prime
X_j+y_2^\prime)}) \bigr],
\end{align}
and
\begin{equation}
x_1^\prime=x_1-x_2,\,x_2^\prime =x_1+x_2;\quad
y_1^\prime=y_1-y_2,\,y_2^\prime= y_1+y_2.
\end{equation}
Therefore
\begin{equation}
\ex(|I_{n2}(\bold X)|^2)\le \frac 1{\pi^4}
 \iiiint\limits_{|x_1|,|x_2|\in
[B/M,\pi/2]\atop y_1,y_2 \in (-\pi/2,\pi/2]}|F(\bold x,\bold
y)|^n\,d\bold x d\bold y, \label{5.38}
\end{equation}
where
\begin{equation}
\begin{aligned} F(\bold x,\bold
y)=&\frac{1}{M}\sum_jp^2(j/M)e^{i(x_1^\prime j+y_1^\prime)}+
\frac{1}{M}\sum_jq^2(j/M)e^{-i(x_1^\prime j+y_1^\prime)}\\
+&\frac{1}{M}\sum_jp(j/M)q(j/M)(e^{i(x_2^\prime
j+y_2^\prime)}+e^{-i(x_2^\prime j+ y_2^\prime)}).
\end{aligned}
\label{5.39}
\end{equation}
To proceed, we need
the following lemma.
 \begin{lemma}
\label{lem5.1} Let $g(u)$ be continuously differentiable on
$[0,1]$. Then, for $x\notin 2\pi\mathbb Z$  and for
all $y$,
\begin{equation}
\left|\sum_{j=1}^M g(j/M)e^{i(xj+y)}\right|\le
\frac{2\|g\|+\|g^\prime\|}{|e^{ix}-1|}, \label{5.40}
\end{equation}
where $\|g\|:=\max\{|g(u)|:u\in [0,1]\}$, and
$\|g^\prime\|:=\max\{|g^\prime(u)|:u\in [0,1]\}$.
\end{lemma}

\noindent  {\bf Proof of Lemma~\ref{lem5.1}.\/} First write
\begin{multline}
\sum_{j=1}^Mg(j/M)e^{i(xj+y)}
=\sum_{j=1}^M\frac{e^{i(x(j+1)+y)}-e^{i(xj+y)}}{e^{ix}-1}
\, g(j/M)\\
=\frac{1}{e^{ix}-1}\,\biggl[
-g(1/M)e^{i(x+y)}+g(M/M)e^{i(x(M+1)+y)}\\
-\sum\limits_{j=1}^{M-1}(g((j+1)/M)-g(j/M))e^{i(x(j+1)+y)}\biggr].
\end{multline}
Since each of the differences in the last sum has absolute value
bounded by $M^{-1}\|g^\prime\|$, while the first two terms are
clearly bounded by $\|g\|$, we obtain \eqref{5.40}.
\qed

 Returning to the proof of the theorem, we observe that since
$|x_t|\le \pi/2$, we have $|x_t^\prime|\le \pi$, $t=1,2$.
Furthermore,
\begin{equation}
B/M\le |x_t|=0.5|x_1^\prime\pm x_2^\prime|,\quad t=1,2,
\end{equation}
implies that $\max\{|x_1^\prime|,|x_2^\prime|\}\ge B/M$. If
$|x_1^\prime|\ge B/M$, then applying Lemma \ref{lem5.1} to the
first two sums in \eqref{5.39}, we obtain
\begin{equation}
\begin{aligned}
|F(\bold x,\bold y)|\le& \frac{2}{M}\sum_jp(j/M)q(j/M)+O(B^{-1})
\\
=&2\int_0^1p(u)q(u)\,du +O(B^{-1}).
\end{aligned}
\label{5.45}
\end{equation}
If $|x_2^\prime|\ge B/M$, then likewise
\begin{equation}
|F(\bold x,\bold y)|\le\int_0^1(p^2(u)+q^2(u))\,du+O(B^{-1}).
\label{5.46}
\end{equation}
And, if $\min\{|x_1^\prime|,|x_2^\prime|\}\ge B/M$, then
\begin{equation}
|F(\bold x,\bold y)|=O(B^{-1}). \label{5.47}
\end{equation}
Since the right hand side of \eqref{5.45} is dominated by the
right hand side of \eqref{5.46} (just use that $2pq\leq p^2+q^2$),
we conclude that there is a constant $c$ independent of $B$ such
that
\begin{equation}
|F(\bold x,\bold y)|\le \rho+cB^{-1}\quad\text{or}\quad |F(\bold
x,\bold y)|\le cB^{-1},
\end{equation}
depending upon whether only one or both $|x^\prime_t|$ exceeds
$B/M$. Here
\begin{equation}
\begin{aligned}
\rho=&\int_0^1(p^2(u)+q^2(u))\,du
=\int_0^1\Big(1-\frac 12 \cosh^{-2}(\zeta u+\eta)\Big)\,du\\
=&1-\frac{\tanh(\zeta+\eta)-\tanh(\eta)}{2\zeta}
\\
&=2^{-\kappa_-(b)}.
\end{aligned}
\label{5.49}
\end{equation}
We conclude then that
\begin{equation}
\ex(|I_{n2}(\bold X)|^2)\le (c/B)^n +
O\big(M^{-1}(\rho+c/B)^n\big) .
\end{equation}
Since $M\rho^n$ is exponentially small
if $\limsup(\kappa-\kappa_-(b))<0$,
we get that for $B$ large enough, there exists a $\delta_4>0$
such that
\begin{equation}
\ex(|I_{n2}(\bold X)|^2)\le e^{-\delta_4 n} (Mn)^{-2}.
\end{equation}
On the other hand,
 $|I_{n1}(\mathbf X)|^2=(Mn)^{-2}e^{O(\log n^2)}$
with probability
$1-O(e^{-\delta_2\log^2 n})-O(e^{-\delta_3\log^2 n})$
by \eqref{tau-n-var} and \eqref{5.32}.
Thus $|I_{n2}(\bold X)|=o(|I_{n1}(\mathbf X)|)$ with
probability at least $1-O(e^{-\delta_5\log^2 n})$, implying
\eqref{5.4}.
To prove \eqref{5.7}, we just note
that the prefactor
in \eqref{5.1} can be rewritten as
\begin{equation}
\label{5.50}
\begin{aligned}
\exp\bigg(&\zeta\frac{\ell}{M}+s\eta +\sum_{j=1}^n
\log\Big(2\cosh\Big(\zeta\frac{X_j}{M}+\eta\Big)\Big)\bigg)
\\
&=\exp\bigg(\zeta\frac{\ell}{M}+s\eta +
n^{1/2} S_n + n\int_0^1\log\big(2\cosh(\zeta x+\eta)\big)dx
+O(nM^{-1})\bigg)
\\
&=\exp\bigg(n L(\zeta,\eta)+n^{1/2}S_n
+O(nM^{-1})+\zeta\ell/M\bigg),
\end{aligned}
\end{equation}
while, with probability $1-O(e^{-\delta\log^2n})$,
\begin{equation}
\label{5.51}
I_n(\bold X)=M^{-1} e^{O(\log^2 n)}=M^{-1}e^{o(n^{1/2})}.
\end{equation}
\qed

\begin{remark}
As the reader may have noticed, the condition that
$\ell$ has the same parity as $\sum_jX_j$ has not
been used in the above proof.  Thus the asymptotics
stated in \eqref{5.4} hold for all $\ell$ with
$\ell=o(Mn^{1/2})$, independent of the parity of $\ell$.
But this does {\em not} mean that the corresponding
asymptotics hold for the number of partitions
$Z_n(\ell, s)$, since \eqref{5.1} is valid only if
the parity of $\ell$ is the same as that of
$\sum_jX_j$.  If this condition is violated, the
left hand side of \eqref{5.1} is zero, while
w.h.p., the random integral
$I_n(\bold X)$ is different from zero.
\end{remark}

 Applied to the special case $|\ell|\le 1$, Theorem~\ref{thm5.1}
asserts, roughly,  that for every point $(b,\kappa)$ such that
$b<b_c$ and $\kappa<\kappa_-(b)<1$, w.h.p.~there are
exponentially many perfect partitions. In fact,
if in the right hand side of \eqref{5.1} the random integral
$I_n(\mathbf X)$ is replaced by the leading term in \eqref{5.4},
then the resulting product remains exponentially large within the
narrow (crescent-shaped) region between $\kappa = \kappa_-(b)$
and $\kappa =\kappa_c(b)$. In principle, this may mean that the
likely number of perfect partitions remains exponentially large in
this extended region! An extensive numerical simulation
(see Section~\ref{sec:openprob}) strongly suggests that the expected
logarithm of the number of perfect partitions at every point of
the region $\kappa<\kappa_c(b)$ is extremely well approximated by
the expected logarithm of the above-mentioned product.
Based on our experience with the unconstrained
problem and these simulations,
 we expect that
 w.h.p. at least the
weaker formula
\begin{equation}
\label{6.1} \log Z_n(\ell,s)=n[L(\zeta,\eta)-\kappa\log 2+o(1)]n,
\end{equation}
cf.\eqref{5.7}, remains valid in the whole region
$\limsup(\kappa-\kappa_c(b)) <0$.

\subsection{Distribution of the Bias}
\label{sec:BiasDist}

 Let us have a closer look at the case $|\ell|\le 1$ and
$s=o(n)$. A simple computation shows that
\begin{equation}
\zeta=6\frac{s}{n}+O((s/n)^2),\quad \eta=-4\frac{s}{n}+O((s/n)^2),
\label{6.2}
\end{equation}
so that
\begin{equation}
p(u)=P(\zeta u+\eta)=\frac{1}{2}+O(|s|/n),
\end{equation}
and the entries of the covariance matrix $K$ are of order
$O(|s|/n)$. Furthermore
\begin{equation}
R=\begin{pmatrix} 1/6+O(|s|/n)&1/4+O(|s|/n)\\
1/4+O(|s|/n)&1/2+O(|s|/n)\end{pmatrix},
\end{equation}
so that $\text{det}R=1/(48)+O(|s|/n)$, and \eqref{5.4} yields
\begin{equation}
I_n(\bold X)=(1+o_p(1))\frac{4\sqrt{3}}{\pi Mn}. \label{6.5}
\end{equation}
In addition, the exponential factor in the formula for
$Z_n(\ell,s)$ becomes $2^ne^{S_n}$, where
\begin{equation}
\begin{aligned}
S_n=&-4\frac{s^2}{n}+O(|s|/(nM)+|s|^3/n^2)\\
&+\frac{1}{2}\sum_{j=1}^n(\zeta X_j/M+\eta)^2+O(s^4/n^3).
\end{aligned}
\label{6.6}
\end{equation}
By the central limit theorem and \eqref{6.2}, the sum can be
written as
\begin{equation}
\begin{aligned}
&\zeta^2\big(n\ex(U^2)+O_p(n^{1/2})\big)
+2\zeta\eta\big(n\ex(U)+O_p(n^{1/2})\big) +\eta^2n
\\
=&4\frac{s^2}{n}+O(|s|^3/n^2)+O_p(s^2/n^{3/2})
\\
=&4\frac{s^2}{n}+o_p(1),
\end{aligned}
\end{equation}
uniformly for $|s|\le n^{2/3}\omega^{-1}(n)$, where $\omega(n)\to
\infty$, however slowly. Thus \eqref{6.6} simplifies to
\begin{equation}
S_n=-2\frac{s^2}{n}+o_p(1).
\end{equation}
Using this formula and \eqref{6.5} we obtain
\begin{equation}
Z_n(\ell,s)=(1+o_p(1))\frac{4\sqrt{3}}{\pi} \frac{2^n}{
Mn}e^{-2s^2/n}, \label{6.9}
\end{equation}
uniformly for $|s|\le n^{2/3}\omega^{-1}(n)$ and $|\ell|\leq 1$;
and, of course, $s\equiv n(\text{mod }2)$, and
$\ell\equiv\sum_jX_j\,(\text{mod }2)$.
In \cite{BCP2}  it was shown that, for $Mn^{1/2}/2^n\to 0$,
$Y_n(\ell)=\sum_{s=-\infty}^{\infty} Z_{\ell,s}$, the total number
of partitions%
\footnote{Note that $Y_n(\ell)$ is equal to the total number of
perfect partitions if $\ell=0$, and equal to half the total number
of perfect partitions if $\ell=1$. In a similar way, $Z_n(\ell,s)$
is the total number of perfect partitions with bias $s$ if
$\ell=0$, and half that number if $\ell=1$.}
 with
$\bss\cdot\bold X=\ell$ is asymptotic, in probability, to
$\frac{2^{n+1}\sqrt{3}}{M\sqrt{2\pi n}}$. Let $s_n$ denote the
bias of a perfect partition $\bss^{(n)}$ chosen uniformly at
random from all perfect partitions. The formula for $Y_n(\ell)$
and \eqref{6.9} prove the following corollary of
Theorem~\ref{thm5.1}.

\begin{corollary}
\label{cor6.1} Assume that $\limsup \frac 1n\log_2M<1$. Then
\begin{equation}
\pr(s_n=s|\bold X)= (1+o_p(1))\frac{2\sqrt{2}}{\sqrt{\pi
n}}e^{-2s^2/n}
\end{equation}
uniformly for $|s|\le n^{2/3}\omega^{-1}(n)$, ($s\equiv
n(\text{mod }2)$), and consequently
\begin{equation}
\pr\left(\left.\frac{|s_n|}{\frac{{n^{1/2}}}{2}}\le a\right|\bold
X\right) \Longrightarrow_p
\sqrt{\frac{2}{\pi}}\int\limits_0^ae^{-u^2/2}\,du. \label{6.11}
\end{equation}
\end{corollary}
\begin{remark}\label{rem6.1} Thus the bias of the randomly
selected (typical) perfect partition is exactly of order $n^{1/2}$,
just like $t_n$, the bias of the sequence of $n$ flips of a fair
coin, i.e.~the difference between number of heads and number of
tails of a fair coin. However,
\begin{equation}
\pr\left(\frac{|t_n|}{{n^{1/2}}}\le
a\right)\Longrightarrow\sqrt{\frac{2}{\pi}}\int
\limits_0^ae^{-u^2/2}\,du,
\end{equation}
i.e. {\em in distribution} the bias of the random perfect
partition is, in the limit, half as large as the bias of the
sequence of $n$ coin flips.
Perhaps we should have anticipated
some reduction of the typical bias, since perfect partitions
are by definition those with the smallest discrepancy, which
should favor smaller bias.
In retrospect, the fact that most perfect partitions turn out to
have small bias may be responsible for the greater mathematical
tractability of the unconstrained problem.
\end{remark}

\section{The Hard Phase}
\label{sec:hard}

\subsection{Lower Bounds on the Minimal Discrepancy}
Cautiously extrapolating the asymptotic formula in
Theorem~\ref{thm5.1}, we expect that w.h.p. there will be
no perfect partitions in the domain where this expression tends to
zero, that is where $\kappa>\kappa_c(b)$. Here we establish this
result outside a small window of width $n^{-1/2}$ above $\kappa_c$.

\begin{theorem}
\label{thm6.1} Suppose that $\limsup |s|/n<b$ and that
$0<\liminf\kappa\leq\limsup\kappa<\infty$.
Let $S_n$ be the random variable defined in \eqref{5.8},
let $\ell_n$ be a sequence of positive integers
with $\ell_n=o(Mn^{1/2})$ and let $|\ell|<\ell_n$.
Then, with probability at least
$1-O(e^{-\log^2 n})$,
\begin{equation}
\label{6.13}
Z_n(\ell,s)
\leq
2^{[\kappa_c(b)-\kappa]n}
e^{n^{1/2}S_{n}}
e^{\log ^2n+O(\ell/M)}
\end{equation}
and
\begin{equation}
\label{6.14}
\sum_{\ell:|\ell|<\ell_n}
Z_n(\ell,s)\leq
\ell_n
2^{[\kappa_c(b)-\kappa]n}
e^{n^{1/2}S_{n}}
e^{\log ^2n+O(\ell_n/M)}.
\end{equation}
If $(\kappa-\kappa _c(b)){n^{1/2}}\to\infty$, we thus have
\begin{equation}
\label{6.15}
d_{opt}\ge 2^{[\kappa-\kappa_c(b)-O_p(n^{-1/2})]n},
\end{equation}
so that w.h.p.~there are no perfect partitions.
\end{theorem}

\begin{remark}
\label{rem6.3}
(i) The proof of \eqref{6.15} can be generalized to
show that there is a constant $\delta>0$ such that,
with probability $1-O(e^{-\delta\log^2 n})$,
\begin{equation}
\label{6.15a}
d_{opt}\ge \lfloor 2^{[\kappa-\kappa_c(b)-n^{-1/2}\log n]n}\rfloor.
\end{equation}
For $\kappa-\kappa_c(b)\geq 1+ n^{-1/2}\log n$,
there are therefore no perfect partitions
with probability $1-O(e^{-\delta\log^2 n})$.

(ii) If $|\kappa-\kappa_c(b)| =O( n^{-1/2})$, the
term $n^{1/2} S_{n}$ is larger than $n|\kappa-\kappa_c(b)|$
with positive probability, implying that with positive
probability, $Z_n(\ell,s)$ is smaller than $1$, and hence zero.
The above theorem therefore implies that for
$|\kappa-\kappa_c(b)|=O(n^{-1/2})$, the probability that there are
no perfect partitions stays bounded away from zero.

(iii) As mentioned in
Remark~\ref{REM2}, we believe
that $d_{opt} =
2^{[\kappa-\kappa_c+o_p(1)]n}$
whenever $\liminf(\kappa-\kappa_c)>0$.
In other words, we believe that for every
$\eps>0$, w.h.p.,
$d_{opt} \leq
2^{[\kappa-\kappa_c+\eps]n}$.
If we assume such a bound, then w.h.p.~the number of
optimal partition $Z_{opt}$ is bounded by the
right hand side of \eqref{6.14} with
$\ell_n=\lfloor 2^{[\kappa-\kappa_c+\eps]n}\rfloor$,
implying that w.h.p.~$Z_{opt}\leq 2^{2\eps n}$.
Since $\eps$ was arbitrary, we get
$Z_{opt} = e^{o_p(n)}$ whenever
$\liminf(\kappa-\kappa_c)>0$.
\end{remark}

\noindent {\bf Proof of Theorem~\ref{thm6.1} and
Remark~\ref{rem6.3} (i).\/} The bounds \eqref{6.15} and
\eqref{6.15a} follow from \eqref{6.14}. Indeed, let $\omega_n$ be
such that $\omega_n\to\infty$ and
$[\kappa-\kappa_c(n)]{n^{1/2}}-\omega_n\to\infty$ as $n\to
\infty$.  Setting
$\ell_n=\lfloor2^{[\kappa-\kappa_c(b)]n-{n^{1/2}}\omega_n}\rfloor$,
the bound \eqref{6.14} immediately implies $d_{opt}\geq \ell_n$,
which gives \eqref{6.15}. To prove \eqref{6.15a}, we set
$\ell_n=\lfloor 2^{[\kappa-\kappa_c(b)-n^{-1/2}\log n]n}\rfloor$
and observe that for $\delta$ sufficiently small, $S_n$ is bounded
by $ \frac 12\log n$ with probability $1-O(e^{-\delta\log^2 n})$.
As a consequence, the r.h.s.~of \eqref{6.14} goes to zero with
probability $1-O(e^{-\delta\log^2 n})$, implying again that
$d_{opt}\geq \ell_n$.

It is thus enough to prove \eqref{6.13} and
\eqref{6.14}. Note that the bound \eqref{6.14} is not just a
consequence of the bound \eqref{6.13} since the intersection of
$2\ell_n-1$ events happening with high probability does not
necessarily happen with high probability if $\ell_n$ is not
bounded.

Using \eqref{5.50}, we rewrite
$Z_n(\ell,s)$ as
\begin{equation}
\label{6.16}
Z_n(\ell,s)=
2^{n\kappa_c(b)}e^{{n^{1/2}}\, S_n+\zeta{\ell}/{M}+O(nM^{-1})}
{\mathcal  Z_n(\ell,s)},
\end{equation}
where $S_n$ is defined in \eqref{5.8}
and
\begin{equation}
\label{6.18}
\mathcal Z_n(\ell,s)= \frac{1}{2\pi^2}
\int\limits_{x\in (-\pi,\pi]\atop{y\in(-\pi/2,\pi/2]}}
e^{-i(\ell x+sy)}f(x,y;\bold X)\,dxdy,
\end{equation}
with $f(x,y;\bold X)$ as defined in \eqref{5.3}.
In contrast to \eqref{5.1}, the relation \eqref{6.16}
holds whether $\ell$ has the same parity as
$\sum_jX_j$ or not, since $x$ is now integrated over $[-\pi,\pi)$
instead of $[-\pi/2,\pi/2)$.
Introducing finally
\begin{equation}
\label{6.19}
\mathcal Z_n(s)=
\sum_{\ell:|\ell|<\ell_n}
\mathcal Z_n(\ell,s),
\end{equation}
we have
\begin{equation}
\label{6.20}
\begin{aligned}
\sum_{\ell:|\ell|<\ell_n}
Z_n(\ell,s)
&=2^{n\kappa_c}e^{n^{1/2} S_n+O(\ell_n/M)+O(nM^{-1})}
\mathcal Z_n(s).
\end{aligned}
\end{equation}
Since $nM^{-1}=o(\log ^2n)$,
the bounds \eqref{6.13} and
\eqref{6.14} are equivalent
to proving that, with probability $1-O(e^{-\log^2 n})$,
we have $\mathcal Z_n(\ell,s)\leq M^{-1}e^{\log^2 n}$ and
$\mathcal Z_n(s)\leq\ell_n M^{-1}e^{\log^2 n}$.

We will prove these bounds by establishing a suitable bound
on the expectation of $\mathcal Z_n(\ell,s)$.
To this end, we first
rewrite $\ex(\mathcal Z_n(\ell,s))$ as
\begin{align}
\ex(\mathcal Z_n(\ell,s))=&\frac{1}{\pi^2}\iint\limits_{x\in
(-\pi,\pi]\atop y\in (-\pi/2, \pi/2]}
e^{-i\ell x-isy}f^n(x,y)\,dxdy,
\label{6.21}
\\
f(x,y)=&\frac{1}{M}\sum_jp(j/M)e^{i(xj+y)}
+\frac{1}{M}\sum_jq(j/M)e^{-i(xj+y)}. \label{6.22}
\end{align}
As in the estimates of $\ex(|I_{n2}(\bold X)|^2)$ in the proofs
of Theorem~\ref{thm5.1}, we need the bounds of $|f(x,y)|$ for
various ranges of $x,y$.

Pick $B>0$. Let $|x|\ge B/M$. Using Lemma \ref{lem5.1} and
\eqref{6.22}, we get
\begin{equation}
|f(x,y)|\le c/B.
\label{6.23}
\end{equation}
Let $|x|\le B/M$. Setting $x=z/M$, we have $|z|\le B$. For these
$x$'s, let us bound $|f(x,y)|$ more sharply. We have
\begin{multline}
\left|\frac{1}{M}\sum_{j=1}^Mp(j/M)e^{i(zj/M)+iy}\right|^2\\
=\frac{1}{M^2} \sum_{j=1}^Mp^2(j/M)+\frac{2}{M^2}
\sum_{j_1<j_2}p(j_1/M)p(j_2/M)\cos(z(j_1/M-j_2/M))\\
=\left(\frac{1}{M}\sum_{j=1}^Mp(j/M)\right)^2
-\frac{2}{M^2}\sum_{j_1<j_2}p(j_1/M) p(j_2/M)
\bigl(1-\cos(z(j_1/M-j_2/M))\bigr).
\label{6.24}
\end{multline}
Here
\begin{equation}
\frac{1}{M}\sum_{j=1}^Mp(j/M)=\int_0^1p(u)\,du+O(M^{-1}).
\end{equation}
Pick a small $\eps$ and consider $j_1<j_2$ such that
\begin{equation}
B\frac{j_2-j_1}{M}\le 2\pi-\eps\Longleftrightarrow j_2-j_1\le \be
M,\,\be= \frac{2\pi-\eps}{B}.
\end{equation}
Clearly there are $\Theta(\beta M^2)$ such pairs $(j_1,j_2)$. For
those $(j_1,j_2)$, and $|z|\le B$, there is a positive constant
$c=c(\eps)$, such that
\begin{equation}
1-\cos(z(j_1/M-j_2/M))\ge c\bigl[z(j_1/M-j_2/M)\bigr]^2
\end{equation}
So the double sum in \eqref{6.24} is bounded below by
\begin{equation}
\begin{aligned}
\frac{cz^2}{M^2}
\sum_{|j_1/M-j_2/M|\le\be}p(j_1/M)&p(j_2/M)(j_1/M-j_2/M)^2
=(c^\prime B^{-1}+O(M^{-1})) z^2,\\
c^\prime =&\frac cB\iint\limits_{u_1,u_2\in [0,1]\atop
|u_1-u_2|\le \be}p(u_1)p(u_2) (u_1-u_2)^2\,du_1du_2.
\end{aligned}
\end{equation}
We thus obtain
\begin{equation}
\left|\frac{1}{M}\sum_{j=1}^Mp(j/M)e^{i(zj/M+y)}
\right|^2\le\left(\int_0^1p(u)\,du+O(1/M) \right)^2-(c^\prime
B^{-1}+O(M^{-1}))z^2,
\end{equation}
so that
\begin{equation}
\left|\frac{1}{M}\sum_{j=1}^Mp(j/M)e^{i(zj/M+y)}\right|
\le\left(\int_0^1p(u)\,du+ O(M^{-1})\right)e^{-\a_1B^{-1} z^2},
\label{6.30}
\end{equation}
for some positive constant $\a_1$. Analogously to \eqref{6.30},
\begin{equation}
\left|\frac{1}{M}\sum_{j=1}^Mq(j/M)e^{-i(zj/M+y)}\right|\le
\left(\int_0^1q(u)\,du+ O(1/M)\right)e^{-\a_2 B^{-1}z^2},\quad
(\a_2>0),
\end{equation}
hence, for $\a=\min\{\a_1,\a_2\}$,
\begin{equation}
|f(x,y)|\le (1+O(M^{-1}))e^{-\a B^{-1} z^2} \le e^{-\a^\prime
B^{-1} z^2},\quad \a^\prime>0, \label{6.32}
\end{equation}
for $ B\ge |z|\ge \Theta(\sqrt{B/M})$. The bounds \eqref{6.23} and
\eqref{6.32} indicate that $|f(x,y)|^n$ is small for large and
moderate values of $z (=xM)$, {\it regardless\/} of $y$. The value
of $y$ begins to matter when $z$ is small. To see how, we need yet
a sharper bound for $|f(x,y)|$ for small $|z|$. By the definition
 \eqref{6.22} of $f(x,y)$, we have
\begin{equation}
\begin{aligned}
|f(x,y)|^2=&\frac{1}{M^2}
\sum_{j_1,j_2}p(j_1/M)p(j_2/M)e^{iz(j_1/M-j_2/M)}
\\
&+\frac{1}{M^2}\sum_{j_1,j_2}q(j_1/M)q(j_2/M)e^{iz(j_2/M-j_1/M)}
\\
&+\frac{1}{M^2}\sum_{j_1,j_2}p(j_1/M)q(j_2/M)e^{i(z(j_1/M+j_2/M)+2y)}
\\
&+\frac{1}{M^2}\sum_{j_1,j_2}q(j_1/M)p(j_2/M)e^{-i(z(j_1+j_2)+2y)},
\end{aligned}
\label{6.33}
\end{equation}
so that
\begin{equation}
\begin{aligned}
|f(x,y)|^2\le&\left(\frac{1}{M}\sum_jp(j/M)\right)^2+
\left(\frac{1}{M}\sum_jq(j/M)\right)
^2\\
&+2\left(\frac{1}{M}\sum_jp(j/M)\right)\,
\left(\frac{1}{M}q(j/M)\right)\\
&-\frac{2}{M^2}\sum_{j_1,j_2}p(j_1/M)q(j_2/M)
\bigl[1-\cos(z(j_1/M+j_2/M)+2y)\bigr].
\end{aligned}
\label{6.34}
\end{equation}
Here the first three terms add up to
\begin{equation}
\left(\frac{1}{M}\sum_j(p(j/M)+q(j/M))\right)^2=1.
\end{equation}
Furthermore, picking $z_0>0$ small enough so that
\begin{equation}
\label{6.36}
2z_0+2|y|\le 2(z_0+\pi/2)< 2\pi-\eps,
\end{equation}
we get: for $|z|\le z_0$,
\begin{equation}
1-\cos(z(j_1/M+j_2/M)+2y)\ge c\bigl[z(j_1/M+j_2/M)+2y\bigr]^2.
\end{equation}
The first inequality in
\eqref{6.36} holds because
$|y|\le\pi/2$, a consequence of $s\equiv n(\mbox{mod }2)$.

So, within the factor $2c$, the double sum in \eqref{6.34} exceeds
\begin{equation}\begin{aligned}
&\frac 1{M^2}\sum_{j_1,j_2}p(j_1/M)q(j_2/M)
\bigl[z^2(j_1/M+j_2/M)^2+4yz(j_1/M+j_2/M)+4y^2\bigr]\\
=&z^2\left(\,\iint\limits_{u_1,u_2\in
[0,1]}p(u_1)q(u_2)(u_1+u_2)^2\,du_1du_2+O(M^{-1})
\right)\\
&+4yz\left(\,\iint\limits_{u_1,u_2\in
[0,1]}p(u_1)q(u_2)(u_1+u_2)\,du_1du_2+O(M^{-1})
\right)\\
&+4y^2\left(\,\iint\limits_{u_1,u_2\in
[0,1]}p(u_1)q(u_2)\,du_1du_2+O(M^{-1})\right),
\end{aligned}
\end{equation}
Since the functions $1$ and $u_1+u_2$ are linearly independent,
there exists $c_0>0$ such that, for $M$ large enough, the
quadratic form is bounded below by $c_0(z^2+y^2)$. So
\begin{equation}
|f(x,y)|^2\le 1-c_0(z^2+y^2)\Longrightarrow |f(x,y)| \le
e^{-c_0(z^2+y^2)/2}, \label{6.39}
\end{equation}
if $|z|\le z_0,\,y\in (-\pi/2,\pi/2],\,\,M\ge M(z_0)$. Putting
\eqref{6.32} and \eqref{6.39} together enables us to conclude that
\eqref{6.39}, with a possibly smaller $c_0$, holds for all $z,y$
with $|z|\le B$ and $|y|\le \pi/2$. Note however, that $c_0$ now
depends on $B$ and goes to zero like $B^{-1}$ as $B\to\infty$.
Making this dependence explicit, we have that
\begin{equation}
|f(x,y)| \le e^{-c_0^\prime(z^2+y^2)/B} \label{6.40}
\end{equation}
whenever $M$ is large enough, $|z|\le B$ and $|y|\le \pi/2$.

 Finally, for $|z|,|y|$ both small
\begin{equation}
e^{\pm i(z(j/M)+y)}=1 \pm i(z\frac jM+y)-\frac{1}{2}(z\frac
jM+y)^2 +O(|z|^3+|y|^3).
\end{equation}
So
\begin{equation}
\begin{aligned}
f(x,y)
=&1+i\left(z\frac{1}{M}\sum_j\frac jM \Big(p(\frac
jM)-q(\frac jM)\Big)+y \frac{1}{M}\sum_j\Big(p(\frac jM)-
q(\frac jM)\Big)\right)
\\
&-\frac{1}{2}\frac{1}{M} \sum_j(z^2(\frac jM)^2+2zy\frac
jM+y^2)+O(|z|^3+|y|^3)\\
&=1+ia_nz +i(b_n+s/n)y
-\frac{1}{2}\frac{1}{M} \sum_j(z^2(\frac jM)^2+2zy\frac
jM+y^2)+O(|z|^3+|y|^3),
\end{aligned}
\end{equation}
where, by the definitions
$p(u)=P(\zeta u+\eta)$ and $q(u)=1-P(\zeta u+\eta)$,
and equations
\eqref{3.8} and \eqref{4.2},
\begin{align}
a_n=\frac{1}{M}\sum_j\frac jM\Big(p(\frac jM)-q(\frac jM)\Big)
=&\int_0^1u(p(u)-q(u))\,du+O(M^{-1})
=O(M^{-1})
\end{align}
\begin{align}
b_n=
\frac{1}{M}\sum_j\Big(p(\frac jM)-q(\frac jM)\Big)-\frac sn
=&\int_0^1(p(u)-q(u))\,du-b+O(M^{-1})
=O(M^{-1}).
\end{align}
Exponentiating the expansion for $f$, we arrive at
\begin{equation}
\begin{aligned}
f(x,y)=&\exp\left(iza_n+iy\left(\frac{s}{n}+b_n
\right)\right)
\\
&\times\exp\left(-\frac{1}{2}(z,y)\mathcal
Q(z,y)^\prime+O(|z|^3+|y|^3)\right),
\end{aligned}
\label{6.45}
\end{equation}
where
\begin{equation}
\begin{aligned}
\mathcal Q_{11}=&\ex[(X/M)^2]+O(M^{-2})\to \ex[(X/M)^2],\\
\mathcal Q_{12}=& \ex(X/M)+O(bM^{-1})\to \ex(X/M),\\
\mathcal Q_{22}=&1-(b+O(M^{-1}))^2\to 1-b^2.
\end{aligned}
\label{6.46}
\end{equation}
Note that the matrix $\mathcal Q$ is positive
definite in the limit since
\begin{equation}
\ex[(X/M)^2]\,(1-b^2)-\ex^2(X/M)\to\frac{1}{3}(1-b^2)-\frac{1}{4}=
\frac{1-4b^2}{12}>0,
\end{equation}
since $b<1/2$. By \eqref{6.45} and \eqref{6.46}
\begin{equation}
e^{-i\ell x-isy}f^n(x,y)
=\exp\left(iz(\tilde a_n+iyb_n-\frac{n}{2}(z,y)\mathcal Q
(z,y)^\prime+ O(n(|z|^3+|y|^3))\right), \label{6.48}
\end{equation}
where $\tilde a_n= n a_n-\ell/M$ and $\tilde b_n=n b_n$.

Let us derive an asymptotic formula for $\ex(\mathcal Z_n(\ell,s))$,
using \eqref{6.23}, \eqref{6.40}, and \eqref{6.45}-\eqref{6.48}.
By \eqref{6.21} and \eqref{6.23}, the contribution
of the $(x,y)$ with $|x|>B/M$ to $\ex(\mathcal Z_n(\ell,s))$
 is of order
$O((c/B)^n)$. Let $|x|\le B/M$. Switch to $z=Mx, y=y$. By
\eqref{6.40}, the contribution of the $(z,y)$'s
with $\|(z,y)\|\ge r_n:=n^{-1/2}\log n$ is of order
\begin{equation}
M^{-1}\int\limits_{r\ge r_n}r e^{-nc_0^\prime r^2/B}\,d r=
M^{-1}e^{-\Theta(B^{-1}\log^2 n)}. \label{6.49}
\end{equation}
By \eqref{6.45}-\eqref{6.48}, the contribution of
the $(z,y)$'s with $\|(z,y)\| \le r_n$ is asymptotic to
\begin{multline}
\label{6.50}
\frac{1}{2M\pi^2} \!\!\!\!
\iint\limits_{\|(z,y)\|\le r_n}\!\!\!\!
\exp\left(iz\tilde a_n+iy\tilde b_n-\frac{n}{2}(z,y)\mathcal Q
(y,z)^\prime\right)
\Big(1+O(n|z|^3+n|y|^3)\Big)\,dzdy\\
=\frac{1}{2M\pi^2}\iint\limits_{\|(z,y)\|\le
r_n}\exp\left(iz\tilde a_n+iy\tilde b_n-\frac{n}{2}(z,y)
\mathcal Q(z,y)^\prime\right)\,dzdy
+O\Big(\frac{1}{Mn}n^{-1/2}\Big).
\end{multline}
The first term on the r.h.s.~of \eqref{6.50} is equal to
\begin{equation}
\frac{1}{Mn\pi\sqrt{\text{det }\mathcal Q}}
\left[\exp\left(-\frac{1}{2n}\|(\tilde a_n,b_n) \mathcal
Q^{-1/2}\|^2\right) +O\Big(\int\limits_{t\ge \Theta(\log
n)}te^{-t^2/2}\,dt\Big)\right],
\end{equation}
where
the exponential term is
\begin{equation}
1+O(n^{-1}(\tilde a_n^2+b_n^2))=1+O(n/M^{2})+O(n^{-1}(\ell/M)^2=
1+O(e^{-\Theta(n)})+O(n^{-1}(\ell/M)^2),
\end{equation}
while the integral is of order $e^{-\Theta(\log^2 n)}$. Combining
the last formula with the estimates of contributions of other
ranges of $(x,y)$, we obtain
\begin{equation}
\begin{aligned}
\ex(\mathcal Z_n(\ell,s))
&=(1+O(n^{-1/2})
+O((\ell/M)^2 n^{-1}))\frac{1}{Mn}\,
\frac{1}{\pi\sqrt{(1-b^2)\ex(U^2)-\ex^2(U)}}
\\
&=(1+o(1))
\frac{2\sqrt 3}{\pi\sqrt{1-b^2}}\frac{1}{Mn},
\end{aligned}
\label{6.53}
\end{equation}
whenever $|\ell|\leq\ell_n=o({n^{1/2}} M)$.
Summing over $\ell\in\{-(\ell_n-1),\dots,\ell_n-1\}$,
the bound \eqref{6.53}
clearly implies a similar bound for $\mathcal Z_n(s)$,
namely
\begin{equation}
\ex(\mathcal Z_n(s))
=(1+o(1))
\frac{2\sqrt 3}{\pi\sqrt{1-b^2}}\frac{2\ell_n-1}{Mn}.
\label{6.54}
\end{equation}
By the bounds \eqref{6.53} and \eqref{6.54}
and Markov's inequality, we have that
$Z_n(\ell,s)\leq M^{-1} e^{\log^2 n}$
and
$\mathcal Z_n(s)\leq \ell_n M^{-1}e^{\log^2 n}$
with probability $1-O(e^{-\log^2 n})$.
Combined with \eqref{6.16} and \eqref{6.20}
this implies \eqref{6.13} and \eqref{6.14}.
\qed

\subsection{A Digression: The Expected Number of Perfect Partitions}
\label{sec:expected}

The reader may have noticed how gingerly we have tiptoed
around the first factor in \eqref{5.1}, concentrating instead on the
asymptotic behavior of the double integral. To see why, we will
show that $\ex(Z_n(\ell,s))$ remains exponentially large above the
curve $\kappa=\kappa_c(b)$, in sharp contrast to the fact
that, in this domain,
w.h.p.~there are no
perfect partitions at all.

 \begin{theorem}
\label{thm6.2} Let $|\ell|\leq 1$. For $s>0$, $s+n\equiv
0(\text{mod }2)$, and $b=s/n$ bounded away from $0$ and $1$,
\begin{equation}
\ex[Z_n(\ell,s)]
\sim\frac{2}{M}\binom{n}{\frac{n+s}{2}}\,\frac{e^{\la
s}\phi^n(\la)} {\sqrt{2\pi n\var(U_{\la})}},\quad
\phi(\la):=\frac{\sinh\la}{\la}. \label{6.56}
\end{equation}
where $U_{\la}\in [-1,1]$ has density
\begin{equation}
\frac{e^{\la x}/2}{\int_{y\in [-1,1]}e^{\la y}/2\,dy},
\end{equation}
and $\la<0$ is such that $\ex (U_{\la})=-b$, or explicitly:
\begin{equation}
\coth\la-\frac{1}{\la}=-b.
\end{equation}
Consequently,
\begin{equation}
\lim_{n\to\infty}\frac{1}{n}\log\ex[Z_n(\ell,s)]=R(\k,b),
\end{equation}
where
\begin{equation}
R(\k,b)=\frac{1+b}{2}\log\frac{2}{1+b}
+\frac{1-b}{2}\log\frac{2}{1-b}+\la b+
\log\frac{\sinh\la}{\la}-\k\log 2.
\end{equation}
\end{theorem}

\noindent {\bf Proof of Theorem~\ref{thm6.2}.\/}
In order to calculate the expectation of
$Z_n(\ell,s)$, we
use the following analogue of
\eqref{3.6}, which is valid whether or not
$\ell$ has the same parity as $\sum_jX_j$:
\begin{equation}
Z_n(\ell,s)
=\frac{2^n}{2\pi^2}
\iint\limits_{x\in(-\pi,\pi]\atop y\in\in(-\pi/2,\pi/2]}
e^{-i(\ell x+sy)}
\prod_{j=1}^n\cos\bigl(xX_j+y\bigr)\,dxdy.
\end{equation}
This gives
\begin{equation}
\ex(Z_n(\ell,s))
=\frac{2^n}{2\pi^2}
\iint\limits_{x\in(-\pi,\pi]\atop y\in\in(-\pi/2,\pi/2]}
e^{-i(\ell x+sy)}\ex^n\bigl(\cos(xX+y)\bigr)\,dx
dy,
\end{equation}
where
\begin{equation}
\begin{aligned}
\ex\bigl(\cos(xX+y)\bigr)
=&\text{Re }
\left[e^{iy}M^{-1}\sum_{j=1}^M e^{ixj}\right]
=\text{Re }
\left[M^{-1}e^{i(x+y)}\,\frac{e^{ixM}-1}{e^{ix}-1}\right]
\\
=&\frac{\sin\Big(\frac{Mx}{2}\Big)}{M\sin\Big(\frac{x}{2}\Big)}
\,\cos\left(\frac{(M+1)x}{2}+y\right).
\end{aligned}
\end{equation}
Notice that the first factor is a function of $x$ only,
and the argument of
the cosine is $y$ shifted by $x(M+1)/2$. Integrating first with
respect to $y$, we can choose $y\in
(-\pi/2+x(M+1)/2,\pi/2+x(M+1)/2]$, thus reducing the $y$-integral
to
\begin{equation}
e^{isx\frac{M+1}{2}}\,\frac{1}{\pi}\int\limits_{y\in
(-\pi/2,\pi/2]}e^{-isy}\cos^n y \,dy.
\end{equation}
Here we can write $\cos y=\ex (e^{iy\sig})$,
$\pr(\sig=\pm 1)=1/2$. So,
introducing the independent copies $\sig_1,\dots,\sig_n$ of
$\sig$, we write the last integral as
\begin{equation}
\pr\left(\sum_{j=1}^n\sig_j=s\right)
=\frac{1}{2^n}\binom{n}{\frac{n+s}{2}}.
\end{equation}
Therefore we get the simpler expression
\begin{align}
\ex(Z_n(\ell,s)) =&\binom{n}{\frac{n+s}{2}}\,\frac{1}{2\pi}
\int\limits_{x\in (-\pi,\pi]}e^{ixt}
\left(\frac{\sin\frac{Mx}{2}}{M\sin\frac{x}{2}}\right)^n\,dx;
\\
t:=&-\ell+s\frac{M+1}{2}.
\end{align}

To simplify the rest of the exposition, let us assume that $\sin x/2$ in
denominator can be replaced asymptotically by $x/2$. (A refined
version of the argument shows that this replacement is
asymptotically correct for $n=o(M)$. We leave it to the interested
reader to check this.) Substituting $z=Mx/2$, we get then
\begin{equation}
\begin{aligned} \ex(Z_n(\ell,s))\sim
&\binom{n}{\frac{n+s}{2}}\,\frac{1}{\pi M}\int\limits_{z\in
(-M\pi/2,M\pi/2]}e^{i\tau z}\left(\frac{\sin z}{z}\right)^n\,dz,
\\
\tau:=&s(1+M^{-1})-2\ell/M,
\end{aligned}
\label{6.67}
\end{equation}
so that $\tau$ is very close to $s$, since $|\ell|\le 1$. Next, we
observe that $y^{-1}\sin y$ is the characteristic function of the
random variable $U$, uniformly distributed on $[-1,1]$. Hence
$\left(\frac{\sin z}{z}\right)^n$ is the characteristic function
of $S_n=\sum_{j=1}^n U_j$, where $U_j$ are the independent copies
of $U$. In particular, using the Fourier inversion formula for
$f_{S_n}$, the density of $S_n$,
\begin{equation}
f_{S_n}(-\tau)=\frac{1}{2\pi}\int\limits_{z\in
(-\infty,\infty)}e^{i\tau z}\left( \frac{\sin z}{z}\right)^n\,dz,
\end{equation}
we get from \eqref{6.67}:
\begin{equation}
\ex(Z_n(\ell,s))\sim\frac{2}{M}\binom{n}{\frac{n+s}{2}}
f_{S_n}(\tau).
\end{equation}
(The error caused by extending the integration
interval in \eqref{6.67} to
$(-\infty,\infty)$ is extremely small, of order $M^{-n}=2^{-\k
n^2}$.) We cannot use the local limit theorem for $f_{S_n}(-\tau)$
directly as $\ex S_n= n\ex U\neq -\tau$. So we introduce an
auxiliary random variable $U_{\la}$ with density proportional to
$e^{\la x}$ on $[-1,1]$, choosing $\la$ such that $\ex
(U_{\la})=-\tau/n$, that is setting $\la$ equal the root of
\begin{equation}
\frac{\phi^\prime(\la)}{\phi(\la)}=-\frac{\tau}{n},\quad
\phi(\la):=\frac{1}{2} \int\limits_{x\in [-1,1]}e^{\la
x}\,dx=\frac{\sinh\la}{\la}.
\end{equation}
Then
\begin{equation}
f_{S_n}(-\tau)=e^{-\la(-\tau)}\phi^n(\la)f_{S_{n,\la}}(-\tau),
\end{equation}
where $S_{n,\la}=\sum_{j=1}^nU_{j,\la}$. By the local limit
theorem (see, e.g., \cite{GrStirz}, Sect.~5.10), the density
$f_{S_{n,\la}}(-\tau)$ of $S_{n,\la}$ is asymptotic to
\begin{equation}
\frac{1}{\sqrt{2\pi n\var{U_{\la}}}},
\end{equation}
where $\var(U_{\la})$ can be explicitly calculated, giving
\begin{equation}
\var(U_{\la})=\frac{\phi^{\prime\prime}(\la)}{\phi(\la)}-(\tau/n)^2.
\end{equation}
Thus
\begin{equation}
\ex(Z_n(\ell,s))
\sim
\frac{2}{M}\binom{n}{\frac{n+s}{2}}
e^{\la\tau}\phi^n(\la)f_{S_{n,\la}}(-\tau)
\sim\frac{2}{M}\binom{n}{\frac{n+s}{2}}\,\frac{e^{\la
s}\phi^n(\la)} {\sqrt{2\pi n\var(U_{\la})}},
\end{equation}
as claimed.
\qed

\subsection{Optimal Imperfect Partitions}
\label{sec:sorted-partitions}

In Theorem~\ref{thm6.1} we proved that, for $\kappa>\kappa_c(b)$
{\it and\/} $b<b_c$, w.h.p.~the minimum discrepancy is at least
$2^{n(\kappa-\kappa_c(b))-O_p({n^{1/2}})}$. The next theorem
provides a complementary upper bound for the minimum
discrepancy.

\begin{theorem}
\label{thm7.1}
Suppose that $\limsup |s|/n<b$ and that
$0<\liminf\kappa\leq\limsup\kappa<\infty$.
Let $\eps>0$,
let $S_n$ be the random variable defined in \eqref{5.8},
and let $\ell_n$ be a sequence
of positive integers with
$\ell_n=o(Mn^{1/2}/\log n)$
and $\ell_n\geq  2^{[\kappa-\kappa_-(b)+\eps]n}$.
Then there is a constant $\delta>0$ such that,
with probability
$1-O(e^{-\delta\log^2 n})$,
\begin{equation}
\label{7.1}
\sum_{\ell:|\ell|<\ell_n}
Z_n(\ell,s)
=\ell_{n,\bold X} 2^{[\kappa_c(b)-\kappa]n}
e^{{n^{1/2}} S_{n}+o(n^{1/2})},
\end{equation}
where $\ell_{n,\bold X}$ is the number of integers
$\ell$ with $|\ell|<\ell_n$ with the same
parity as $\sum_jX_j$.
With probability
$1-O(e^{-\delta\log^2 n})$, we thus have
\begin{equation}
\label{7.2}
 d_{opt}\le
\lceil 2^{(\kappa-\kappa_-(b)+\eps)n}\rceil,
\end{equation}
implying in particular that
$M^{-1}d_{opt}$ is exponentially small in $n$.
\end{theorem}

%
Note that for $\limsup(\kappa-\kappa_-(b))<0$,
we recover that $d_{opt}\le 1$.
Not astonishingly, the proof of Theorem~\ref{thm7.1}
follows closely the proof of Theorem~\ref{thm5.1},
which established that, w.h.p., there are exponentially
many perfect partitions for $\limsup(\kappa-\kappa_-(b))<0$.

\bigskip
\noindent {\bf Proof of Theorem~\ref{thm7.1}.\/}
Let us first note that depending on the parity of
$\sum_jX_j$, the number $\ell_{n,\bold X}$ is either
$\ell_n$ or $\ell_n-1$, implying in particular
that $\ell_{n,\bold X}\geq \ell_n-1$.  Using
this fact, it is not hard to see that the bound
\eqref{7.2} follows from \eqref{7.1}.
Indeed, let
$\ell_n=\lceil 2^{(\kappa-\kappa_-+\eps)n}\rceil+1$.
Since $\ell_{n\bold X}\geq \ell_n-1\geq
2^{(\kappa-\kappa_-+\eps)n}$,
the right hand side of \eqref{7.1} goes to infinity,
implying that $d_{opt}\geq \ell_n-1$.  It is
therefore enough to prove \eqref{7.1}.
Also, since both sides of \eqref{7.1}
are identically zero if
$\ell_{n,\bold X}=0$, it is enough to consider
$\ell_{n,\bold X}>0$.

Analogously to
the proof of Theorem~\ref{thm6.1}, we set
\begin{equation}
\label{7.3}
\mathcal Z_n(s)=
\sum_{\ell:|\ell|<\ell_n}
\mathcal Z_n(\ell,s),
\end{equation}
where $\mathcal Z_n(\ell,s)$ is the random integral defined
in \eqref{6.18}.  Let us note, however, that
$\mathcal Z_n(\ell,s)$ is equal to the
random integral defined in \eqref{5.2}
if $\ell$ is of the same parity as $\sum_jX_j$,
and equal to zero otherwise.  The sum in \eqref{7.3}
can therefore be replaced by the sum over all
$\ell$ with the same parity as $\sum_jX_j$,
and
\begin{equation}
\mathcal Z_n(s)=\frac{1}{\pi^2}\iint\limits_{x\in
(-\pi/2,\pi/2]\atop y\in (-\pi/2,\pi/2]}
\bigg[{\sum_{\ell:|\ell|<\ell_n}}^\prime e^{-i\ell x} \bigg]
e^{-isy}f(x,y;\bold X)\,dx dy,
\label{7.4}
\end{equation}
where the sum $\sum^\prime$ indicates the sum over all $\ell$ with the
same parity as $\sum_jX_j$.
Obviously, we have
\begin{equation}
\label{7.5a}
\bigg|{\sum_{\ell:|\ell|<\ell_n}}^\prime e^{-i\ell x}
\bigg|
\leq \ell_{n,\bold X}
\end{equation}
and
\begin{equation}
\label{7.5}
{\sum_{\ell:|\ell|<\ell_n}}^\prime e^{-i\ell x}
=\ell_{n,\bold X} \big(1+O\big((\ell_n|x|)^2\big)\big)
\quad\text{as}\quad \ell_n|x|\to 0.
\end{equation}
Recalling that, depending on the parity of
$\sum_jX_j$, the number $\ell_{n,\bold X}$ of terms in the sum
$\sum^\prime$ is either $\ell_n$ or $\ell_n-1$, we also have
\begin{equation}
\label{7.6}
\bigg|{\sum_{\ell:|\ell|<\ell_n}}^\prime e^{-i\ell x}\bigg|
=\bigg|\sum_{k=0}^{\ell_{n,\bold X}-1}
e^{-i2kx}\bigg|
\leq 1+\bigg|\frac{\sin((\ell_n-1)x)}{\sin x}\bigg|
\leq \ell_n.
\end{equation}

As in the proof of Theorem~\ref{thm5.1}, we split
the right hand side of \eqref{7.4}
into two parts, $ J_{n1}(\bold X)$ and $J_{n2}(\bold X)$, for
$|x|\le B/M$ and $|x|>B/M$ respectively.
We start with $J_{n1}(\bold X)$.
As in \eqref{5.21}, we restrict ourselves
to the event $A_n$ defined in
\eqref{5.19}.
With the help of \eqref{7.5a},
we then bound the contributions of all $x$ with
$\|(Mx,y)\|\ge r_n=n^{-1/2}\log n$ by
\begin{equation}
\begin{aligned}
\frac {\ell_{n,\bold X}}{\pi^2}
\iint\limits_{\|(Mx,y)\|\ge r_n\atop|Mx|\leq B}
|f(x,y;\bold X)|\,dxdy
\,\leq&\,\frac{\ell_{n,\bold X}}{Mn}
\iint\limits_{\|\bold z\|\ge \log n}
e^{-c_4\|\bold z\|^2}
\,dz_1dz_2
=O\left(\frac{\ell_{n,\bold X}}{Mn}e^{-c_4(\log^2n)}\right).
\end{aligned}
\label{7.7}
\end{equation}
For $\|(Mx,y)\|\le r_n$,
we may use \eqref{7.5}
since
$\ell_n|x|\leq M^{-1}\ell_n r_n=o(r_n {n^{1/2}}\log^{-1}n)\to 0$.
And, analogously to \eqref{5.32},
we have that with probability at least $1-O(e^{-\delta_3\log^2 n})$
\begin{multline}
\frac 1{\pi^2}\iint\limits_{\|(Mx,y)\|\le r_n}
\bigg[{\sum_{\ell:|\ell|<\ell_n}}^\prime e^{-i\ell x}\bigg]
e^{-isy}f(x,y;\bold X)
\,dxdy\\
=\frac{\ell_{n,\bold X}}{Mn}
\left[\frac{1}{\pi\sqrt{\text{det }R}}\,
\exp\bigl(-\frac{1}{4} \boldsymbol{\tau}
_nR^{-1}\boldsymbol{\tau}_n^{\prime}\bigr)
+o(1) \right].
\label{7.8}
\end{multline}
Comparing \eqref{7.7} and \eqref{7.8} (and using again the bound
\eqref{tau-n-var}, this time to get a lower bound on the right
hand side of \eqref{7.8}), we see that
the last expression is a
sharp asymptotic formula for $J_{n1}(\bold X)$.

Next, we use \eqref{7.6} to bound
\begin{equation}
\ex(|J_{n2}(\bold X)|^2)\le \frac 1{\pi^4}
 \iiiint\limits_{|x_1|,|x_2|\in
[B/M,\pi/2]\atop y_1,y_2 \in (-\pi/2,\pi/2]}
(1+S(x_1))(1+S(x_2))
|F(\bold x,\bold y)|^n\,d\bold x d\bold y,
\end{equation}
where $F(\bold x,\bold y)$ is defined in \eqref{5.39}
and
\begin{equation}
S(x)=\frac{|\sin((\ell_n-1)x)|} {|\sin(x)|}.
\end{equation}
Arguing as in the case of $I_{n2}(\bold X)$, we
thus
obtain that
$\ex(|J_{n2}(\bold X)|^2)$
is of order
\begin{multline}
(c/B)^n\left(1+\,\int\limits_{x\in
(-\pi/2,\pi/2]}\frac{|\sin((\ell_n-1)x)|}{|\sin(x)|}
\,dx\right)^2\\
+(\rho+c/B)^n
\iint\limits_{x_1,x_2\in (-\pi/2,\pi/2]\atop |x_1+x_2|\le B/M}
\Big(1+\frac{|\sin((\ell_n-1) x_1)|}{|\sin(x_1)|}
\Big)\Big(1+\frac{|\sin((\ell_n-1) x_2)|}{|\sin(x_2)|}\Big)
\, dx_1dx_2,
\label{7.11}
\end{multline}
with, as before, $\rho=2^{-\kappa_-}$.
Here the single integral is of order
\begin{equation}
\int\limits_{x\in (-\pi/2,\pi/2]}
\frac{|\sin((\ell_n-1)x)|}{|x|}
\,dx=
\int\limits_{|t|\le \pi(\ell_n-1)/2}\frac{|\sin t|}{|t|}\,dt
=O(\log \ell_{n})
=O(n),
\label{7.12}
\end{equation}
and the double integral is of order
\begin{equation}
\frac{B}{M}
\int\limits_{x\in (-\pi/2,\pi/2]}
\ell_{n}\Big(1+\frac{|\sin((\ell_n-1)x)|}{|\sin (x)|}\Big)
\,dx
=O(M^{-1}\ell_{n}\log\ell_n)
=O(n\ell_{n}M^{-1}).
\label{7.13}
\end{equation}
Combining the contributions from \eqref{7.12} and \eqref{7.13},
we get that for $B$ sufficiently large
\begin{equation}
\begin{aligned}
\ex(|J_{n2}(\bold X)|^2)
&=O\Big(n^2 (c/B)^n\Big) + O\Big(n\ell_{n}M^{-1}(\rho+c/B)^n\Big)
\\
&=O\Big(\frac{\ell_n}{Mn^2}\rho^ne^{(\eps/2) n}\Big)
=O\Big(\Big(\frac{\ell_n}{Mn}\Big)^2e^{-(\eps/2)n}\Big).
\end{aligned}
\end{equation}
Here, in the second step we used that
$\rho$ is bounded away from
$0$ and $1$, and that $B$ is large enough,
while in the third step, we used that
$\ell_n/M$ is assumed to be at
least $\rho^ne^{\eps n}$.
By Markov's inequality (and the fact
that $\ell_n\leq \ell_{n,\bold X}+1\leq 2\ell_{n,\bold X}$
whenever $\ell_{n,\bold X}\neq 0$), we conclude that
with probability $1-O(e^{-(\eps/4)n})$,
\begin{equation}
|J_{n2}(\bold X)|\leq \frac{\ell_n}{Mn} e^{-(\eps/8)n}
\leq \frac{2\ell_{n,\bold X}}{Mn} e^{-(\eps/8)n}
.
\end{equation}
By \eqref{7.7}, \eqref{7.8} and
\eqref{tau-n-var}, $J_{n1}(\bold X)$ is of order at least
$\frac{\ell_{n,\bold X}}{Mn} e^{-\log^2 n}$ with probability
$e^{-\delta_4\log^2 n}$ for some $\delta_4>0$, much
larger than the above order of $|J_{n2}(\bold X)|$.
We thus have shown that there is a constant
$\delta=\delta(\eps)>0$ such that
with probability at least
$1-O(e^{-\delta\log^2 n})$,
\begin{equation}
\mathcal Z_n(s)
=(1+o(1))
\frac{\ell_{n,\bold X}}{Mn}\frac{1}{\pi\sqrt{\text{det}R}} \,
\exp\bigl(-\boldsymbol\tau_n(4R)^{-1}
\boldsymbol\tau_n^{\prime}\bigr).
\label{7.14}
\end{equation}
Together with \eqref{6.20}, this implies the bound \eqref{7.1}.
\qed

The following corollary follows immediately from
Remark~\ref{rem6.3}~(i) and
Theorem~\ref{thm7.1}.

 \begin{corollary}
\label{cor7.1}
Assume that $\limsup |s|/n<b$,  $\limsup\kappa<\infty$
and $\liminf[\kappa-\kappa_c(b)]>0$, and let $\eps>0$.
Then there exists a constant $\delta>0$ such that
with probability $1-O(e^{-\delta\log^2 n})$,
\begin{equation}
2^{[\kappa-\kappa_c(b)-n^{-1/2}\log n ]n}
\le
d_{opt}
\le
2^{[\kappa-\kappa_-(b)+\eps]n}.
\end{equation}
\end{corollary}

As argued in Remark~\ref{rem6.3}(iii), we expect that for
$\liminf\kappa>\kappa_c$, the number of optimal partitions
$Z_{opt}$ grows subexponentially. But so far, we only can
prove the following corollary to Theorem~\ref{thm7.1}.

\begin{corollary}\label{cor7.2}
Suppose that $\limsup |s|/n<b$ and that
$0<\liminf\kappa\leq\limsup\kappa<\infty$.
Then for every $\eps>0$, there exists a constant
$\delta>0$ such that
\begin{equation}
\label{Znopt-upper}
Z_{opt} \le
2^{(\kappa_c+\eps)n}\max\{2^{-\kappa n}, 2^{-\kappa_-n}\}
\end{equation}
with probability
$1-O(e^{-\delta\log^2 n})$.
\end{corollary}

\noindent {\bf Proof.} Let $\ell_n=\lceil
2^{[\kappa-\kappa_-(b)+\eps/2]n}\rceil +1$. By the bound
\eqref{7.2} of Theorem~\ref{thm7.1}, we have that
there exists some $\delta>0$ such that
$d_{opt}\leq
\ell_n-1$ with probability $1-O(e^{-\delta\log^2 n})$.
Using the bound \eqref{7.1}, we therefore get
\begin{equation}
Z_{opt}
\le \ell_n 2^{[\kappa_c(b)-\kappa]n}e^{n^{1/2}S_n+o(n^{1/2})}
\le \ell_n 2^{[\kappa_c(b)-\kappa]n}e^{O(n^{1/2}\log n)},
\end{equation}
again with probability $1-o(e^{-\delta\log^2 n})$. Bounding
$\ell_n\leq 3\max\{1,2^{[\kappa-\kappa_-(b)+\eps/2]n}\}$,
we obtain the bound \eqref{Znopt-upper}. \qed

\section{The Sorted Phase}
\label{sec:sorted}
It remains to study the minimum discrepancy for $b>b_c$.

\subsection{Characteristics of the Sorted Phase}
\label{sec:opt-sort}
\nobreak
 \begin{theorem}
\label{thm7;3} Suppose that
$M\gg n^2$ and
$\liminf s/n>b_c$. Then w.h.p.~the
optimal partition $\bss^*$ is the {\it sorted partition\/}
obtained as follows. Order $X_j$ in the increasing order, so that
$X_{\pi(1)}\le\cdots\le X_{\pi(n)}$ for some permutation $\pi$ of
$\{1,\dots,n\}$. W.h.p.~there will be no ties, and the ordering $\pi$
will be uniquely defined. Denoting $j_n=(n+s)/2=n(1+b)/2$,
$b=s/n$,
\begin{equation}
\sigma_j^*=\left\{\begin{aligned}
&1,\quad&&1\le j\le j_n,\\
&-1,\quad&&j_n<j\le n,\end{aligned}\right.
\end{equation}
and w.h.p.~the minimum discrepancy is asymptotic to
$\frac{Mn}{4}[(1+b)^2-2]$, i.e.~of order $Mn$.
\end{theorem}

\noindent {\bf Proof of Theorem~\ref{thm7;3}.\/}
 We begin with the
following observation. If
\begin{equation}
\delta_{s}({\boldsymbol X})=
\sum_{j=1}^{j_n}X_{\pi(j)}- \sum_{j=j_n+1}^nX_{\pi(j)}\ge 0,
\label{7.18}
\end{equation}
then the sorted partition is optimal, and
$d_{opt}=\delta_s(\bold X)$. Indeed, let $\bss$ be
any feasible partition $\bss$.
Then $s(\bss)=|\{j:\sig_j=1\}|-|\{j:\sig_j=-1\}|=2|\{j:\sig_j=1\}|-n$
so
that $|\{j:\sig_j=1\}|=j_n$. Thus
\begin{align}
\bss\cdot\bold
X=&\sum_{\{j:\sig_j=1\}}X_j-\sum_{\{j:\sig_j=-1\}}X_j
\\
\ge&\sum_{j=1}^{j_n}X_{\pi(j)}-\sum_{j=j_n+1}^nX_{\pi(j)}\ge 0,
\end{align}
which implies optimality of the partition $\bss^*$ and
$d_{opt}=\delta_s(\bold X)$.

In light of this property, all we need to do is to show that, for
$b>b_c$ and bounded away from $b_c$, w.h.p.
\begin{equation}
\delta_s(\bold X)=\frac{Mn}{4}\Big[(1+b)^2-2\Big](1+o(1))\ge 0.
\end{equation}

 Let $U$ be $[0,1]$-uniform, and $U_1,\dots,U_n$ be the
independent copies of $U$. Then the sequence
$\{X_j^\prime\}:=\{\lceil MU_j\rceil\}$ has the same distribution
as our sequence $\{X_j:1\le j\le n\}$. So we will consider
$\{X_j^\prime\}$ instead. Since $M\gg n^2$, it follows easily that
w.h.p.~$X_{\pi(1)}^\prime<\cdots<X_{\pi(n)}^\prime$ if and only if
$U_{\pi(1)}<\cdots<U_{\pi(n)}$. Furthermore
\begin{equation}
0\le \lceil MU_j\rceil -MU_j\le 1,
\end{equation}
so it suffices to show that w.h.p.~
\begin{equation}
\sum_{j=1}^{j_n}U_{\pi(j)}-\sum_{j=j_n+1}^nU_{\pi(j)}
=\frac{n}{4}\Big[(1+b)^2-2\Big](1+o(1)).
\label{7.23}
\end{equation}
A second simplification is based on a fact that the sequence
$\{U_{\pi(j)}\}_{1\le j \le n}$ has the same distribution as
$\{\frac{S_j}{S_{n+1}}\}_{1\le j\le n}$, where
$S_j=\sum_{k=1}^jZ_k$, and $Z_1,\dots,Z_n$ are independent copies
of $Z$, the Exponential $(\lambda)$, $\lambda>0$ being arbitrary.
Choose $\lambda=1$ for certainty. Since $S_{n+1}$ is
w.h.p.~asymptotic to $(n+1)\ex Z=n+1$, it suffices to show that
w.h.p.~
\begin{equation} \sum_{j=1}^{j_n}S_j-\sum_{j=j_n+1}^nS_j
=\frac{n^2}{4}\Big[(1+b)^2-2\Big](1+o(1)),
\label{7.24}
\end{equation}
or, in terms of $Z_t$'s, that w.h.p.~
\begin{equation}
\Sigma_1-\Sigma_2
=\frac{n^2}{4}\Big[(1+b)^2-2\Big](1+o(1)),
\end{equation}
where
\begin{equation}
\begin{aligned}
\Sigma_1=&\sum_{k=1}^{j_n}(j_n-k+1)Z_k,\\
\Sigma_2=&(n-j_n)\sum_{k=1}^{j-n}Z_k+\sum_{k=j_n+1}^n(n-k+1)Z_k.
\end{aligned}
\label{7.26}
\end{equation}
Finally, introduce $\{Z_k^\prime\}$, the truncated version of
$\{Z_k\}$, namely
\begin{equation}
Z_k^\prime=\min(Z_k,2\log n),\quad 1\le k\le n.
\end{equation}
Noticing that
\begin{align}
\pr(\exists\,k\le n:\,Z_k^\prime\neq Z_k)\le& n\pr(Z_1^\prime\neq
Z_1)
\\
=&n\pr(Z_1>2\log n)=n^{-1}\to 0,
\end{align}
we can and will replace $Z_k$ by $Z_k^\prime$ in \eqref{7.26},
denoting the corresponding sums by $\Sigma_i^\prime$. Observe that
\begin{equation}
\ex(Z_k^\prime)=\ex(Z_k)-\int\limits_{2\log
n}^{\infty}\!\!(y-2\log n)e^{-y}\,dy=1-n^{-2},
\end{equation}
so that
\begin{equation}
\ex(\Sigma_i^\prime)=\ex(\Sigma_i)+O(1),\quad i=1,2.
\end{equation}
By the Azuma-Hoeffding inequality (see, e.g.,
\cite{GrStirz}, Section 12.2) and $Z_k^\prime\le 2\log
n$, we have: for every $\a>0$,
\begin{equation}
\pr(|\Sigma_i^\prime-\ex(\Sigma_i^\prime)|>\a)\le
2\exp\left(-\frac{\a^2} {2n(2n\log
n)^2}\right)=2\exp\left(-\frac{\a^2}{8n^3\log^2 n}\right).
\end{equation}
Using this bound with
$\a=n^{7/4}$,
we obtain that
w.h.p.~
\begin{equation} |\Sigma_i^\prime-\ex(\Sigma_i^\prime)|\le
n^{7/4}\ll n^2,
\end{equation}
implying that w.h.p.~
\begin{equation}
\Sigma_1-\Sigma_2=
\Sigma_1^\prime-\Sigma_2^\prime
=\ex(\Sigma_1^\prime)-\ex(\Sigma_2^\prime)+O(n^{7/4}).
\end{equation}
It remains to observe that
\begin{equation}
\label{7.35}
\begin{aligned}
\ex(\Sigma_1^\prime-\Sigma_2^\prime)=&\ex(\Sigma_1)-\ex(\Sigma_2)+O(1)
\\
=&\frac{j_n(j_n-1)}{2}-(n-j_n)j_n-\frac{(n-j_n)(n-j_n+1)}{2}
+O(1)
\\
=&\frac{2j_n^2-n^2}{2}+O(n)\\
=&\frac{n^2}{4}[(1+b)^2-2]+O(n),
\end{aligned}
\end{equation}
which completes the proof.
\qed

\begin{remark}\label{rem:sort?} Consider a point $(\kappa,b)$ such that
$b<b_c$. From the above proof, it is easy to show that
w.h.p.~the sorted partition $\bss^*$ cannot be optimal.
Indeed, by \eqref{7.35}, we see that
\begin{align}
\ex(\Sigma_1^\prime-\Sigma_2^\prime)
=\frac{n^2}{4}\bigl[(1+b)^2-2\bigr]+O(n)
\le-\Theta(n^2).
\end{align}
This means that w.h.p.~
\begin{equation}
\bss^*\cdot\bold
X\le-\Theta(Mn)\Longrightarrow |\bss^*\cdot\bold X|\ge\Theta(Mn),
\end{equation}
and we know that $d_{opt}=o(Mn)$ for every $(\kappa,b)$,
with $b<b_c$.
\end{remark}

\subsection{The Dual Role of $b_c$.} \label{sec:Sort-Saddle}

In Theorem~\ref{thm4.1}, we proved that $b_c$ is the threshold of
the values of $b$ for solvability of the saddle point equations
\eqref{3.17}. Then, in Theorem~\ref{thm7;3}, we proved that the
same $b_c$ is also the threshold for optimality of the sorted
partition. At this point, while the results are complete, the
equality of these two thresholds seems to be nothing but a
numerical coincidence. In this subsection, we give an explanation
of this coincidence. It turns out that the coincidence reflects a
{\em deterministic} property of the integer partitioning problem
which holds for a broad set of $\mathbf X$, see
Theorem~\ref{thm9.1} below.

We caution the reader that Theorem~\ref{thm9.1} does not directly
imply the two Theorems~\ref{thm4.1} and \ref{thm7;3}.
Theorem~\ref{thm9.1} holds only for a given instance $\mathbf X$,
and states that for $s/n$ greater than some $b_c(\mathbf X)$, the
sorted partition is optimal, while for $s/n < b_c(\mathbf X)$, the
saddle point equations \eqref{3.13} have an unique solution. In
order to apply this theorem to the random integer partitioning
problem, we would have to address two issues. First, we would have
to show that if $b>b_c$, then w.h.p., $b > b_c(\mathbf X)$.  This
follows in a relatively straightforward fashion using the
techniques of the last subsection.  The more difficult issue is to
relate existence of solutions of the saddle point equations
\eqref{3.13} to that of the averaged saddle point equations
\eqref{3.17}.  This requires that we establish existence and
commutation of the limits $n \to \infty$ and $s/n \to b$, and deal
with the fact that $n b_c$ is generally not an integer so that, as
$b \to b_c$, the solution to the saddle point equations gives a
$\bss$ with some $\sigma_j$ in the interval $[-1,+1]$ rather than
all $\sigma_j \in \{-1,+1\}$, see below.  Since the coincidence is
already proved in Theorems~\ref{thm4.1} and \ref{thm7;3}, and
since the purpose of this section is simply to elucidate the
coincidence, we do not deal with these, admittedly difficult,
issues here.  We just present the result for a given
$\mathbf X$.

Consider $n$ arbitrary numbers $X_1,\dots,X_n\in\{1,\dots,
M\}$, subject to the constraints that no two of them are equal,
and that their sum is even (the odd case is similar and is left to
the reader).  Without loss of generality, further assume
that the $X_i$ are ordered in increasing order, so that
$X_1<X_2<\dots<X_n$. Consider the equations \eqref{3.13} with
$\ell=0$, and define $s_c({\bold X})$ as the supremum over all
$s$ for which the equations \eqref{3.13} have a solution.

Let $\boldsymbol\sigma$ be a fractional partition, i.e., let
$\boldsymbol\sigma\in[-1,1]^n$. We say that $\boldsymbol\sigma$
has bias $s$, if $\sum_{i=1}^n\sigma_i=s$, and we say that it is
sorted, if $\sigma_i\leq\sigma_{i+1}$ for all $i$ and
$|\sigma_i|=1$ for all but at most one $i$.  Note that there is
exactly one sorted partition with bias $s$ for any real
$s\in[-n,n]$. We finally introduce the {\em critical sorted
partition} as the sorted (fractional) partition
$\tilde{\boldsymbol\sigma}$ that obeys the condition
\begin{equation}
\label{9.1}
 \sum_{j=1}^n \tilde\sigma_j X_j =0.
\end{equation}
There is at most one such partition for a fixed set of weights
$X_1,\dots,X_n$. With slight abuse of notation, we
say that a probability distribution $\mathbb P(\boldsymbol\sigma)$
on partitions is concentrated on a fractional sorted partition
$\tilde{\boldsymbol\sigma}$ if $\mathbb
P(\sigma_i=\tilde\sigma_i)=1$ whenever $|\tilde\sigma_i|=1$, and
$\mathbb P(\sigma_i=1)=p$ when $\tilde\sigma_i$ takes the
fractional value $2p-1$.

The following theorem shows that the critical bias for the
existence of a solution to the random saddle point equations and
the critical bias for the optimality of the fractional sorted
partition are identical. For brevity, we drop the term ``fractional''
throughout.

\begin{theorem}\label{thm9.1}
Let $X_1<X_2<\dots<X_n\in\{1,\dots,M\}$, and assume that the sum
$\sum_{i=1}^n X_i$ is even.

i) If $\ell=0$ and $0< s<s_c(\bold X)$, then the saddle point
equations \eqref{3.13} have a unique solution $(\xi,\eta)$ with
$-\infty<\eta<0$, $0<\xi<\infty$ and $|\eta|/\xi<M$.

ii) For $\ell=0$, $s<s_c(\bold X)$, and a solution $(\xi,\eta)$ of
the saddle point equations \eqref{3.13}, let $\mathbb P_s(\cdot)$
be the probability distribution on partitions defined in
\eqref{3.10}.  If $s\nearrow s_c(\bold X)$, then
$\xi\nearrow\infty$, $\eta\searrow -\infty$, and the distribution
$\mathbb P_s(\cdot)$ gets concentrated on the critical sorted
partition.

iii) If $\ell=0$ and $s\geq s_c(\bold X)$, then the saddle point
equations \eqref{3.13} have no solution, and the sorted partition
with bias $s$ is optimal; if $s>s_c(\bold X)$, this partition has
non-zero discrepancy, implying that there are no perfect
partitions with bias $s>s_c(\bold X)$.
\end{theorem}

\begin{remark}
Statement ii) clearly implies that the critical sorted partition
has bias $s_c(\bold X)$.  As a consequence, a sorted partition
with bias $s>s_c(\bold X)$ has non-zero discrepancy.  By an easy
extension of the argument given for non-fractional partitions,
this in turn implies that sorted partitions with bias $s\geq
s_c(\bold X)$ are optimal.  Except for the statement that the
saddle point equations \eqref{3.13} have no solution for $\ell=0$
and $s=s_c(\bold X)$, statement iii) is therefore an immediate
consequence of statement ii).
\end{remark}

\noindent  {\bf Proof of Theorem~\ref{thm9.1}:\/} i)
Given $X_1<X_2<\dots<X_n\in\{1,\dots,M\}$, let
\begin{equation}
\begin{aligned}
F(\xi,\eta)&=\sum_{j=1}^nX_j\tanh (\xi X_j+\eta),
\\
G(\xi,\eta)&=\sum_{j=1}^n\tanh(\xi X_j+\eta).
\end{aligned}
\end{equation}
Since the partial derivatives $\partial F(\xi,\eta)/\partial\xi$
and $\partial F(\xi,\eta)/\partial\eta$ are strictly positive for
all $(\xi,\eta)\in {\mathbb R}^2$, the equation
\begin{equation}
F(\xi,\eta(\xi))=0
\end{equation}
has a well defined, unique solution $\eta(\xi)$ for all
$\xi\in{\mathbb R}$, and $\eta(\xi)$ is strictly decreasing on
$\mathbb R$.  Let
\begin{equation}
g(\xi)=G(\xi,\eta(\xi)).
\end{equation}
Each solution $(\xi,\eta)$ of the saddle point equations
\eqref{3.13} is then a solution of $g(\xi)=-s$ and
$\eta=\eta(\xi)$, and vice versa.  Using the fact that the
derivatives of $F$ and $G$ are second order derivatives of the
strictly convex function $L_n(\xi,\eta)$, one easily shows that
$g(\cdot)$ is strictly decreasing. Combined with the fact that
$\eta(0)=0$ so that $g(0)=0$, we easily complete that proof of i).
Indeed, by the monotonicity of $g$, the equation $g(\xi)=-s$ has a
unique solution $\xi\in(0,\infty)$ whenever
\begin{equation}
g(0)=0<s<s_c=-\lim_{\xi\nearrow\infty}g(\xi).
\end{equation}
But $\xi>0$ implies $\eta=\eta(\xi)<0$, so we are just left with
the proof of the inequality $|\eta|<\xi M$. To this end, we just
observe that $F(\xi,\eta)=0$, $\xi> 0$ and the fact that not all
$X_j\in\{1,\dots, M\}$ are equal imply that
\begin{equation}
0=\sum_{j=1}^nX_j\tanh (\xi X_j+\eta) <\tanh (\xi
M+\eta)\sum_{j=1}^nX_j,
\end{equation}
which in turn gives $\xi M+\eta>0$, as desired.

ii) By the strict monotonicity of $g$, $\xi\nearrow\infty$ as
$s\nearrow s_c$ (otherwise, the equation $g(\xi)=-s_c$ would have
a finite solution $\xi<\infty$, which contradicts the strict
monotonicity of $g$ on $(0,\infty)$).  If $\eta=\eta(\xi)$ stayed
bounded away from $-\infty$
 as $\xi\nearrow\infty$, the function $F(\xi,\eta(\xi))$ would
converge to the sum of the weights $X_j$, which is not compatible
with $F(\xi,\eta(\xi))=0$.  Thus $\eta\searrow -\infty$ as
$s\nearrow s_c$. We now set $\sigma_j(\xi)=\tanh(\xi
X_j+\eta(\xi))$, so that
\begin{equation}
F(\xi,\eta(\xi))=\sum_{j=1}^n X_j\sigma_i(\xi).
\end{equation}
In order to complete the proof of ii), we have to show that
$\boldsymbol\sigma(\xi)$ converges to the critical sorted
partition as $\xi\to\infty$.

To this end, we first note
\begin{equation}
|1-|\sig_j(\xi)||\le 2e^{-\xi},
\label{9.8}
\end{equation}
for all but at most one $j$. Indeed, let $X(\xi)=-\eta(\xi)/\xi$,
so that $\sigma_j(\xi)=\tanh(\xi(X_j-X(\xi)))$.  Let $j_0$ be such
that $|X_{j_0}-X(\xi)|$ is minimal.  Since two consecutive weights
$X_j$, $X_{j+1}$ differ by at least $1$, we conclude that
$|X_{j}-X(\xi)|\geq 1/2$ for all $j\neq j_0$.  Together with the
bound $||\tanh x|-1|\leq 2e^{-2|x}|$ this proves~\eqref{9.8}
for $j\neq j_0$.

Consider now a sequence $(\xi_r)$ with $\xi_r\to\infty$ as
$r\to\infty$.  By compactness, we can always find a subsequence
such that $\boldsymbol\sigma(\xi_r)$ converges to some
$\tilde{\boldsymbol\sigma}$.  Due to \eqref{9.8}, we
must have that $|\tilde\sigma_j|=1$ for all but at most one $j$.
Since $F(\xi_r,\eta(\xi_r))=\sum_j\sigma_j(\xi_r)X_j=0$ and
$\sigma_1(\xi)\leq\dots\leq\sigma_n(\xi)$, the same holds for the
limiting sequence $\tilde\sigma_1,\dots,\tilde\sigma_n$. As a
consequence, $\tilde{\boldsymbol\sigma}$ is the critical sorted
partition defined in \eqref{9.1} (recall that the critical
sorted partition is unique). Thus any convergent subsequence of
$\boldsymbol\sigma(\xi_r)$ converges to the critical sorted
partition $\tilde{\boldsymbol\sigma}$, implying that
$\boldsymbol\sigma(\xi_r)$ itself converges to
$\tilde{\boldsymbol\sigma}$.  This concludes the proof of ii).

iii) We already showed above that the equation $g(\xi)=-s_c$ has
no finite solution $\xi<\infty$.  As pointed out in the remark
following the theorem, this is the only statement in iii) which
does not follow directly from the statements in ii).

\section{Relaxed Version of the Integer Partitioning Problem}
\label{sec:LP}

 It is a rather common idea to approximate an optimization problem
defined with integer-valued variables by its relaxed version,
where the variables are now allowed to assume any value within the
real intervals whose endpoints are the admissible values of the
original integer variables. In our case, the relaxed version is a
linear programming problem (LPP) which can be stated as follows.
Find the minimum value $d_{opt}$ of $d$, subject to linear
constraints
\begin{equation}
\begin{aligned}
&-d\le\sum_j\sig_jX_j,\quad \sum_j\sig_jX_j\le d,\\
&\sum_j\sig_j=s,\\
&-1\le\sig_j\le 1,\quad (1\le j\le n).
\end{aligned}
\label{8.1}
\end{equation}
As usual, the LPP has at least one basis solution, i.e.~a solution
$(\bss, d_{opt})$, which is an extreme
(vertex) point of the polyhedron defined by
the constraints \eqref{8.1}.  Let
$N(\bss):=|\{j:\sig_j\in (-1,1)\}|$ be the number of
components of $\bss$ which are non-integer.
It is easy for the reader to verify that $N(\bss) \le 2$
for all basis solutions $\bss$.
In fact, $N(\bss)$ cannot be $1$ either, since in this case the
exceptional $\sig_{j_0}\neq \pm 1$ must be zero, which contradicts
to $s\equiv n(\text{mod }2)$. Thus, for a basis solution,
$N(\bss)\in \{0,2\}$. $N(\bss)=0$ signals that $\bss$ is an
optimal partition. Suppose $N(\bss)=2$, and let
$\{j_1,j_2\} =\{j:\sig_j\in (-1,1)\}$.
Using the second line in \eqref{8.1}, we
see that
\begin{equation}
\sig_{j_1}+\sig_{j_2}\in \{-1,0,1\}.
\end{equation}
Moreover, the second condition in \eqref{8.1},
combined with $s\equiv n(\text{mod }2)$,
rules out the values $\pm 1$. Therefore,
$\{\sig_j\}_{j\neq j_1, j_2}$ is a partition of
$\{X_j\}_{j\neq j_1,j_2}$, of bias $s$.

 Our last theorem shows that the horizontal line $b=b_c$ is a
phase boundary for the LPP as well. For $b>b_c$ the solutions of
the initial partition problem and of its LPP version coincide.
For $b<b_c$ they are very far apart,
in terms of the {\it ratio\/} of respective optimal discrepancies.
To state this precisely, we introduce
the fraction of basis
solutions $\bss$ with a property that the deletion of the $N(\bss)$
components of $\bss$ with values in $(-1,1)$ produces an optimal
integer partition for the remaining weights $X_j$.
We denote this fraction by $F_n(\kappa,b)$.

 \begin{theorem}
\label{thm8.1}
Let $\limsup\kappa<\infty$ and $0<\liminf\kappa$.
\begin{description}

\item{(i)} If
$\liminf s/n>b_c$, then w.h.p.~the sorted partition
$\bss^*$ is an unique solution of the LLP \eqref{8.1}, so
that in particular $F_n(\kappa,b)=1$ and
$d_{opt}=\Theta(Mn)$.

Let $\limsup s/n<b_c$.

\item{(ii)} W.h.p. $d_{opt}=0$, and there
are $2^{\kappa_c(b)n+O({n^{1/2}}\log n)}$ basis solutions
$(\bss,0)$.

\item{(iii)} If $\liminf(\kappa-\kappa_c(b))>0$, all basis solutions are
fractional (i.e. have $N(\bss)=2$), and if
$\limsup(\kappa-\kappa_-(b))<0$, the number of basis solutions
with $N(\bss)=0$ is at most
$2^{(\kappa_c(b)-\kappa)n+O_p({n^{1/2}})}$.

\item{(iv)}  If $0<\eps<\kappa_-(b)$ and $\liminf(\kappa-\kappa_-(b))>0$,
then, w.h.p.,
$2^{-\kappa_c(b)n-\eps n}\leq F_n(\kappa,b)
\leq 2^{-\kappa_-(b)n+\eps n}$.
If $\limsup(\kappa-\kappa_-(b))<0$, then, w.h.p.,
$F_n(\kappa,b) = 2^{-\kappa n +O({n^{1/2}}\log n)}$.
\end{description}
\end{theorem}

\begin{remark}\label{rem8.1}
(i) As discussed in Remark~\ref{rem6.3}~(iii), we believe that the
number of optimal partitions above $\kappa_c$ grows subexponentially
in $n$. Let us assume such a bound, more precisely, assume
that for $\eps>0$, for $b<b_c$, and for $\kappa$ with
$\liminf(\kappa-\kappa_c(b))>0$,
we have that the number of
optimal partitions is bounded by $2^{\eps n}$,
with probability at least $1-o(n^{-2})$. Under this
assumption, the proof of
Theorem~\ref{thm8.1} can be easily generalized to show that,
w.h.p., $2^{-\kappa_c(b)n-\eps n}\leq
F_n(\kappa,b) \leq  2^{-\kappa_c(b)n+\eps n}$ whenever $b<b_c$
and $\liminf(\kappa-\kappa_c(b))>0$.

(ii) If, on the other hand, one believes that the asymptotics of
Theorem~\ref{thm5.1} hold up to $\kappa_c$, more precisely, if we
assume that the bound \eqref{6.1} holds with probability at least
$1-o(n^{-2})$, whenever $b<b_c$ and
$\limsup(\kappa-\kappa_c(b))<0$, then one can prove that, w.h.p.,
$F_n(\kappa,b) = 2^{-\kappa n +o(n))}$ for all $(b,\kappa)$ with
$b<b_c$ and $\limsup(\kappa-\kappa_c(b))<0$.

\end{remark}

\noindent {\bf Proof of Theorem~\ref{thm8.1}
and Remark~\ref{rem8.1}.\/} (i) Suppose
$\liminf s/n>b_c$. Then
w.h.p.~$\sum_j\sig_j^*X_{\pi(j)}=\Theta(Mn)>0$, where $\bss^*$ is
the sorted partition, so that $X_{\pi(1)}<\cdots<X_{\pi(n)}$, and
\begin{equation}
\sig_j^*=\left\{\begin{aligned}
&1,\quad&&j\le j_n,\\
&-1,\quad&&j>j_n.\end{aligned}\right.
\label{8.2}
\end{equation}
Let $\bss\in[-1,1]^n$ be an arbitrary relaxed partition
with $\bss\cdot\bold e=\bss^*\cdot\bold e$.
The proof of (1) then reduces to the proof of the
statement that
\begin{equation}
\sum_j\sig_jX_{\pi(j)}>\sum_j\sig_j^*X_{\pi(j)}
\quad\text{if}\quad \bss\neq \bss^*.
\label{8.3}
\end{equation}
If $\bss\neq \bss^*$, then there exists a pair of indices $i<j$
such that $\sigma_i<1$ and $\sigma_j>-1$.  Assume that $i$ is the
first such index, and that $j$ is the last such index.  Since
$X_{\pi(i)}<X_{\pi(j)}$, we can strictly lower the value of
$\sum_j\sig_j X_{\pi(j)}$ by raising $\sigma_i$ and lowering
$\sigma_j$, and preserving $\sigma_i+\sigma_j$, until at least one
of them has absolute value 1, i.e. either $\sigma_i=\sigma_i^*$ or
$\sigma_j=\sigma_j^*$. Repeating this procedure with the lowest
$i$ and the largest $j$ such that $\sigma_i<1$ and $\sigma_j>-1$
in the new configuration, we will eventually arrive at the
configuration $\bss^*$.  Since the value of
$\sum_j\sig_jX_{\pi(j)}$ was strictly lowered in each step, this
proves \eqref{8.3}, and thus statement (i).

(ii) Let $\limsup s/n<b_c$, and consider the partitioning problem
for $\{X_j\}_{j\ge 3}$.
Let $a>0$, and set
$\ell_n=\lceil Mn^{-a}\rceil$.  By Theorem~\ref{thm7.1},
we have that with probability $1-O(e^{-\delta\log^2 n})$,
there are at least
$2^{n\kappa_c-O({n^{1/2}}\log n)}$ tuples $\{\sigma_j\}_{j\ge 3}$ such
that
\begin{equation}
|\sum_{j\ge 3}\sig_jX_j|\le \ell_n.
\label{8.4}
\end{equation}
We will denote the event expressed in \eqref{8.4} by $\cal A$.
On the other hand, introducing the event ${\cal {B}}=
\{|X_1-X_2|>Mn^{-a_1}\}$, $a_1\in (0,a)$, we have
\begin{equation}
\label{8.5}
\begin{aligned}
\pr({\cal {B}})
=&1-\frac{1}{M^2}\sum\limits_{1\le i,j\le M\atop|i-j|\le Mn^{-a_1}}
1\\
=&1-O(n^{-a_1}).
\end{aligned}
\end{equation}
Introducing $\cal C=\cal A\cap\cal B$, we have then that
$1-\pr({\cal C})\to 0$. On $\cal C$, we can define
$\sigma_1=-\sigma_2$ where
\begin{equation}
\sig_1=\frac{\sum_{j\neq
1,2}\sig_jX_j}{X_1-X_2}=O(n^{-(a-a_1)})\in [-1,1].
\end{equation}
Clearly $\sum_{j=1}^n\sig_jX_j=0$, and denoting $\bss=\{\sigma_j\}_
{1\le j\le n}$, we have that $(\bss, 0)$ is a basis solution of the
LPP.  Thus w.h.p.~the LPP has at least
$2^{n\kappa_c-O({n^{1/2}}\log n)}$
basis solutions.

Conversely, suppose $\bss$ is a basis solution, with
$\sigma_1,\sig_2\in (-1,1)$, and $\sig_1+\sig_2=0$. Then
\begin{equation}
\left|\sum_{j\ge 3}\sig_jX_j\right|\le |X_1-X_2|\le M=o(Mn^{1/2}\log^{-1}
n).
\end{equation}
Again by Theorem~\ref{thm7.1}, we know that
with probability $1-O(e^{-\delta\log^2 n})$,
the total number
of all such $(n-2)$-tuples  with bias $s$ is
$2^{n\kappa_c+O({n^{1/2}}\log n)}$, and for each such tuple, the
feasible values of $\sig_1,\sig_2$ are determined uniquely. Since
there are $\binom{n}{2}$ ways to select $j_1,j_2$ as indices of
components $\sig_j \in (-1,1)$,
and since the union of $\binom n2$ events happening
with probability $1-O(e^{-\delta\log^2 n})$
happens with probability $1-O(n^2e^{-\delta\log^2 n})$,
we get that
$2^{n\kappa_c+O({n^{1/2}}\log n)}$ is
also w.h.p.~an upper bound on the
number of basis solutions $\bss$ with $N(\bss)=2$.

If $N(\bss)=0$ then,
since $b<b_c$, $\bss$ is a perfect partition, and
thus of zero discrepancy.
The number of such partitions can clearly be bounded by
the number of partitions with discrepancy smaller than
$\ell_n=M$.  By Theorem~\ref{thm7.1},
this is in turn bounded by $2^{n\kappa_c+O({n^{1/2}}\log n)}$,
completing the proof that the number of basis solutions
with $N(\bss)=0$ or $N(\bss)=2$ is $2^{n\kappa_c+O({n^{1/2}}\log n)}$.

(iii)  If $\liminf(\kappa-\kappa_c(b))>0$
or $\limsup(\kappa-\kappa_-(b))<0$, we know a little bit
more.  In the first case, w.h.p.~there are no
perfect partitions, and thus no basis solutions
with $N(\bss)=0$.  In the latter case, w.h.p.~the
number of perfect partitions of zero discrepancy (and thus
the number of basis solutions with $N(\bss)=0$)
is at most $2^{(\kappa_c(b)-\kappa)n+O_p({n^{1/2}})}$,
which is negligible relative to $2^{\kappa_c(b)n+
O({n^{1/2}}\log n)}$.

(iv)
If $\eps>0$ and
$\liminf (\kappa-\kappa_-(b))>0$, then
with probability $1-O(n^2e^{-\delta\log^2 n})$,
for every subset
of $(n-2)$ weights $X_j$, there are at most
$2^{(\kappa_c-\kappa_-+\eps)n}$ optimal integer partitions by
Corollary~\ref{cor7.2}. As in (ii), once the $\pm 1$ values of
the corresponding $\sig_j$ are known, the values of the two remaining
$\sig_j$ are determined uniquely. So w.h.p.~the LPP may have at
most $\binom{n}{2}2^{(\kappa_c-\kappa_-+\eps)n}$ basis solutions
with $N(\bss)=2$ that have the optimal subpartition property.
Similarly, the number of $\bss$ with $N(\bss)=0$ is at
most $2^{(\kappa_c-\kappa_-+\eps)n}$. So for every $\eps$,
w.h.p.~
\begin{equation}
F_n(\kappa,b)\le 2^{-\kappa_-(b)n+2\eps n}.
\end{equation}
If we assume that for $\liminf(\kappa-\kappa_c(b))>0$, the number
of optimal partitions obeys the bound $Z_{opt}\leq 2^{\eps n}$
with probability at least $1-o(n^{-2})$, so that w.h.p.~for every
subset of $(n-2)$ weights $X_j$ there are at most $2^{\eps n}$
many optimal integer partitions, the above argument gives that for
$\liminf(\kappa-\kappa_c(b))>0$, w.h.p.,
\begin{equation}
F_n(\kappa,b)\le 2^{-\kappa_c(b)n+2\eps n}.
\end{equation}
Combined with the bound \eqref{8.13} below, this proves
Remark~\ref{rem8.1}.

Conversely, given
$j_1,j_2$, with probability
$1-O(e^{-\delta\log^2 n})$
there exists
at least one
optimal
partition $\{\sig_j\}_{j\neq j_1,j_2}$
of the subset $\{X_j\}_{j\neq j_1,j_2}$, with discrepancy
\begin{equation}
\left|\sum_{j\neq j_1,j_2}\sig_jX_j\right|
\leq M2^{-\kappa_-(b)n +\epsilon n}
\end{equation}
by the bound  \eqref{7.2} of Theorem \ref{thm7.1}. Given such an
optimal subpartition (and assuming that $\eps$ has been chosen
smaller than $\kappa_-(b)$), with sufficiently high probability we
can find an unique pair $\sig_{j_1},\sig_{j_2}\in (-1,1)$ such that
$\{\sig_j\}_{1\le j\le n}$ is a basis solution of the full
LPP, by solving
\begin{eqnarray*}
\sig_{j_1}X_{j_1}+\sig_{j_2}X_{j_2}&=&-\sum_{j\neq j_1,j_2}\sig_jX_j,
\nonumber\\
\sig_{j_1}+\sig_{j_2}&=&0.\nonumber
\end{eqnarray*}
Indeed, with probability
$1-O(e^{-\delta\log^2 n})$,
the sum on the right is
bounded by $M2^{-\kappa_-(b)n+\eps n}$,
and with probability $1-O(2^{-\kappa_-(b)n+\epsilon n})$,
\begin{equation}
|X_{j_1}-X_{j_2}|\ge M2^{-\kappa_-(b)n+\epsilon n}.
\end{equation}
Thus w.h.p.~there exist
at least
$\binom{n}{2}$ basis solutions with the
optimal subpartition property, so that
\begin{equation}
\label{8.13}
2^{-\kappa_c(b)n-\eps n}
\le F_n(\kappa,b)
\end{equation}
as long as $0<\eps<\kappa_-(b)$.

Consider finally $\limsup(\kappa-\kappa_-(b))<0$. Then,
by Theorem \ref{thm5.1}, on the event ``$\sum_j X_j$ is even,''
w.h.p.~we have
$2^{(\kappa_c(b)- \kappa)n+O({n^{1/2}}\log n)}$
perfect
partitions $\bss$ of discrepancy zero, each being a basis solution
of the LPP with $N(\bss)=0$. On the complementary event
``$\sum_jX_j$ is odd'', w.h.p.~there are
$2^{(\kappa_c(b)- \kappa)n+O({n^{1/2}}\log n)}$
perfect partitions
$\{\sig_j\}_{j\neq 1,2}$ of $\{X_j\}_{j\neq 1,2}$, of discrepancy
one, and we can find, uniquely, $\sig_1,\sig_2\in (-1,1)$ such
that
\begin{equation}
\sig_1+\sig_2=0,\quad \sig_1X_1+\sig_2X_2=-\sum_{j\neq 1,2}\sig_jX_j.
\end{equation}
Indeed $|X_1-X_2|$ is w.h.p.~of order, say, $Mn^{-a}$ at least
$(a>0)$, and the last sum is in $[-1,1]$. Thus, regardless of the
parity of $\sum_jX_j$, the LPP has w.h.p.~at least
$2^{(\kappa_c(b)- \kappa)n+O({n^{1/2}}\log n)}$
basis solutions with the optimal subpartition
property.

On the other hand, the total number of
the optimal subpartition basis
solutions $\bss$ is at most
$2^{(\kappa_c(b)- \kappa)n+O({n^{1/2}}\log n)}$
with probability $1-O(n^2e^{-\delta\log^2 n})$.
 Indeed, for $N(\bss)=0$, $\bss$
is a perfect partition, and for $N(\bss)=2$, the $\pm 1$-valued
$\sig_j$ form an optimal, hence perfect partition of the
corresponding weights $X_j$. In either case, the number of
corresponding perfect partitions is asymptotic to
$2^{(\kappa_c(b)- \kappa)n+O({n^{1/2}}\log n)}$,
and $F_n(k,b)=2^{-\kappa n+O(\sqrt{n}\log n)}$.

Finally, if we
assume that the bound \eqref{6.1} holds with probability at least
$1-o(n^{-2})$, whenever $b<b_c$ and
$\limsup(\kappa-\kappa_c(b))<0$, then the above arguments immediately
give that, w.h.p., $F_n(k,b)=2^{-\kappa n+o(n)}$.
\qed

\section{Open Problems and Numerical Experiments}
\label{sec:openprob}
\subsection{Open Problems}

Many problems are left open in our analysis.  Most important is
the question of how we characterize the phase transition from the
perfect phase to the hard phase. In the unconstrained case
\cite{BCP2}, our theorems would have allowed us to give at least
three equivalent definitions.

First, in the unconstrained problem, we defined the phase
transition to be the point $\kappa=\kappa_c=1$ at which the
probability of a perfect partition decreases abruptly from $1$ to
$0$. For the constrained case considered in this paper, we have
proved only that such a transition occurs somewhere within the
interval $(\kappa_-(b),\kappa_c(b))$.  In particular, we have
not proved that such a transition is sharp.

Second, in the constrained case, we could have characterized the
phase transition as the point up to which the expected number of
perfect partitions remains exponential, which again would have
led to the value $\kappa=\kappa_c=1$ \cite{BCP2}.  In contrast,
in the unconstrained case, here we have shown that the expected number
of perfect partitions remains exponential until some $\kappa =
\kappa_e(b)> \kappa_c(b)$, because up to $\kappa =\kappa_e(b)$, with
vanishingly small probability, there are very many perfect
partitions.  Thus we cannot use the second definition in the
constrained case. Alternatively, one could ask whether the {\em
typical} (say median) number of perfect partitions changes from
exponentially large to $0$ at $\kappa = \kappa_c(b)$, a
definition that would have given the same transition point
$\kappa_c=1$ in the unconstrained case, and that might also work
here.

A third possible definition of the transition point is the point
above which there is an unique optimal partition, a definition
which would have again led to $\kappa_c = 1$ in the unconstrained
case \cite{BCP2}.  Here we cannot prove uniqueness until we are
in the sorted phase ($b > b_c$), far above $\kappa = \kappa_c(b)$.

A natural conjecture would be that at least the first and third
definitions coincide and that both lead to a sharp transition
along the line $\kappa=\kappa_c(b)$.

Assuming we could establish a sharp transition, we could then
examine some other open problems. In the unconstrained problem, we
were able to determine the finite-size scaling window around the
transition point $\kappa = \kappa_c$, i.e.~the region
in which the probability of a perfect partition has nontrivial
distribution.  In \cite{BCP2}, we showed that the
window has width of order $n^{-1}$, and is centered at
$\kappa_c+\Theta(n^{-1}\log n)$ .
In the constrained case, as discussed in Remark~\ref{REM4}, the
fluctuations in the number of perfect partitions should be large
enough to lead to a nontrivial probability distribution within a
window of width of order $n^{-1/2}$ about $\kappa_c$.  Hence we
expect the width of the window to be larger here than it was in
the unconstrained case, at least of order $n^{-1/2}$.
Our numerical experiments, reviewed in the next
subsection, support this expectation.

Next, detailed estimates in the unconstrained case proved that the
$k$ smallest discrepancies (i.e., for a given set of $n$ random
integers, the $k$ smallest absolute values of the difference in
the sums of the integers in the two subsets) have a Poisson joint
distribution \cite{BCP2}.  This is the behavior that would be
observed if the discrepancies of $2^{n-1}$ partitions (with
$\sigma_1$ fixed) were independent random variables. This result
confirmed the validity of the so-called Random Energy Model or REM
approximation for the continuous case, proposed earlier by
Mertens~\cite{Mer2}. We have no analogous estimates for the
constrained problem.

Finally, for both the constrained and unconstrained problems, it
would be very useful to have theorems which establish the
relevance of our phase transition results (and associated results
like the Poisson joint distribution) to the complexity of the
number partitioning problem, and to the performance of widely used
algorithms for the problem.  In particular, is the perfect phase
easy, i.e., in this phase, is it possible to find some perfect
partition in polynomial time?  Is the so-called hard phase
actually computationally difficult (under the usual assumptions,
e.g., P $\neq$ NP)?  What are the changes in the behavior of
commonly used algorithms at $\kappa = \kappa_c(b)$?
Is
$\kappa_-(b)$
an artifact of our
proofs or
does it
reflect some change in the complexity of the
problem?  For example, is there a change in the (admittedly
non-rigorous, but often very instructive) replica solution at
$\kappa = \kappa_-(b)$?
Do commonly used algorithms experience any slowdown across
this curve?
And finally, can our LPP results be extended to establish
genuine average-case complexity results for a class of
partitioning algorithms?

\subsection{Numerics}

In this subsection, we present some simulations which address the
question of sharpness of the transition and finite-size scaling.
We begin with a brief discussion of methods, then go on to finite-size effects, since this is important for the interpretation of
the results.  Most of our simulations concern the number of
perfect partitions in the crescent-shaped region from $\kappa =
\kappa_-(b)$ to $\kappa = \kappa_c(b)$.

\subsubsection{Methods and Accessible System Sizes}

The general experimental setup is this: generate a random
instance, i.e.\ $n$ random integers $X_j$, uniformly drawn from
the interval $[1,M]$ and calculate the optimum discrepancy and the
number of optimal solutions of the corresponding constrained
partitioning problem. Loop over many such instances to get the
{\it empirical\/} mean and the empirical standard deviation of
quantities like the logarithm of the number of optimal partitions.

The numbers $X_j$ are constructed from the output of {\em
pseudorandom number generators}; we use LCG64, a 64 bit linear
congruential random number generator from the TRNG collection
\cite{trng}. It is known that the least significant bits of linear
congruential recurrences are correlated, so it is dangerous to use
the elements of such sequences directly as a random instance.
Instead, we use a random number for {\em each bit} in $X_j$. The
bits are set to 1 or 0 with probability $\frac12$, depending on
the most significant bits of the corresponding random number,
thereby minimizing the influence of hidden correlations in the
pseudorandom sequence. On the other hand, this restricts us to
values of $M$ that are integer powers of 2.

We can afford these extra calls to the random number generator,
since the generation of the random instances is not the part of
the simulation that limits the accessible system sizes.  The hard
part of the simulations is the solution of the particular
instance. (After all, we are dealing with an NP-hard problem.)
Even in the perfect phase, where a smart heuristic algorithm
might find one of the exponentially many perfect partitions
quickly, we want to find them all, so there is no obvious way to
avoid an {\em exhaustive enumeration} of all partitions. The
corresponding $O({2^n})$ time complexity is the limiting factor
for our accessible system sizes.

Horowitz and Sahni \cite{horowitz:sahni:74} presented an algorithm
that solves the unconstrained integer partitioning problem in time
$O(n\cdot2^{\frac{n}{2}})$, and this algorithm can easily be
modified to count all perfect partitions for the constrained
problem within the same time complexity. The Horowitz-Sahni
algorithm achieves its prodigious speed-up by dividing the set of
numbers $X_j$ in two halves and tabulating all $2^{\frac{n}{2}}$
possible discrepancies of the two half-sized sets. Each
discrepancy of the original problem can then be represented by the
sum or the difference of two elements, one from each table. The
tables are sorted (this is the origin of the complexity
$O(n\cdot2^{\frac{n}{2}})$), and due to the monotonicity of
the sorted tables, the perfect partitions of the original problem
can be found with only one single scan through both lists.

The drawback of the Horowitz-Sahni algorithm is that it trades
time against space. For $\kappa=1$, the $X_j$'s are $n$-bit
numbers, hence the tables require $2 n 2^{\frac{n}{2}}$ bits of
computer memory. Equipped with 512 MByte of main memory, this
means $n\leq 50$. For a single instance of size $n=50$, the
optimum discrepancies for all (discrete) values of $b$ can be
found in about 4 minutes on a Pentium III CPU with 800 Mhz clock
rate. Averages over $10^3$ random instances can be calculated in
less than half an hour using 156 CPUs of a Beowulf cluster
\cite{tina}. Counting all perfect partitions in
the perfect phase can take
considerably longer if there are many such partitions, i.e.\ for
small values of $\kappa$.

\subsubsection{Concentration and Finite-Size Effects}

For our first numerical experiment, we study the finite-size
effects of the central quantity in the perfect phase, namely the
number $Z$ of perfect partitions for given $b$ and
$\kappa$.

Optimistically extending the formula~\eqref{5.1} -- \eqref{5.6}
(Theorem~\ref{thm5.1}), we expect that for large $n$
\begin{equation}
  \label{eq:logZ-finite}
  \ex\Big(\frac{1}{n}\log Z\Big) \approx L(\zeta,\eta) -
\kappa\log2 -
  \frac{1}{n}\log\Big(\pi n\sqrt{\det R}\Big) -
\frac{1}{4n}\mbox{Tr}R^{-1}K
\end{equation}
is a valid approximation for all $\kappa < \kappa_c(b)$.
It is illuminating to check accuracy of this
approximation numerically, even in absence of an error term bound,
better than $o(n^{-1})$, which is implicit in
\eqref{eq:logZ-finite}.

\begin{figure}[htb]
  \centering \includegraphics[width=0.75\columnwidth]{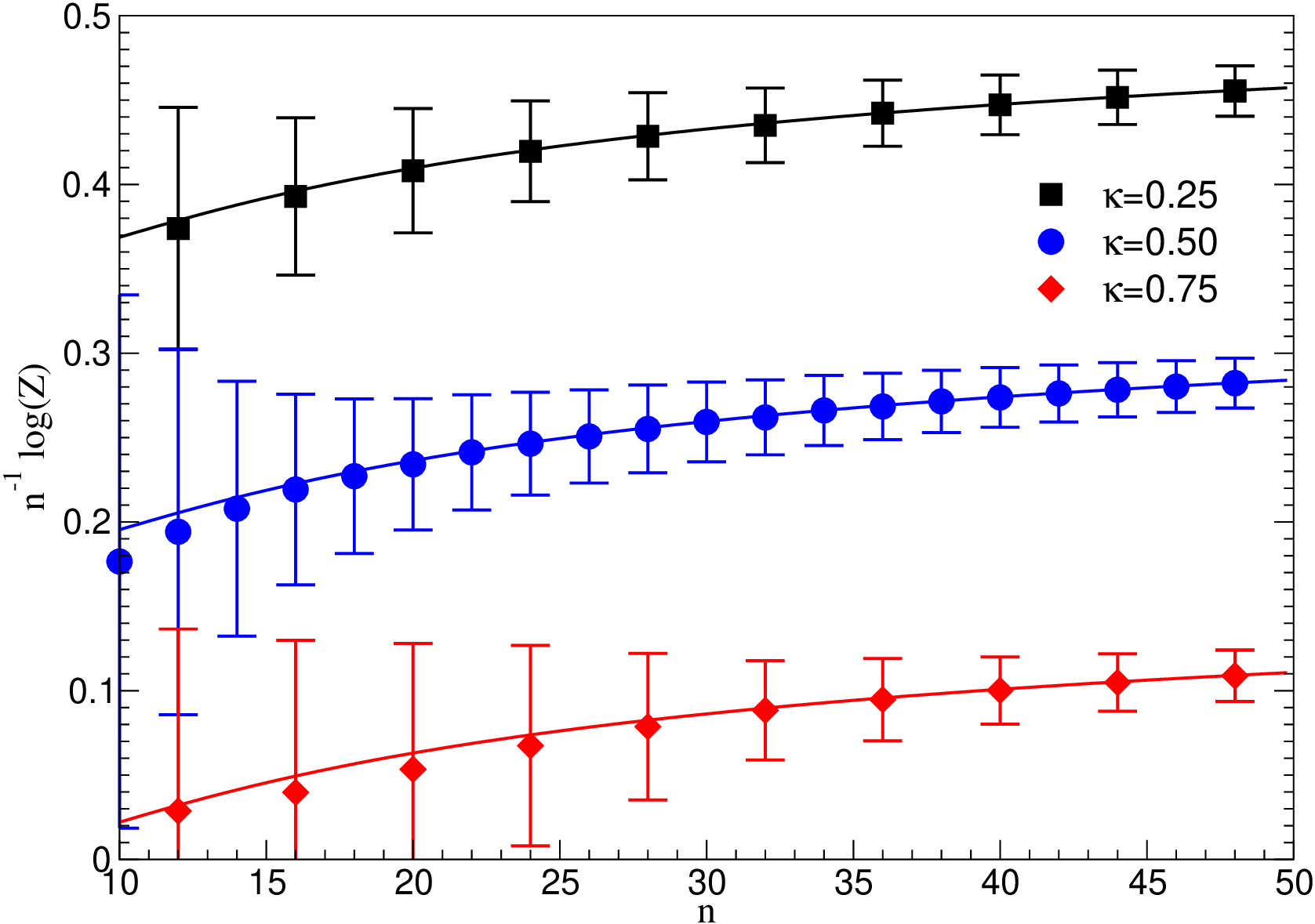}\\[5mm]
  \includegraphics[width=0.75\columnwidth]{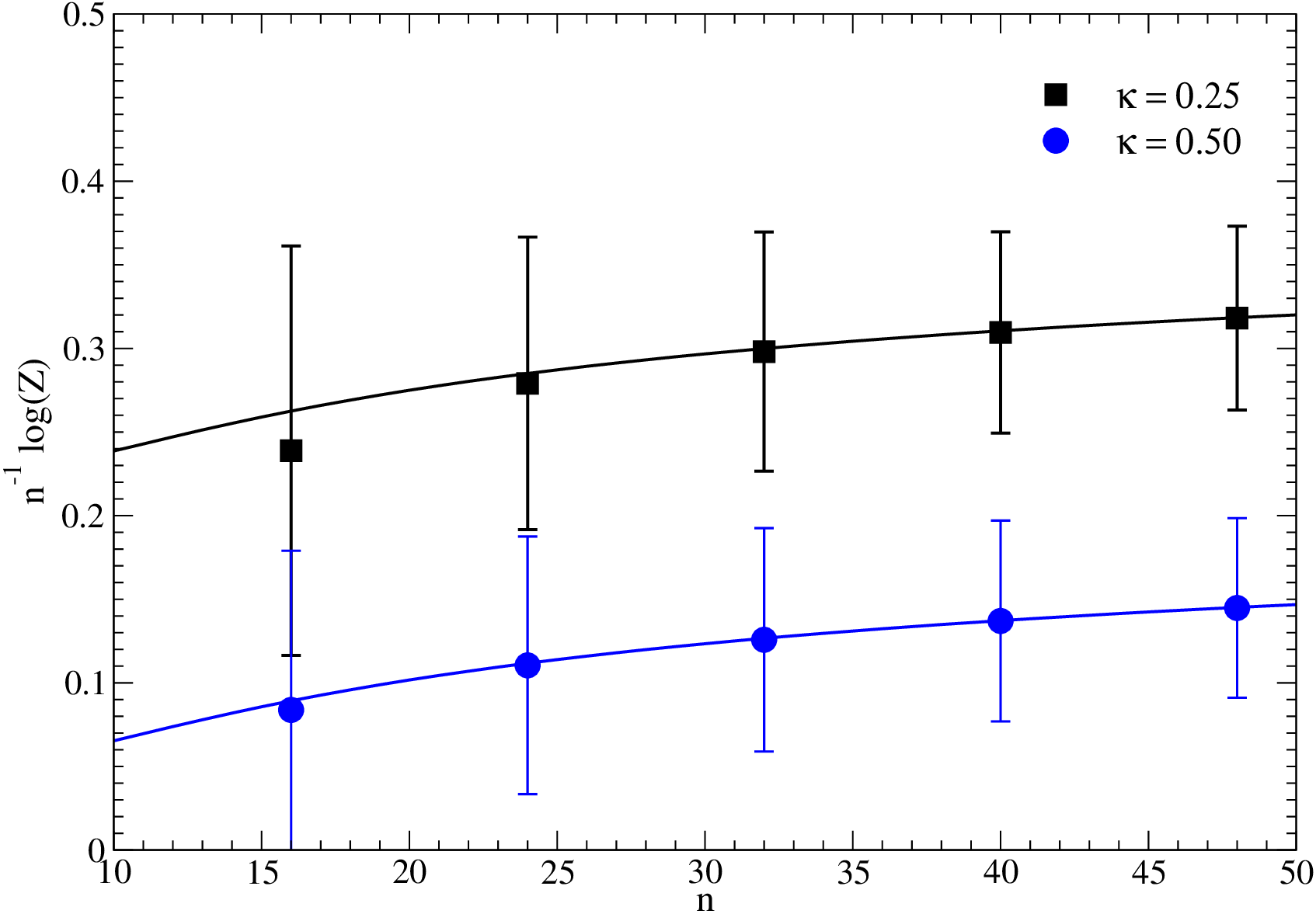}
  \caption{Comparison of eq.~\ref{eq:logZ-finite} (lines) for
    $b=0$ (top)
    and $b=0.25$ (bottom) with simulations. Symbols denote averages over
    $10^3$ random samples of uniform numbers $X$, errorbars indicate
    $\pm 1$ standard deviation.
  \label{fig:logz-n}}
\end{figure}

Figure \ref{fig:logz-n} shows the results of the simulation. For
each data point we generated $10^3$ instances of $n$ random
$\kappa n$-bit integers and counted the number $Z$ of perfect
partitions using the Horowitz-Sahni algorithm.  Each symbol in
Figure \ref{fig:logz-n} denotes the empirical average over the
$10^3$ values of $n^{-1}\log Z$, and the error bars indicate $\pm1$
empirical standard deviation.

For $n \leq 50$, finite-size effects are clearly visible. On the
other hand, the error bars decrease with increasing $n$,
indicating a concentration of $n^{-1}\log Z$ around its expected
value. Note that the size of the error bars does not decrease if
the number of random samples increases. This indicates that the
error bars are a measure of the inherent fluctuations in
$n^{-1}\log Z$.  Note also that for $b >0$, the statistical
fluctuations are much larger than in the $b = 0$ case, consistent
with our rigorous results in Theorem~\ref{thm1} and Remark~\ref{REM4}.

The most surprising observation is that Eq.~\ref{eq:logZ-finite} is a
very good approximation for finite $n > 30$.  Note that the
``finite-size corrections'' in Eq.~\ref{eq:logZ-finite}, i.e.\ the
$O({n^{-1}\log n})$-term  and $O({n^{-1}})$-term,  are essential for
a good approximation: even for $n=50$, the measured values of
$n^{-1}\ex(\log Z)$ are about $20\%$ below the predicted {\em
asymptotic} values.

\subsubsection{Sharpness and Location of the Perfect to Hard Transition}
\label{NumericalSharpness}

The major open problem with the phase diagram is the behavior of
the system inside the crescent-shaped region between
$\kappa = \kappa_-(b)$ and $\kappa =\kappa_c(b)$.  From
Theorems~\ref{thm5.1} and \ref{thm6.1}, we know that w.h.p.~there are
exponentially many perfect partitions for $\kappa < \kappa_-$ and
no perfect partitions for $\kappa > \kappa_c$. What happens to the
number of perfect partitions  between $\kappa=\kappa_-$
and $\kappa=\kappa_c$? Is there a sharp transition from ``no'' to
``exponentially many'' perfect partitions, like in the
unconstrained case or the case $b=0$? If so, where does it occur?

For all numerical experiments shown in this section, we have chosen
$b=0.25$ because here the gap between $\kappa_- = 0.674\ldots$ and
$\kappa_c = 0.799\ldots$ is relatively large. Setting $b=0.25$
means that $n$ must be a multiple of $8$.

\begin{figure}[htb]
  \centering \includegraphics[width=0.75\columnwidth]{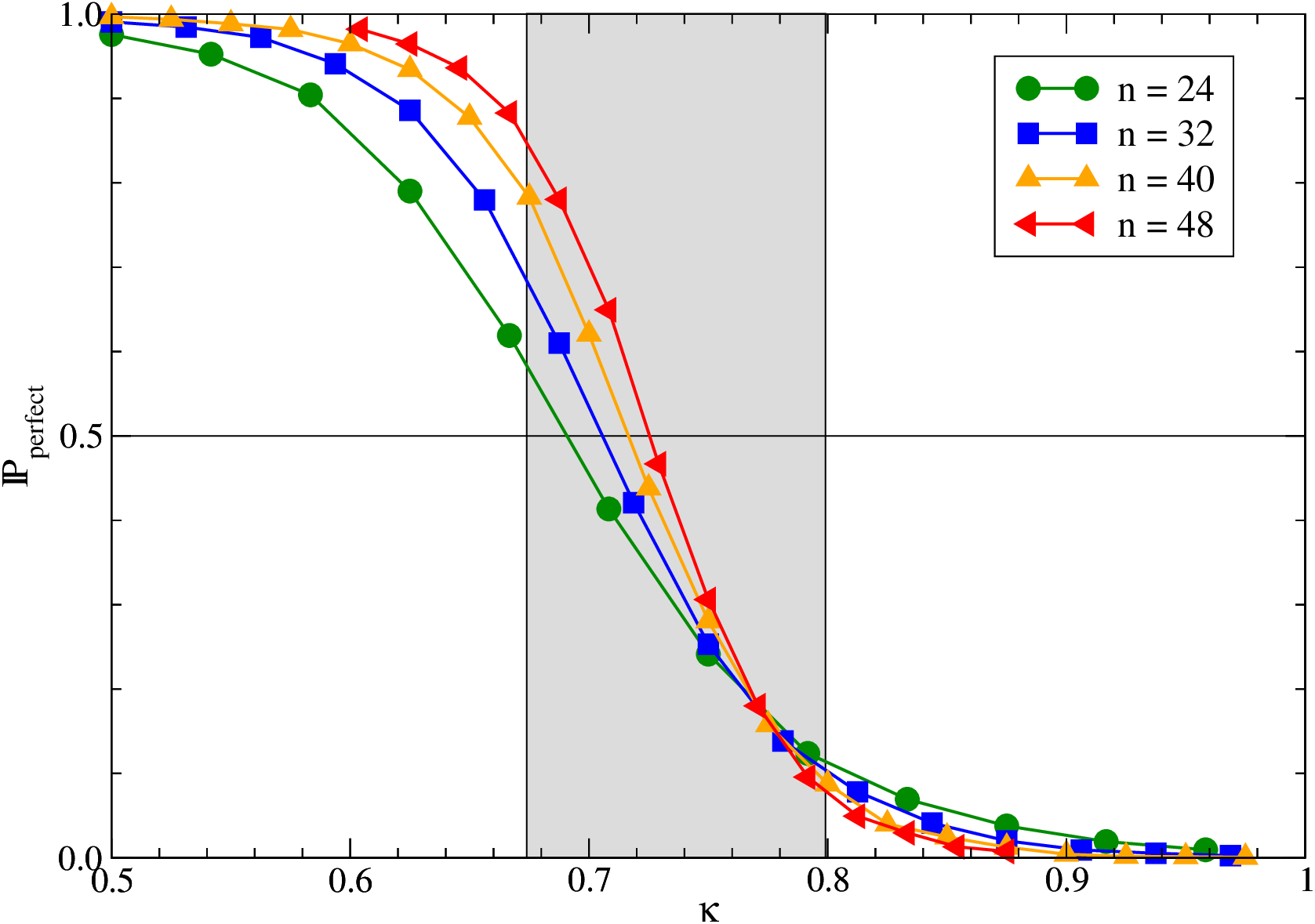}
  \caption{Probability of having a perfect partition in a random
    instance of the constrained partitioning problem with $b=0.25$.
    Symbols are empirical probabilities found in
    $10^3$ random samples. The shaded area is the
    crescent-shaped region from $\kappa_-$ to $\kappa_c$.
  \label{fig:pperf-b-k}}
\end{figure}

In the first experiment, we determine $\Bbb{P}_{\text{perfect}}$,
the fraction of randomly generated instances that have a perfect
partition. According to Theorems~\ref{thm5.1} and \ref{thm6.2},
as $n\to\infty$, this fraction should tend to $1$ for $\kappa <
\kappa_-$ and to $0$ for $\kappa > \kappa_c$ .  Figure
\ref{fig:pperf-b-k} shows the results for $n=32,40,48$. For these
finite $n$, the decay of $\Bbb{P}_{\text{perfect}}$ from $1$ to
$0$ extends over an interval larger than the crescent-shaped
region, but the values outside this region seem to converge to
their limits $1$ and $0$ as $n$ gets larger.

\begin{figure}[htb]
  \centering \includegraphics[width=0.75\columnwidth]{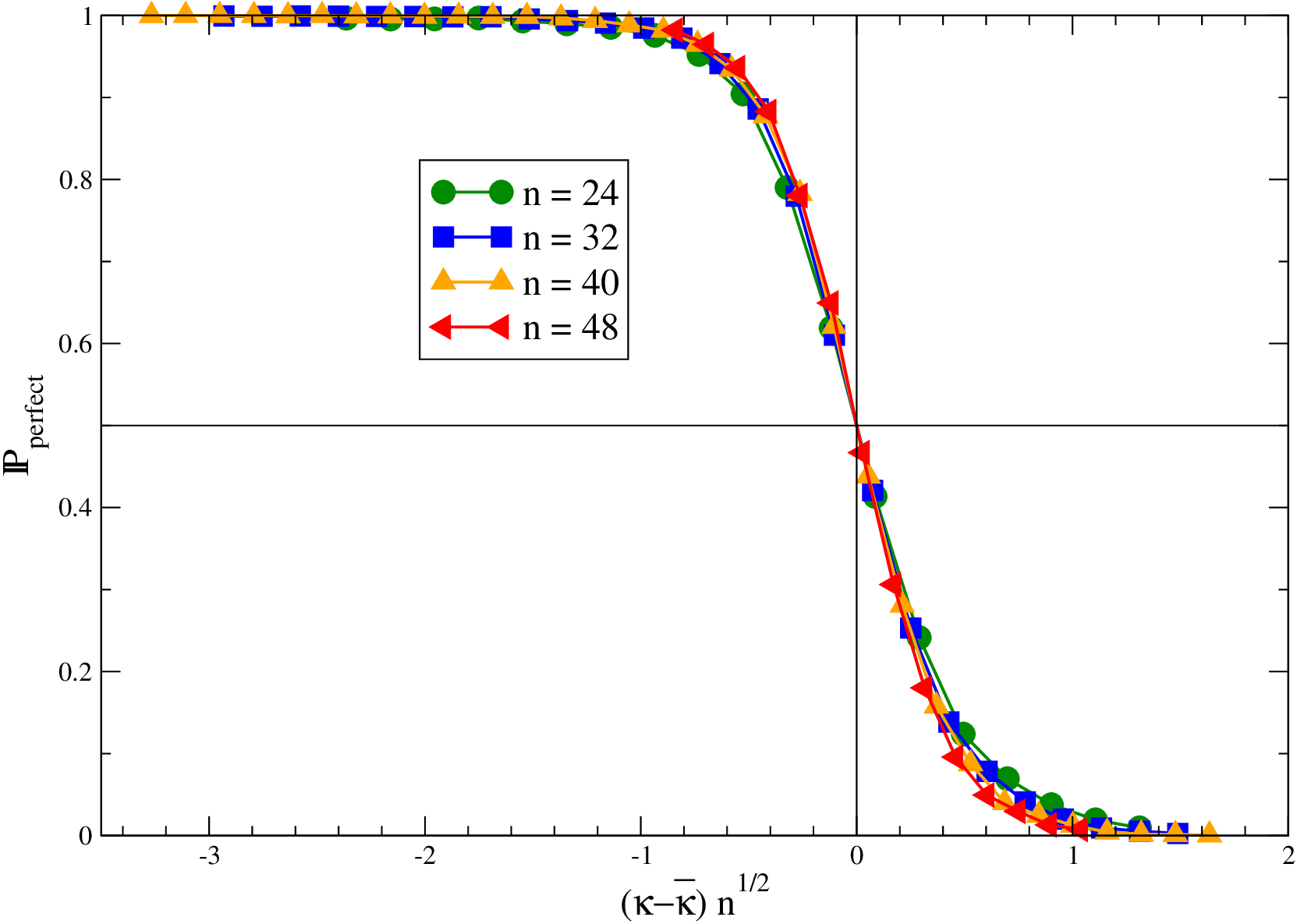}
  \caption{Same data as in Fig.~\ref{fig:pperf-b-k}, but this time
plotted
   versus the
    scaled control parameter $\Delta\kappa = (\kappa -
  \overline{\kappa}) n^{1/2}$. The data collapse indicates a sharp
  transition as $n\to\infty$ and a width $O(n^{-1/2})$ of the
  scaling window. \label{fig:dcollaps}}
\end{figure}

To see more clearly what is happening with
$\Bbb{P}_{\text{perfect}}$, we define%
\footnote{
Of course, for finite $n$, there is in general no solution
$\overline{\kappa}$ to the equation
$\Bbb{P}_{\text{perfect}}(\overline{\kappa}) = \frac{1}{2}$.
Instead, we take $\overline{\kappa}_n$ to be the closest linear
interpolation of solutions with parameters $n$ and
$M = 2^{\overline{\kappa}_n n}$.
}
$\overline{\kappa}$ by
$\Bbb{P}_{\text{perfect}}(\overline{\kappa}) = \frac{1}{2}$ and
plot $\Bbb{P}_{\text{perfect}}$ versus the rescaled control
parameter $\Delta\kappa = (\kappa-\overline{\kappa})n^{1/2}$
for various values of $n$.
By definition, all these curves intersect at
$\Delta\kappa=0$, but the simulation shows that the curves coincide
for all values of $\Delta\kappa$ (Figure \ref{fig:dcollaps}). This
{\em data collapse} indicates that $\Bbb{P}_{\text{perfect}}$ is not a
function of the two parameters $\kappa$ and $n$, but of one single
parameter $\Delta\kappa = (\kappa-\overline{\kappa})n^{1/2}$,
\begin{equation}
  \label{eq:scaling}
  \Bbb{P}_{\text{perfect}} (\kappa, n) =
   f \big((\kappa-\overline{\kappa}) n^{1/2}\big).
\end{equation}
Validity of this {\em scaling hypothesis} in particular would imply
that the transition $\Bbb{P}_{\text{perfect}} = 1$ to
$\Bbb{P}_{\text{perfect}} = 0$ becomes sharp as $n\to\infty$ and that
the width of the transition region scales like $O({n^{-1/2}})$.
The transition point $\overline{\kappa}$ itself depends on $n$ but
seems to converge to $\kappa_c$ as $n\to\infty$ (see Figure
\ref{fig:kappa-b-n}). Our simulation results  support the
conclusion that the probability of a perfect partition shows a
sharp transition at $\kappa_c$:
\begin{equation}
  \label{eq:pperf-transition}
  \lim_{n\to\infty}\Bbb{P}_{\text{perfect}} = \left\{
    \begin{array}{cc}
      1 & \kappa < \kappa_c \\
      0 & \kappa > \kappa_c
    \end{array}
  \right.
\end{equation}
So far, the crescent-shaped region is not visible in the simulations,
but this
may change if we look at other quantities like the number  of
perfect partitions for $\kappa < \kappa_c$.

\begin{figure}[htb]
  \centering \includegraphics[width=0.75\columnwidth]{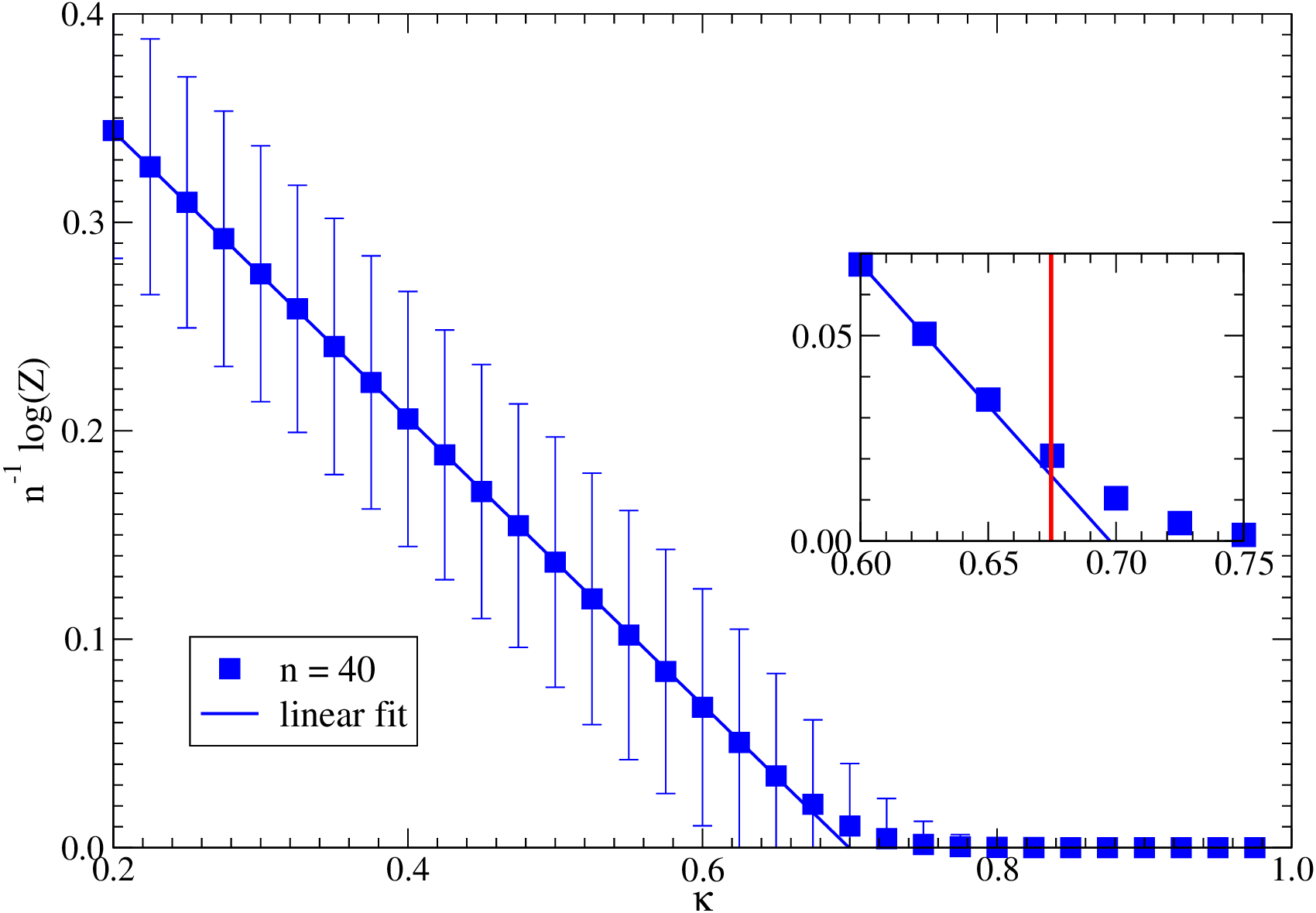}
  \caption{Logarithm of the number of perfect partitions in random
    instances of the constrained partitioning problem with $b=0.25$.
    Symbols are averages over
    $10^3$ random samples, errorbars indicate $\pm1$ standard deviation.
    Inset: Magnification of the ``critical'' region, the vertical line
is
    $\kappa_-$.
  \label{fig:logz-b-k}}
\end{figure}

Figure \ref{fig:logz-b-k} shows the result of a simulation with
$n=40$. The mean value of $n^{-1}\log Z$  is very well
approximated by a piecewise linear function
\begin{equation}
  \label{eq:logZ-linear}
  n^{-1}\cdot\ex\big(\log Z(\kappa)\big) = \left\{
    \begin{array}{cc}
      (\tilde{\kappa} - \kappa)\log 2 & \kappa < \tilde{\kappa} \\
      0                               & \kappa > \tilde{\kappa}
    \end{array}
  \right.
\end{equation}
where $\tilde{\kappa}$ is a fit parameter that depends on $n$.
Eq.~\ref{eq:logZ-linear} is a very accurate description of the
numerical data even for rather small values of $n$.  The value of
$n$ influences only the size of the fluctuations of
$n^{-1}\cdot\log Z(\kappa)$ around its expected value and the
value of $\tilde{\kappa}$.  As $n$ becomes larger, the
fluctuations decrease and $\tilde{\kappa}$ increases, moving
beyond $\kappa_-$ and towards $\kappa_c$ (see Figure
\ref{fig:kappa-b-n}). Note that Theorem~\ref{thm5.1} implies that
$\lim_{n\to\infty} n^{-1}\log Z$ is linear function of $\kappa$
for $\kappa < \kappa_-$.

\begin{figure}[htb]
  \centering \includegraphics[width=0.75\columnwidth]{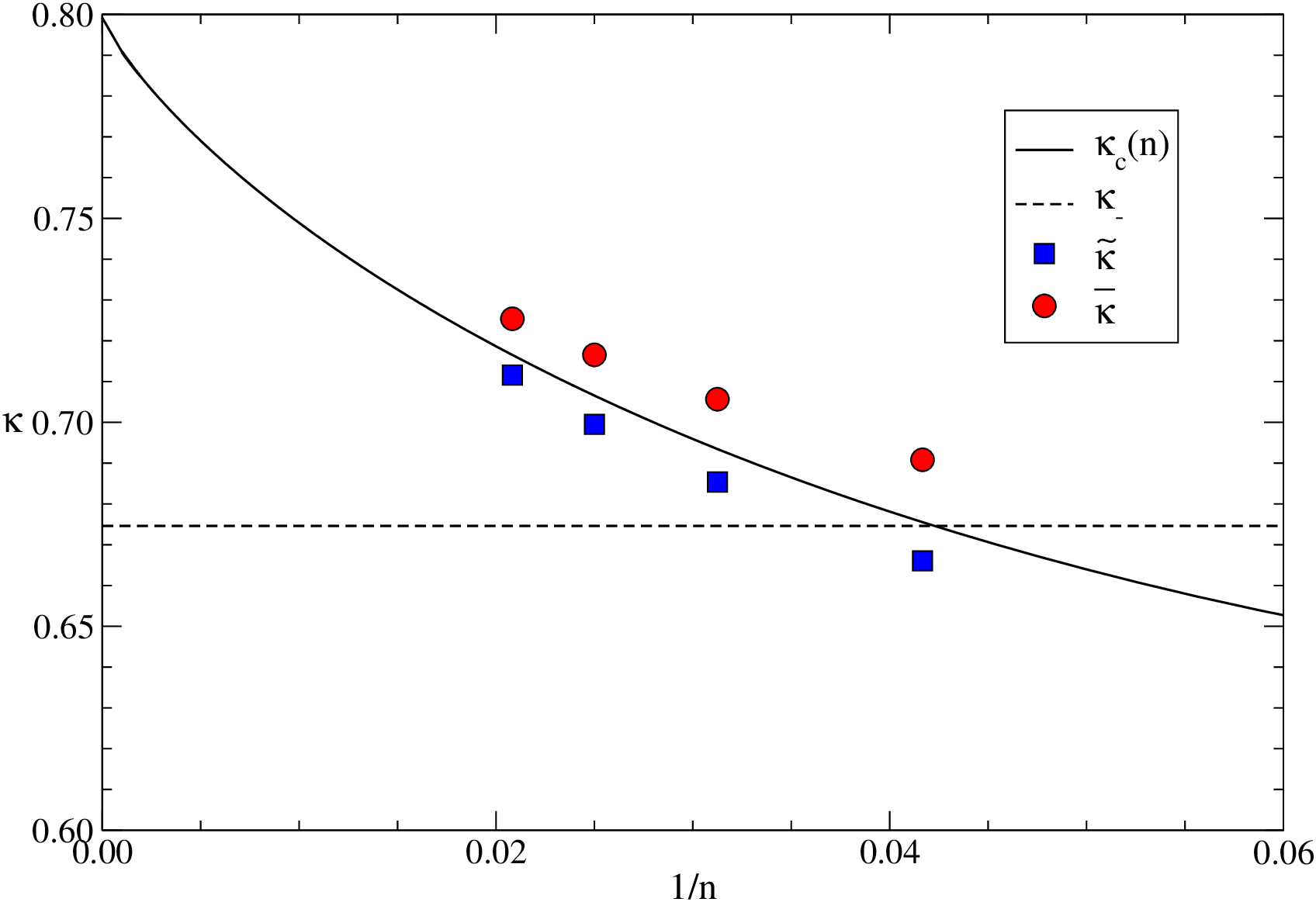}
  \caption{Values of $\kappa$ that indicate a transition from the perfect
  to the hard phase.
  \label{fig:kappa-b-n}}
\end{figure}

Of course, the numerical data on $\tilde{\kappa}$ and
$\overline{\kappa}$ do not allow us to conclude  that both values will
converge to $\kappa_c$ as $n\to\infty$, but it is obvious that
both values increase with increasing $n$ and are larger than
$\kappa_-$.  Again we can use Theorem~\ref{thm5.1} to estimate the
finite-size corrections to $\kappa_c$,
\begin{equation}
  \label{eq:kappac-finite}
  \kappa_c(n) = \frac{L(\zeta,\eta)}{\log 2} -
  \frac{\log\Big(n\pi\sqrt{\det R}\Big)}{n\log 2}
  - \frac{\mbox{Tr} R^{-1}K}{4n\log 2}.
\end{equation}
Figure \ref{fig:kappa-b-n} shows that both $\overline{\kappa}$ and
$\tilde{\kappa}$ are close to the finite-size estimate of
$\kappa_c$.

In all our simulations, we did not find any trace of a critical
line $\kappa_-$. The properties of the system do change for values
below $\kappa_c$, but above $\kappa_-$, but there is some evidence
that this is a finite-size effect and that for really large
systems, it is only $\kappa_c$ that matters.

\bibliographystyle{plain}
\bibliography{intpartbib}

\begin{thebibliography}{10}

\bibitem{tina}
See http://tina.nat.uni-magdeburg.de for a description of the 156 cpu selfmade
  parallel computer {TINA}.

\bibitem{trng}
Heiko Bauke.
\newblock {TRNG} - a portable random number generator for parallel computing.
\newblock http://tina.nat.uni-magdeburg.de/{TRNG}.

\bibitem{BCP2}
C.~Borgs, J.T. Chayes, and B.~Pittel.
\newblock Phase transition and finite-size scaling for the integer partitioning
  problem.
\newblock {\em Rand.~Struc.~Alg.}, 19:247--288, 2001.

\bibitem{BCP1}
C.~Borgs, J.T. Chayes, and B.~Pittel.
\newblock Sharp threshold and scaling window for the integer partitioning
  problem.
\newblock {\em Proc. $33^{\text{rd}}$ ACM Symp. on Theor. of Comp.}, pages
  330--336, 2001.

\bibitem{Y}
B.Yakir.
\newblock The differencing algoritm $\text{LDM}$ for partitioning; a proof of a
  conjecture of $\text{Karmakar and Karp}$.
\newblock {\em Math. of Operations Res.}, 21:85--99, 1996.

\bibitem{Der}
B.~Derrida.
\newblock Random-energy model: An exactly solvable model of disordered systems.
\newblock {\em Phys.~Rev.~B (3)}, 24:2613--2626, 1981.

\bibitem{FF1}
F.F. Ferreira and J.F. Fontanari.
\newblock Probabilistic analysis of the number partitioning problem.
\newblock {\em J.~Phys.~A: Math.~Gen.}, 31:3417--3428, 1998.

\bibitem{FF2}
F.F. Ferreira and J.F. Fontanari.
\newblock Statistical mechanics analysis of the continuous number partitioning
  problem.
\newblock {\em Physica A}, 269:54--60, 1999.

\bibitem{Fu}
Y.~Fu.
\newblock The use and abuse of statistical mechanics in computational
  complexity.
\newblock In {\em Lectures in the Science of Complexity; Proceedings of the
  1988 Complex Systems Summer School, Santa Fe, New Mexico, 1988, edited by
  D.L.~Stein}. Addison-Wesley, Reading, MA, 1989.

\bibitem{GW}
I.P. Gent and T.~Walsh.
\newblock In {\em Proc.~of the 12th European Conference on Artificial
  Intelligence, Budapest, Hungary, 1996, edited by W.~Wahlster}, pages
  170--174. John Wiley {\&} Sons, New York, NY, 1996.

\bibitem{GrStirz}
G.R. Grimmett and D.R. Stirzaker.
\newblock {\em Probability and Random Processes}.
\newblock Oxford University Press, 1992.

\bibitem{Hay}
B.~Hayes.
\newblock The easiest hard problem.
\newblock {\em American Scientist}, 90:113--117, 2002.

\bibitem{horowitz:sahni:74}
E.~Horowitz and S.~Sahni.
\newblock Computing partitions with applications to the {K}napsack problem.
\newblock {\em Journal of the {ACM}}, 21(2):277--292, 1974.

\bibitem{KK}
N.~Karmarkar and R.M. Karp.
\newblock The differencing method of set partitioning.
\newblock Technical Report UCB/CSD 82/113, Computer Science Division (EECS),
  University of California, Berkeley, 1982.

\bibitem{KKLO}
N.~Karmarkar, R.M. Karp, G.S. Lueker, and A.M. Odlyzko.
\newblock Probabilistic analysis of optimum partitioning.
\newblock {\em J.~Appl.~Prob.}, 23:626--645, 1986.

\bibitem{Lue}
G.S. Lueker.
\newblock Exponentially small bounds on the expected optimum of the partition
  and subset sum problem.
\newblock {\em Rand.~Struc.~Alg.}, 12:51--62, 1998.

\bibitem{Mer1}
S.~Mertens.
\newblock Phase transition in the number partitioning problem.
\newblock {\em Phys.\ Rev.\ Lett.}, 81:4281--4284, 1998.

\bibitem{Mer2}
S.~Mertens.
\newblock Random costs in combinatorial optimization.
\newblock {\em Phys.\ Rev.\ Lett.}, 84:1347--1350, 2000.

\end{thebibliography}

\end{document}